\documentclass[structabstract]{aa}  
\usepackage{graphicx}
\usepackage{txfonts}
\usepackage{natbib}
\usepackage{longtable}
\usepackage{longtable,lscape}
\begin{document}
   \title{FORS2/VLT survey of Milky Way globular
     clusters}
   \subtitle{I. Description of the method for derivation of metal
     abundances in the optical and application to NGC~6528, NGC~6553, M~71, NGC~6558,
     NGC~6426 and Terzan~8 \thanks{Based on Observations collected at
       the European Southern Observatory/Paranal, Chile, under
       programmes 077.D-0775(A) and 089.D-0493(B).}}

   \author{B. Dias\inst{1,2}
          \and B. Barbuy\inst{1}
         \and I. Saviane\inst{2}
          \and E. V. Held\inst{3}
          \and G. S. Da Costa\inst{4}
          \and S. Ortolani\inst{3,5}
          \and S. Vasquez\inst{2,6}
          \and M. Gullieuszik\inst{3}
          \and D. Katz\inst{7}
%          \fnmsep
          }
   \institute{Universidade de S\~ao Paulo, Dept. de Astronomia, Rua do Mat\~ao 
     1226, S\~ao Paulo 05508-090, Brazil\\ \email{bdias@astro.iag.usp.br}
         \and European Southern Observatory, Alonso de Cordova 3107, Santiago, Chile
         \and INAF, Osservatorio Astronomico di Padova, Vicolo dell'Osservatorio 5,
35122 Padova, Italy
         \and Research School of Astronomy \& Astrophysics, Australian National University, Mount Stromlo Observatory,
    via Cotter Road, Weston Creek, ACT 2611, Australia 
         \and Universit\`a di Padova, Dipartimento di Astronomia, Vicolo
 dell'Osservatorio 2, 35122 Padova, Italy
         \and Instituto de Astrofisica, Facultad de Fisica, Pontificia Universidad Catolica de Chile, 
Casilla 306, Santiago 22, Chile
         \and  GEPI, Observatoire de Paris, CNRS, Universit\'e Paris
         Diderot, 5 Place Jules Janssen 92190 Meudon, France
             }

   \date{Received: ; accepted: }

% \abstract{}{}{}{}{} 
% 5 {} token are mandatory
 
  \abstract
  % context heading (optional)
  % {} leave it empty if necessary  
  {We have observed almost 1/3 of the globular clusters in the Milky
    Way, targeting distant and/or highly reddened objects, besides a few
    reference clusters.
A large sample of red giant stars was observed with FORS2@VLT/ESO at
R$\sim$2,000. The method for derivation of stellar parameters is presented
 with application to six reference clusters.} 
  % aims heading (mandatory)
  {We aim at deriving the stellar parameters effective
    temperature, gravity, metallicity and alpha-element enhancement,
    as well as radial velocity, for membership confirmation of
    individual stars in each cluster.  We analyse the spectra
    collected for the reference globular clusters NGC~6528
    ([Fe/H]$\sim$-0.1), NGC~6553 ([Fe/H]$\sim$-0.2), M~71
    ([Fe/H]$\sim$-0.8), NGC~6558 ([Fe/H]$\sim$-1.0), NGC~6426
    ([Fe/H]$\sim$-2.1) and Terzan~8 ([Fe/H]$\sim$-2.2). They
    cover the full range of globular cluster metallicities, and are
    located in the bulge, disc and halo.}
  % methods heading (mandatory)
   {Full spectrum fitting techniques are applied, by comparing each
     target spectrum with a stellar library in the optical region
     at 4560-5860 {\rm \AA}. We employed the library of observed
     spectra MILES,
     and the synthetic library by Coelho et
     al. (2005).
     Validation of the method is achieved through recovery of the known
     atmospheric parameters for 49 well-studied stars that cover a
     wide range in the parameter space. We adopted as final
         stellar parameters (effective temperatures, gravities,
         metallicities) the average of results using MILES and Coelho
         et al. libraries. }
  % results heading (mandatory)
      {We identified 4 member stars in NGC~6528, 13 in NGC~6553, 
      10 in M~71, 5 in NGC~6558, 5 in NGC~6426 and 12 in Terzan~8. 
      Radial velocities, T$_{\rm eff}$, log($g$), [Fe/H] and
      alpha-element enhancements were derived. 
We derived $<$v$_{\rm helio}$$>$ = -242$\pm$11~km/s, [Fe/H] = -2.39$\pm$0.04,
   [Mg/Fe] = 0.38$\pm$0.06 for NGC~6426 from spectroscopy for
   the first time.} 
  % conclusions heading (optional), leave it empty if necessary 
    {The method proved to be reliable for red giant stars observed with
   resolution R$\sim$2,000, yielding results compatible with
     high-resolution spectroscopy. The derived $\alpha$-element
     abundances show [$\alpha$/Fe] vs. [Fe/H] consistent with that of field
     stars at the same metallicities.}


   \maketitle
%
%________________________________________________________________

\section{Introduction}

The derivation of stellar metallicities and abundances are best
defined when based on high spectral resolution and high
signal-to-noise (S/N) data. \cite{cayrel88} showed that higher
resolution carries more information than higher S/N. Such kind of data
require however substantial telescope time. For this reason, very
large samples of stellar spectra have been gathered in recent years,
or are planned to be collected in the near future, with multi-object
low and medium-resolution instruments. A few examples are the Sloan
Digital Sky Survey (SDSS, \citealp{york+00}), at a resolution
R$\sim$1800, the Radial Velocity Experiment survey (RAVE,
\citealp{steinmetz+06}) of R$\sim$7500 in the CaT region, and other
large ongoing surveys such as LAMOST at the Guoshoujing telescope
\citep[GSJT,][]{wu+11} of R$\sim$2,000, and future ones such as GAIA
\citep{perryman+01}. Large data sets of low/medium-resolution spectra
are reachable for extragalactic stars, such as presented in \cite{kirby+09}.
 A few recent surveys are able to use medium/high-resolution spectra
 focused on specific targets such as provided by the 
APOGEE \citep[R$\sim$22,500,][]{meszaros+13}, GAIA-ESO using the
FLAMES-GIRAFFE spectrograph (R$\sim$22,000)  at the Very Large
Telescope \citep[VLT,][]{gilmore+12} and HERMES \citep[R$\sim$28,000
or 45,000,][]{wylie+10} at the AAT. More complete reviews of
available, ongoing and future surveys, as well as automated methods
for stellar parameter derivation can be found in \cite{allende+08},
\cite{lee+08}, \cite{koleva+09}, \cite{meszaros+13}, and \cite{wu+11},
among others.

In most analyses of medium to low-resolution spectra, the 
least squares ($\chi^2$ minimization), or ``euclidian distance'', also called 
minimum distance method, such as 
Universit\'e de Lyon Spectroscopic Analysis Software
\citep[ULySS,][]{koleva+09}, and the k-means clustering
described in \cite{sanchez-almeida+13}, are employed.

In the present work we analyse spectra in the optical,
in the range 4560-5860 {\rm \AA}, obtained at the FORS2/VLT
at a  resolution R$\sim$2,000, carrying out
 full spectrum fitting.
This spectral region, in particular from  H$_{\beta}$ to Na I lines, is 
sensitive to metallicity and temperature, to
gravity due to MgH molecular bands (as part of the Mg$_2$ index), 
and it includes the Lick indices
Fe5270, Fe5335 and Mg$_2$, that are usual Fe and Mg abundance indicators 
(\citealp{katz+11}; \citealp{cayrel+91}; 
\citealp{faber+85}; \citealp{worthey+94}). 

The same sample was observed in the near-infrared (CaT), as presented in
  \cite{saviane+12}, \cite{dacosta+09} and Vasquez et al. (in prep),
where two among the triplet Ca II lines were used to derive velocities and
metallicities. A comparison of their results with the present ones
show good consistency, as will be discussed in the present paper.

In this work we study six reference globular clusters, spanning
essentially the full range of metallicities of globulars:
the metal-poor halo clusters \object{NGC~6426} and \object{Terzan~8}
([Fe/H]$\sim$-2.1 and -2.2, respectively), the moderately metal-poor
\object{NGC~6558}  ([Fe/H]$\sim$-1.0) in the bulge, the template ``disc'' metal-rich cluster  
\object{M~71} (\object{NGC~6838}, [Fe/H]$\sim$-0.7), and the metal-rich bulge clusters
 \object{NGC~6528} and \object{NGC~6553} ([Fe/H]$\sim$-0.1 and -0.2, respectively).
 
These reference clusters are analysed with the intent of testing
and improving the method, and verifying the metallicity range of 
applicability of each library of template spectra. 
In all cases, member stars and surrounding field
stars are analysed. For some of these clusters previous
high-resolution spectroscopic and photometric data of a few member
stars are available. 

The minimum distance method was adopted by \cite{cayrel+91},
 by measuring residuals in each of the stellar parameters
effective temperature, gravity, and metallicity; the method required
the input of reference parameters. In the present work, we
adopt the code ETOILE \citep{katz+11} that uses the minimum distance
method, where the reliability and coverage of T$_{\rm eff}$, log($g$), [Fe/H],
[$\alpha$/Fe] of the template stars are important to find well-founded
parameters for the target stars.
We adopted two different libraries of spectra, the MILES\footnote{http://miles.iac.es/}
library of low-resolution spectra (R$\sim$2,000) and the grid of
synthetic spectra by
\cite{coelho+05}\footnote{http://www.mpa-garching.mpg.de/PUBLICATIONS/DATA/
SYNTHSTELLIB/synthetic\_stellar\_spectra.html}.
 
In Sect. 2 the observations are described. In Sect. 3 the method of
stellar parameter derivation is detailed. In Sect. 4 the method is
applied to six cluster as a validation of the procedures. In Sect. 5 the
results are discussed, and in Sect. 6 a summary is given.

%__________________________________________________________________
\section{Observations and data reduction}

\begin{figure*}[!htb]
\centering
\includegraphics[width=0.33\textwidth]{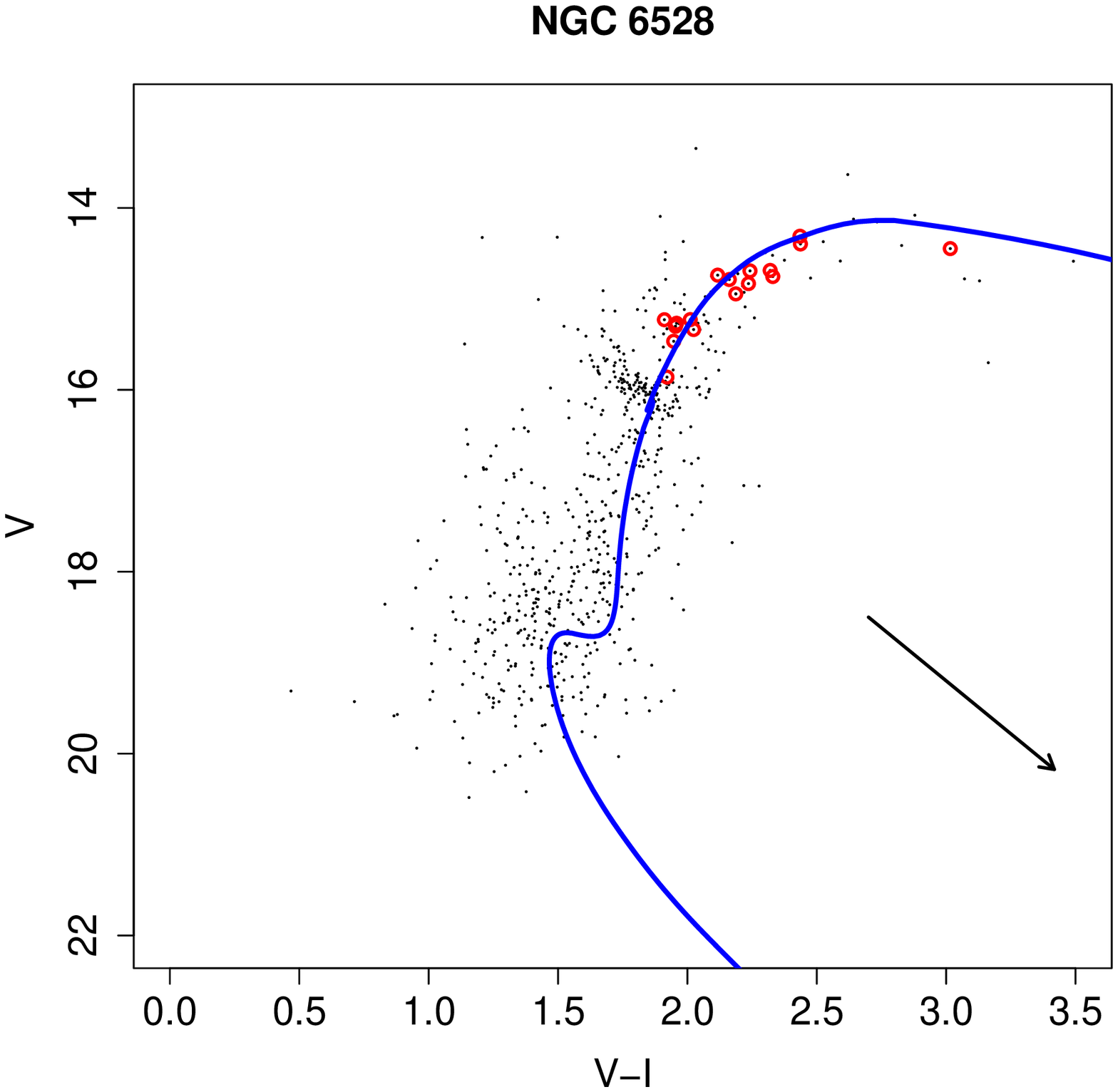}
\includegraphics[width=0.33\textwidth]{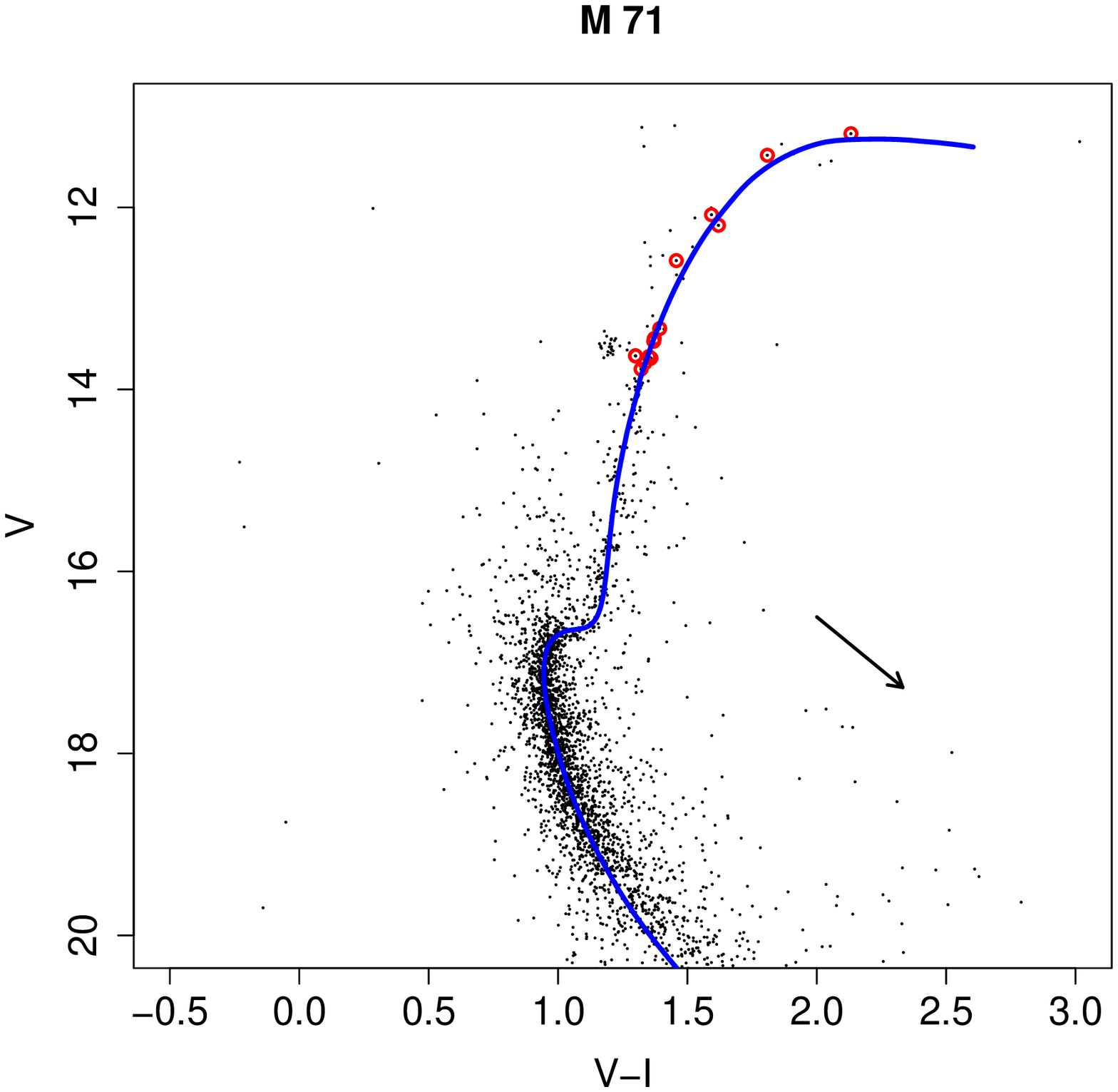}
\includegraphics[width=0.33\textwidth]{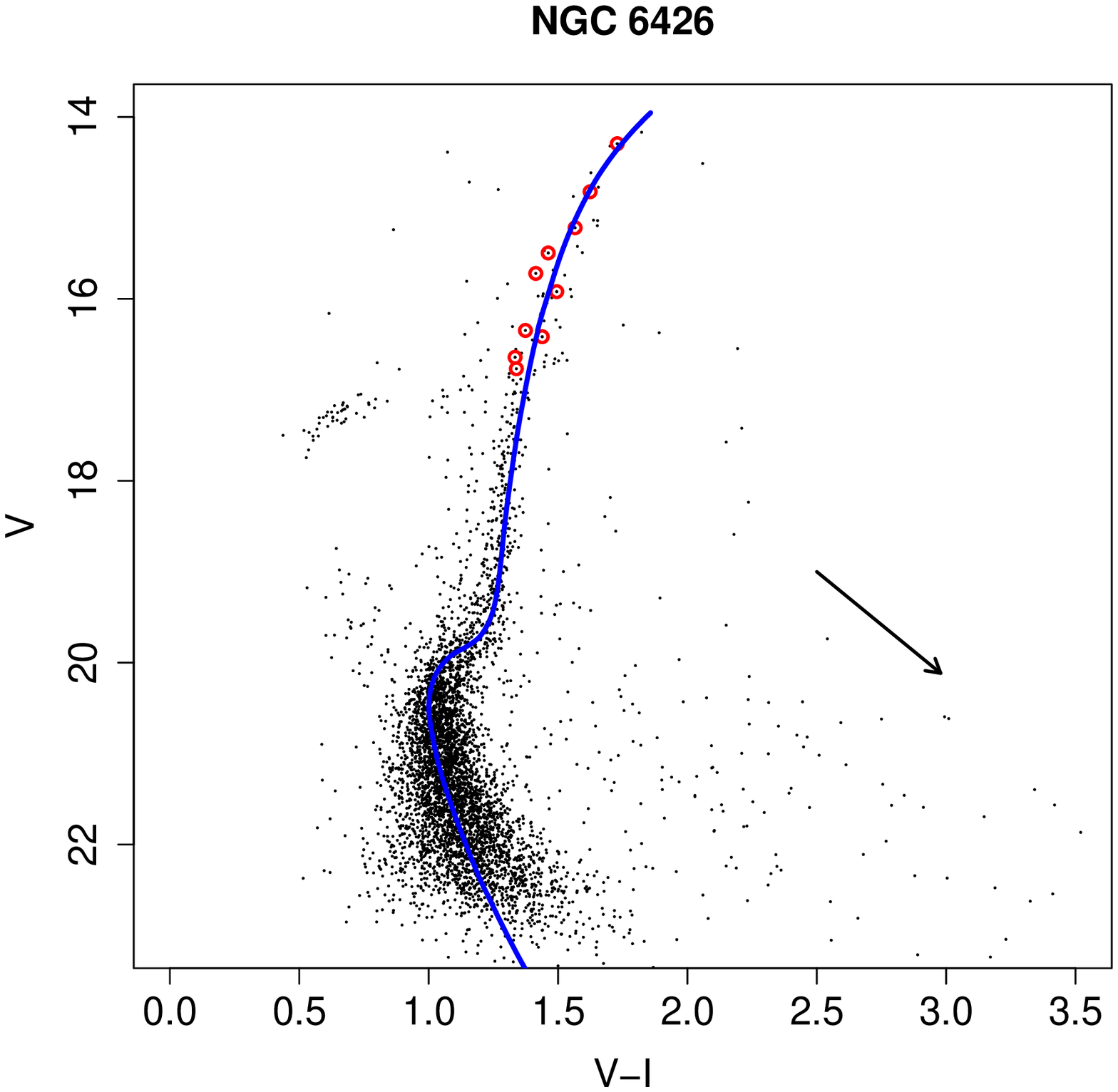}
\includegraphics[width=0.33\textwidth]{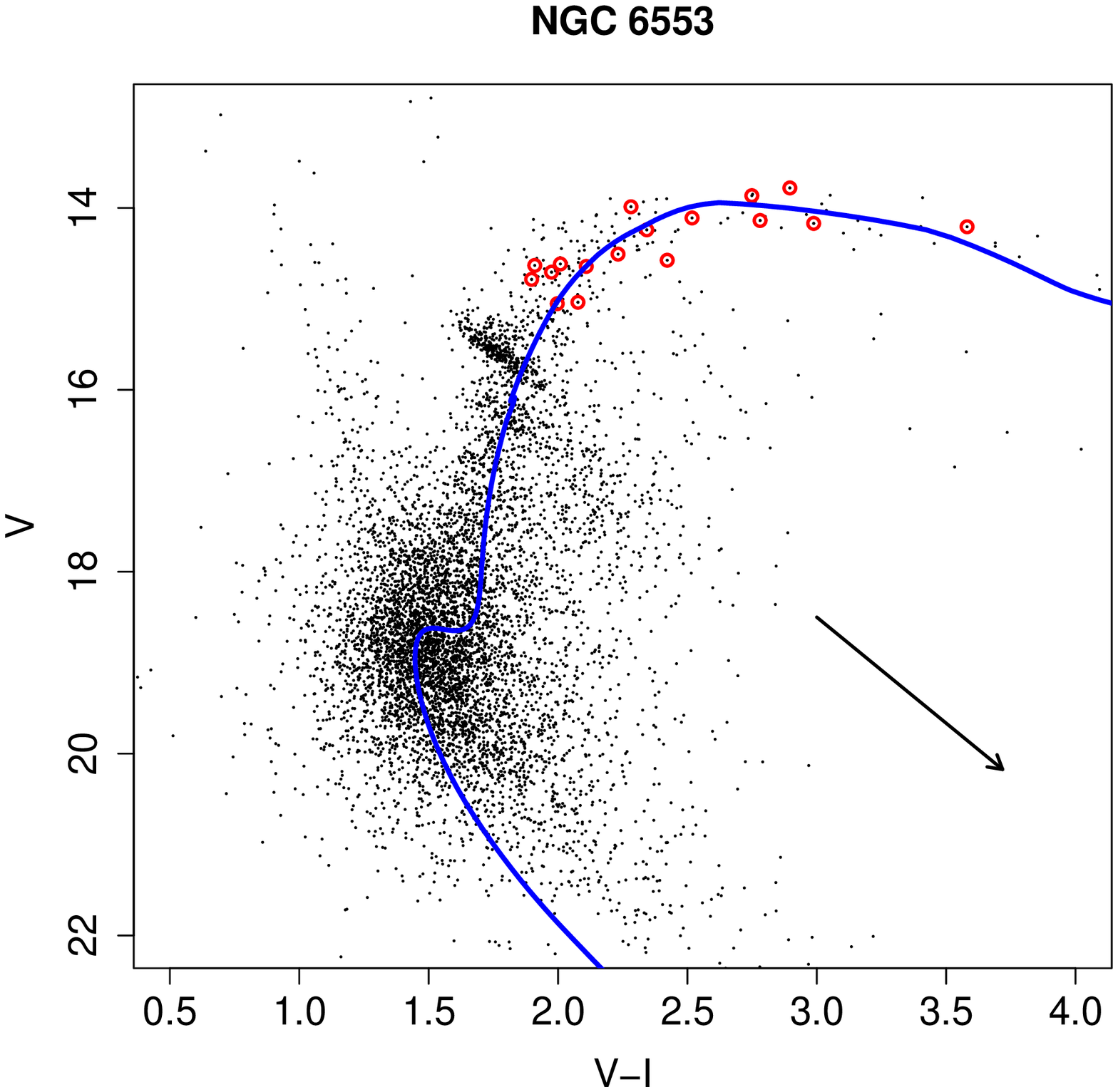}
\includegraphics[width=0.33\textwidth]{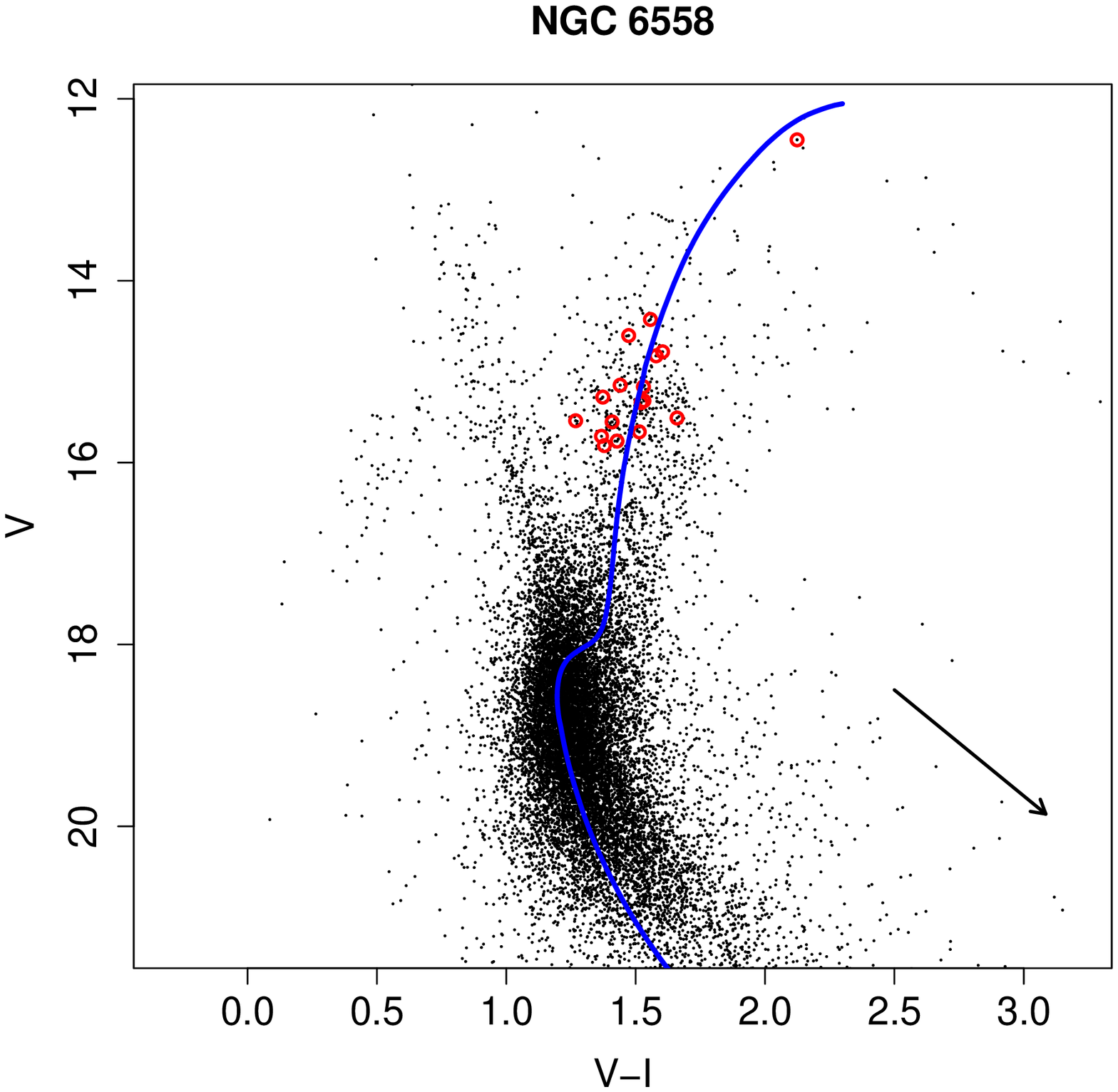}
\includegraphics[width=0.33\textwidth]{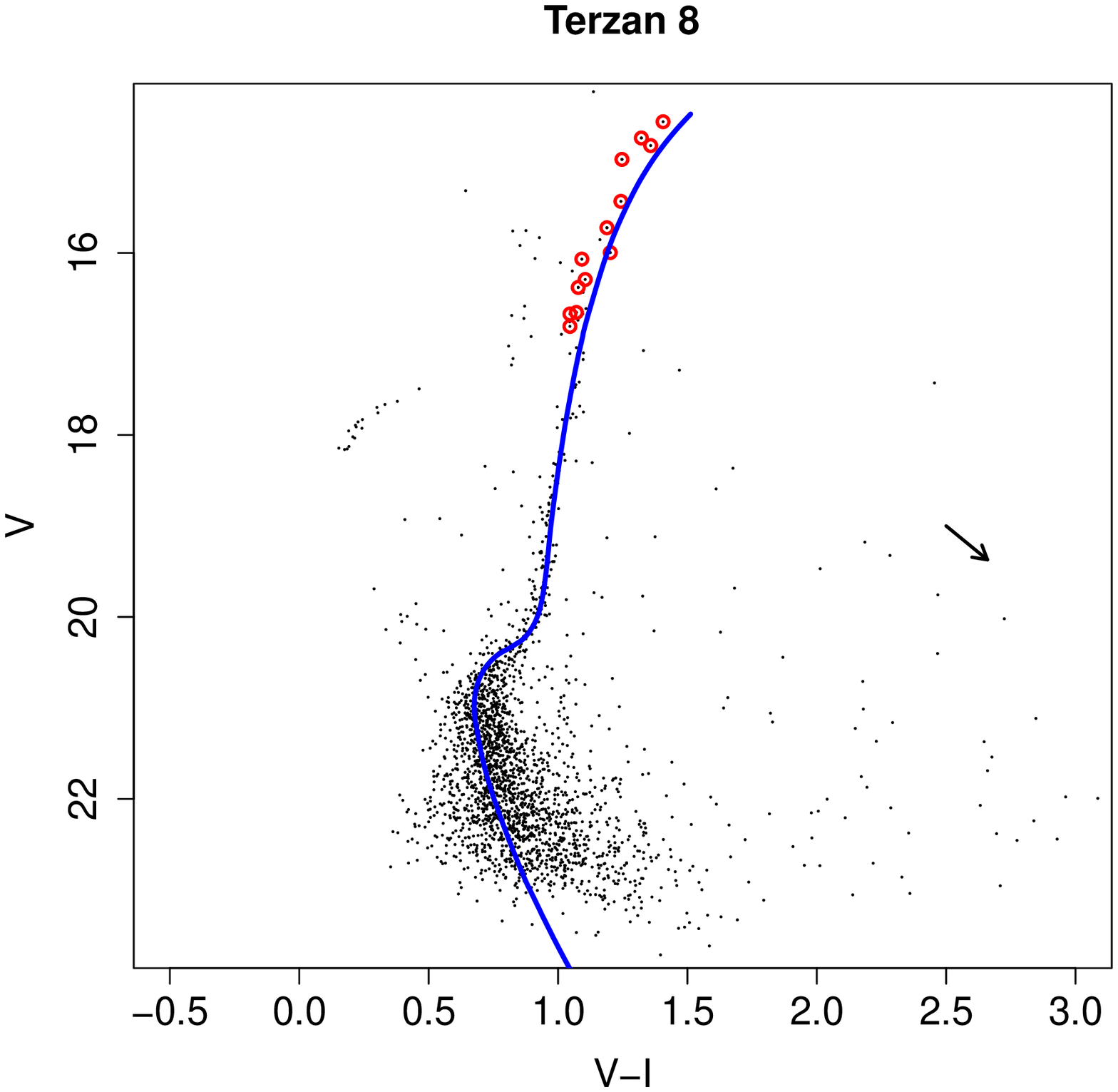}
\caption{Colour-Magnitude diagrams of all clusters analysed in the
  present work. Left panels are metal-rich clusters, middle panels are
  intermediate metallicity clusters and right panels correspond to the
  more metal-poor ones. All stars within 2~$\times$~r$_{half-light}$ are
  plotted, without any cleaning procedure. Dartmouth isochrones with literature
  parameters (Table \ref{clusterparam}) are overplotted.
  Selected RGB stars for spectroscopic observations are in
  red. Reddening vectors are shown in each CMD based on E(B-V) listed in
  Table \ref{clusterparam}.}
\label{cmds}
\end{figure*}

\begin{figure*}[!htb]
\centering
\includegraphics[width=0.33\textwidth]{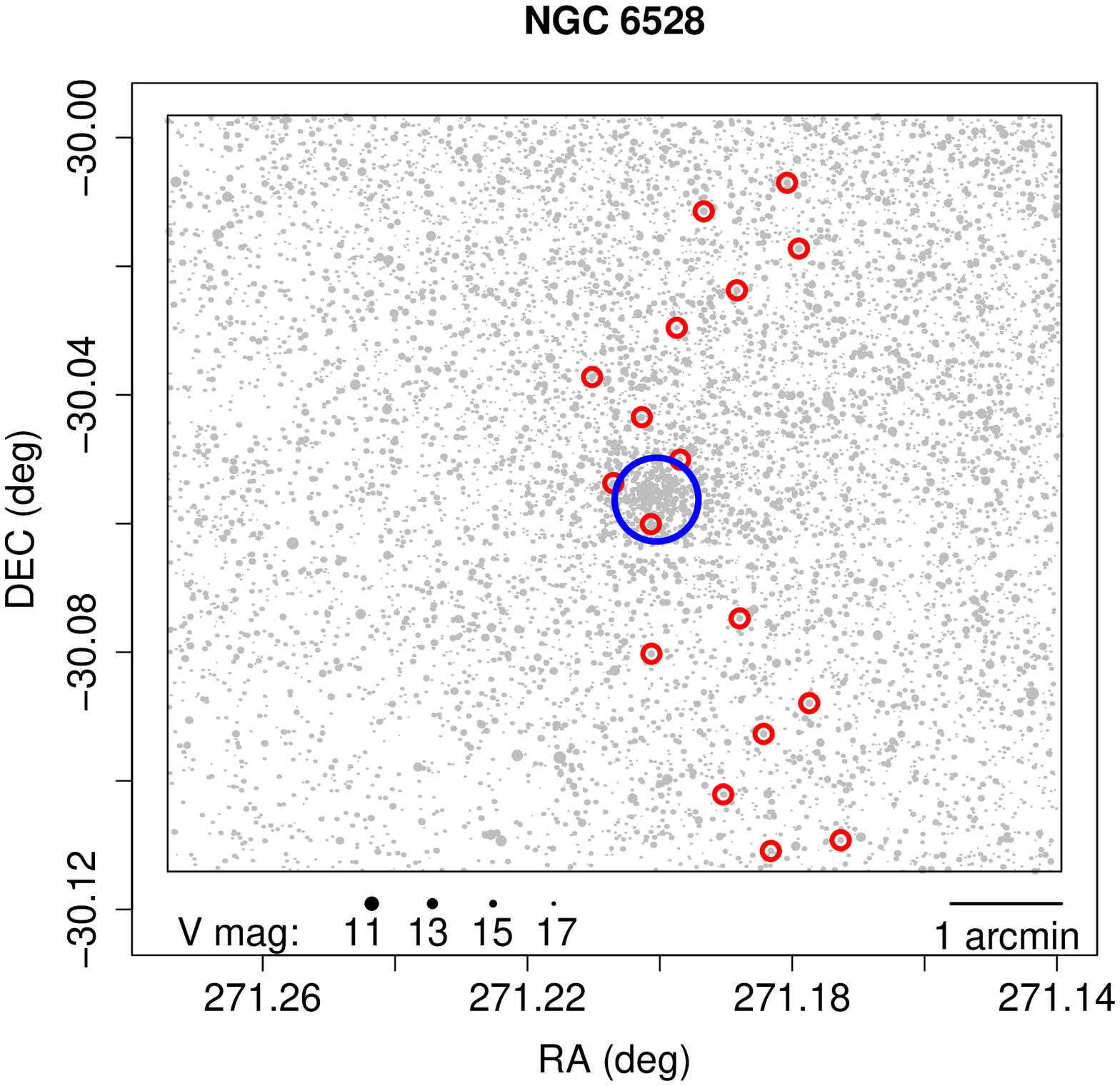}
\includegraphics[width=0.33\textwidth]{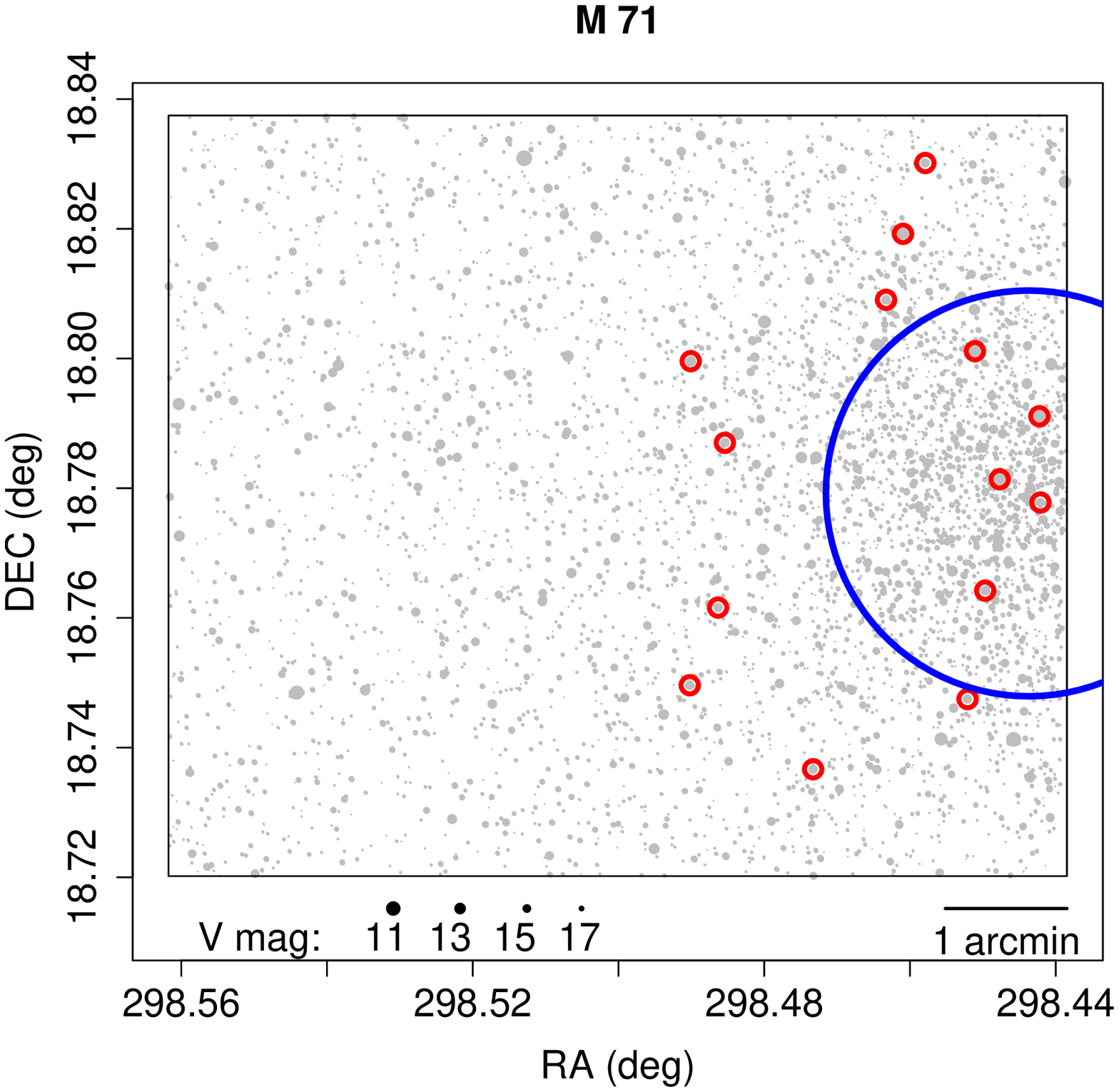}
\includegraphics[width=0.33\textwidth]{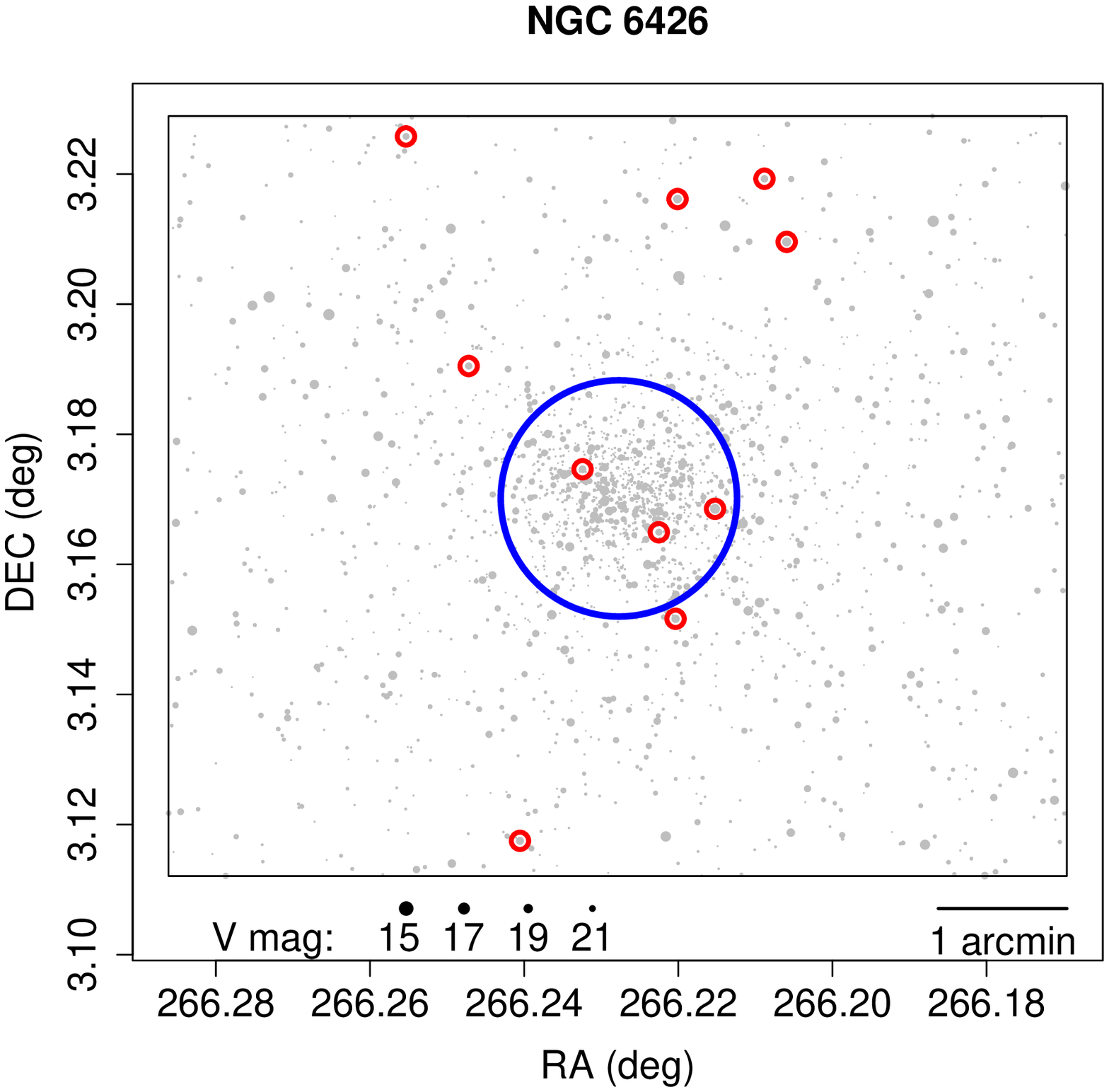}
\includegraphics[width=0.33\textwidth]{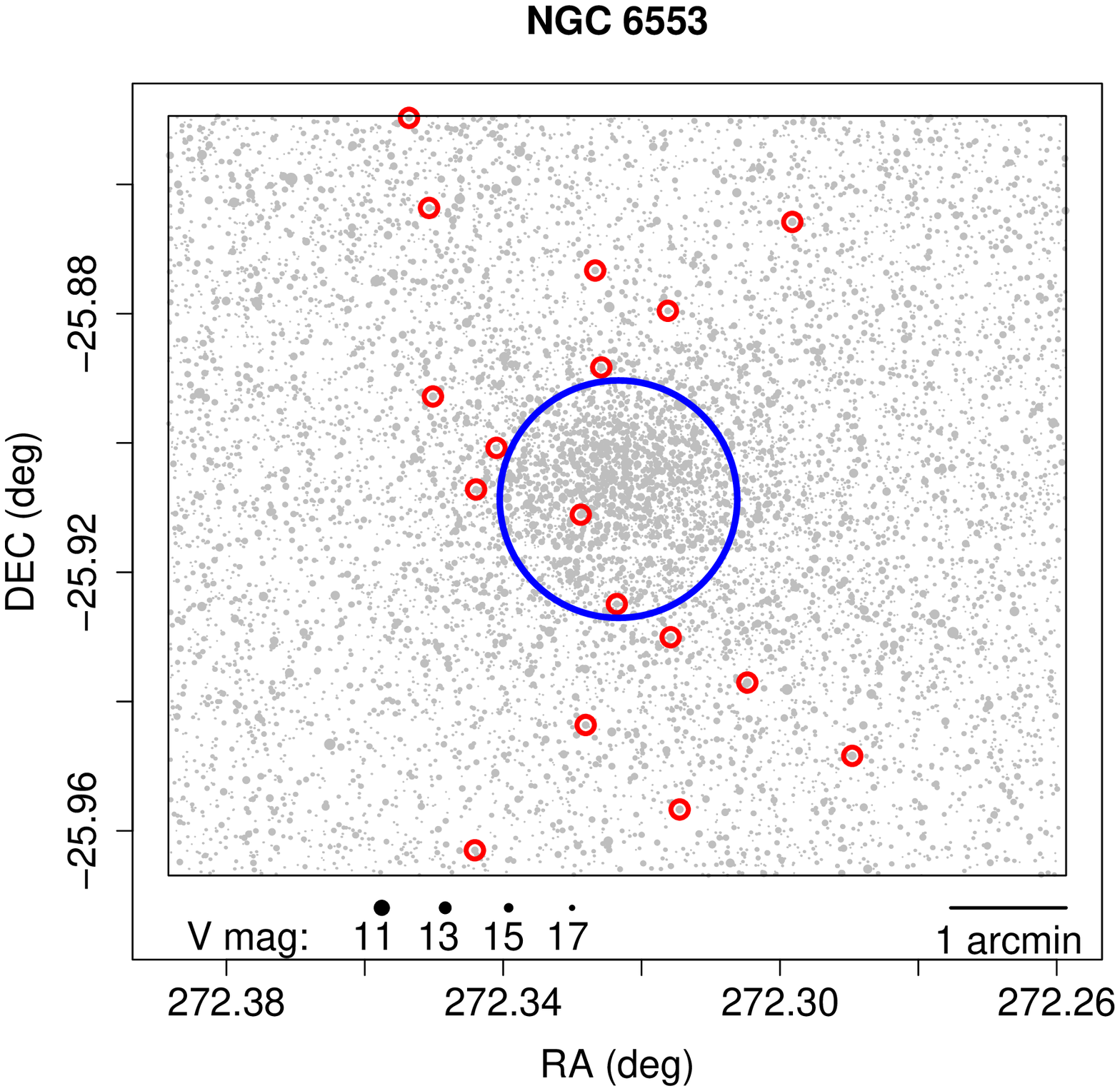}
\includegraphics[width=0.33\textwidth]{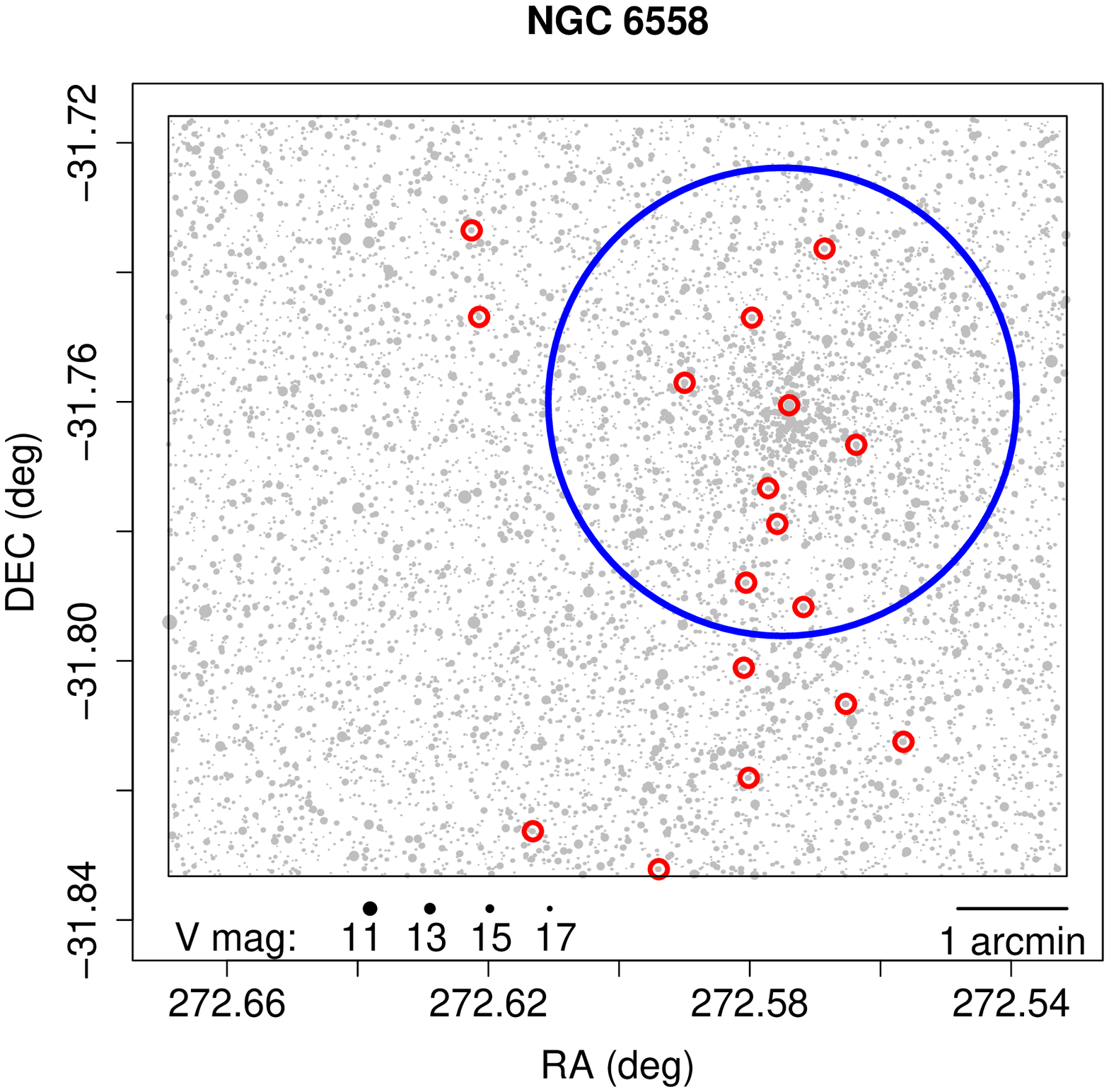}
\includegraphics[width=0.33\textwidth]{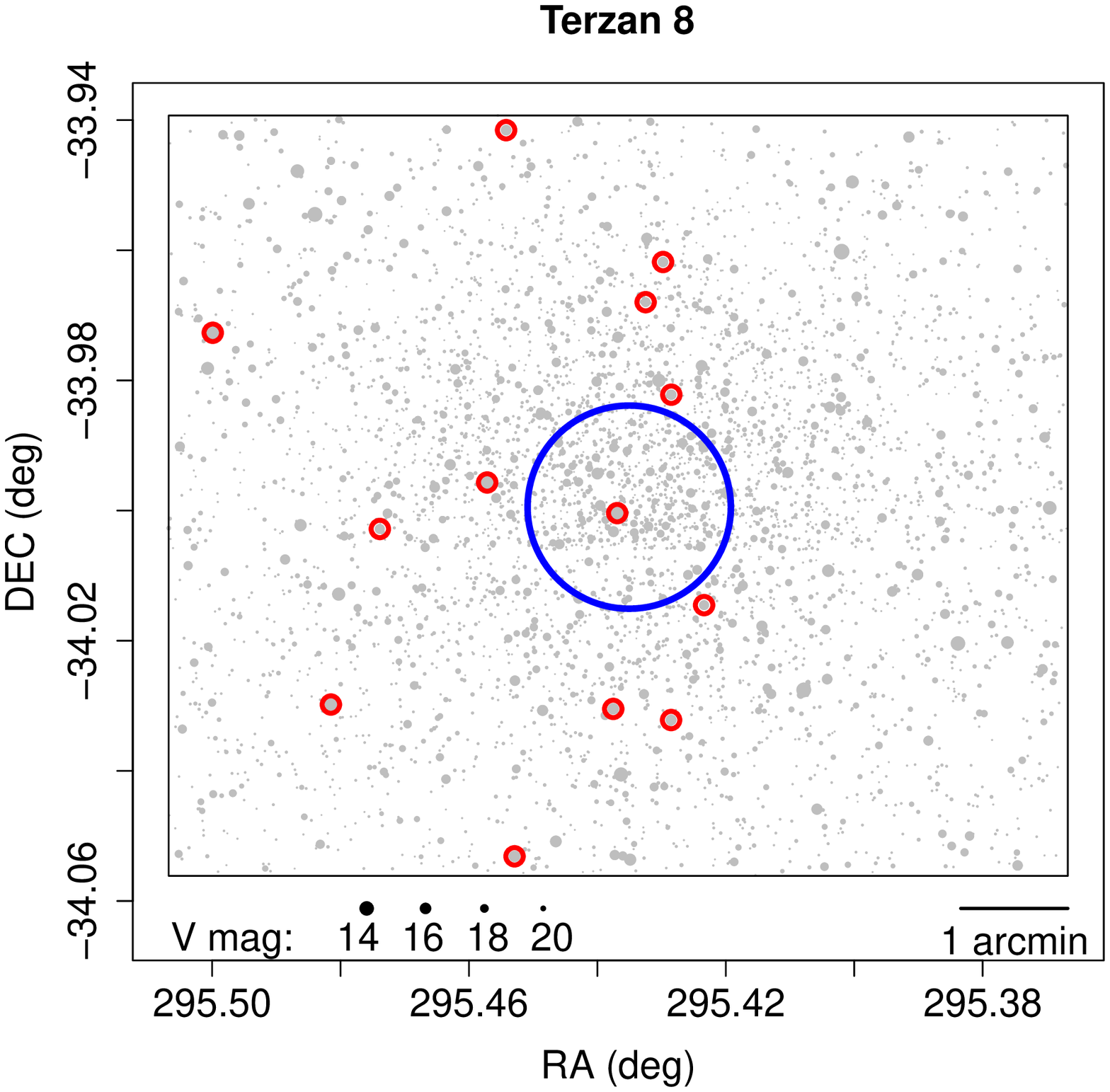}
\caption{Sky map of all clusters analysed in the
  present work. Panels are displayed as in Figure \ref{cmds}. Only
  brightest stars are shown, and the size of the dots are scaled with
  the stars magnitudes as indicated in each plot. Selected RGB stars
  for spectroscopic observations are in red. The blue circle corresponds to the half-light
  radius of each cluster from Table \ref{clusterparam}.}
\label{skymaps}
\end{figure*}

We observed respectively 17, 17, 12, 17, 10 and 13 red giant stars of
the globular clusters NGC 6528, NGC~6553, M~71, NGC~6558, NGC~6426,
Terzan 8, and surrounding fields, using FORS2@VLT/ESO (\cite{appenzeller+98}, 
under projects
077.D-0775(A) and 089.D-0493(B).
Table \ref{log} summarizes the setup of the
observations. Pre-images were taken using filters Johnson-Cousins V
and I in order to select only stars in the red giant branch (RGB)
brighter than the Red Clump (RC) level. Zero points in colours and
magnitudes were fitted to match isochrones with
parameters from Table \ref{clusterparam} (see Colour-Magnitude
Diagrams, CMDs, in Figure \ref{cmds}). 
We selected stars covering the whole interval in colour of the RGB, 
and when possible trying to avoid Asymptotic Giant Branch (AGB) stars. 
These stars are spatially distributed as shown in Figure
\ref{skymaps}, partly due to the slitlet configuration.
Cluster parameters and log of observations are given
in Table \ref{clusterparam}. The list of individual stars, their coordinates
and $VI$ magnitudes from the present FORS2 observations
 are given in Table \ref{starinfo}.

\begin{table}[!htb]
  \caption{Telescope and spectrograph}
\label{log}
  \centering
  \begin{tabular}{l|c}
\hline\noalign{\smallskip}
\multicolumn{2}{c}{Observing information} \\
\noalign{\smallskip}\hline \hline \noalign{\smallskip}
Telescope & Antu/UT1-VLT@ESO \\
Instrument & FORS2 \\
Grism & 1400V \\
FoV & $6\farcm8 \times 6\farcm8$\\
Pixel scale & 0.25$\arcsec$/pixel\\
Slit width & 0.53 mm \\
Spec. resolution  & 2,000 \\
Dispersion  & 0.6 {\rm \AA}/pix \\
\noalign{\smallskip}\hline
  \end{tabular}  
\end{table}

\begin{table*}[!htb]
  \caption{Log of observations and clusters parameters from literature. Main reference is \citet[2010 edition]{harris96}, when not indicated explicitly.}
\label{clusterparam}
  \centering
\scriptsize
  \begin{tabular}{c||c|c||c|c||c|c}
\noalign{\smallskip} \hline
Parameter     & NGC~6528                                                & NGC~6553                                                 & M~71                                                        & NGC~6558                                                & NGC~6426                                               & Terzan~8                                                 \\
\hline \hline \noalign{\smallskip}
Date of obs. &  29.05.2006                                                & 29.05.2006                                              & 29.05.2006                                                & 29.05.2006                                                & 13.07.2012                                             & 12.07.2012                             \\
UT                & 08:36:22                                                     & 08:57:50                                                     & 09:14:32                                                   & 06:55:32                                                    & 02:31:12                                                   & 07:47:29.346                                             \\
$\tau$          & 149.4 s                                                      & 79.4 s                                                        & 17.2 s                                                        & 148.3 s                                                      & 500.0 s                                                          & 360 s                                                         \\
\noalign{\smallskip} \hline \noalign{\smallskip}
RA                & 18$^h$ 04$\arcmin$ 49.64$\arcsec$       &  18$^h$ 09$\arcmin$ 17.60$\arcsec$       &  19$^h$ 53$\arcmin$ 46.49$\arcsec$     &  18$^h$ 10$\arcmin$ 17.60$\arcsec$      &  17$^h$ 44$\arcmin$ 54.65$\arcsec$        &  19$^h$ 41$\arcmin$ 44.41$\arcsec$   \\
DEC              & -30$^{\circ}$ 03$\arcmin$ 22.6$\arcsec$  &  -25$^{\circ}$ 54$\arcmin$ 31.3$\arcsec$ & +18$^{\circ}$ 46$\arcmin$ 45.1$\arcsec$ & -31$^{\circ}$ 45$\arcmin$ 50.0$\arcsec$ & +03$^{\circ}$ 10$\arcmin$ 12.5$\arcsec$   & -33$^{\circ}$ 59$\arcmin$ 58.1$\arcsec$ \\
age               & 13 Gyr$^{(1)}$                                              & 13 Gyr$^{(1)}$                                              & 11.00 $\pm$ 0.38 Gyr$^{(2)}$                     & 14 Gyr$^{(3)}$                                            &  13.0 $\pm$ 1.5 Gyr$^{(4)}$                        & 13.00 $\pm$ 0.38 Gyr$^{(2)}$                            \\
{[Fe/H]}         & -0.11 dex                                                   & -0.18 dex                                                   & -0.78 dex                                                  & -0.97 $\pm$ 0.15 dex$^{(3)}$                   & -2.15 dex                                                   & -2.16 dex                                                 \\
{[Mg/Fe]$^a$ or 
[$\alpha$/Fe]$^b$}  & 0.24$^{(b,5)}$                                   & 0.26 dex$^{(b,6)}$                                        &   0.19$\pm$0.04$^{(a,7)}$, 0.40$^{(b,5)}$                                               & 0.24$^{(a,3)}$                                                & 0.4$^{(b,4)}$                                              & 0.47 $\pm$ 0.09 $^{(a,8)}$                          \\
E(B-V)           & 0.54                                                            & 0.63                                                            & 0.25                                                          & 0.44                                                           & 0.36                                                          & 0.12                                                          \\
(m-M)$_{\rm V}$ & 16.17                                                      & 15.83                                                          & 13.80                                                        & 15.70                                                         & 17.68                                                         & 17.47                                                         \\
R$_{\rm Sun}$  & 7.9 kpc                                                        & 6.0 kpc                                                       & 4.0 kpc                                                      & 7.4 kpc                                                      & 20.6 kpc                                                      & 26.3 kpc                                                     \\
R$_{\rm GC}$   & 0.6 kpc                                                        & 2.2 kpc                                                       & 6.7 kpc                                                      & 1.0 kpc                                                      & 14.4 kpc                                                     & 19.4 kpc                                                     \\
$<$v$_{\rm helio}>$ & 206.6 $\pm$ 1.4 km/s                      & -3.2 $\pm$  1.5 km/s                               & -22.8 $\pm$ 0.2 km/s                             & -197.3 $\pm$ 4 km/s$^{(3)}$                      & -162.0 km/s                                            & 130.0 km/s                                                  \\
r$_{\rm core}$ & 0.13$\arcmin$                                            & 0.53$\arcmin$                                            & 0.63$\arcmin$                                          & 0.03$\arcmin$                                           & 0.26$\arcmin$                                          & 1.00                                                           \\
r$_{\rm tidal}$ & 4.11$\arcmin$                                            & 7.66$\arcmin$                                            & 8.90$\arcmin$                                          & 9.49$\arcmin$                                           & 13.03$\arcmin$                                        & 3.98                                                           \\
r$_{\rm half-light}$ & 0.38$\arcmin$                                      & 1.03$\arcmin$                                          & 1.67$\arcmin$                                          & 2.15$\arcmin$                                           & 0.92$\arcmin$                                          & 0.95                                                             \\
\noalign{\smallskip}\hline
\end{tabular}  
\tablefoot{ 
\tablefoottext{1}{\cite{zoccali+01}}
\tablefoottext{2}{\cite{vandenberg+13}}
\tablefoottext{3}{\cite{barbuy+07}}
\tablefoottext{4}{\cite{dotter+11}}
\tablefoottext{5}{\cite{carretta+10}}
\tablefoottext{6}{\cite{cohen+99}}
\tablefoottext{7}{\cite{melendez+09}}
\tablefoottext{8}{\cite{carretta+14}}
}
\end{table*}

%% Table \ref{starinfo}. #3
%% The list of individual stars, their coordinates, S/N and $VI$ magnitude
\addtocounter{table}{1}

\begin{figure*}[!htb]
\includegraphics[width=\columnwidth]{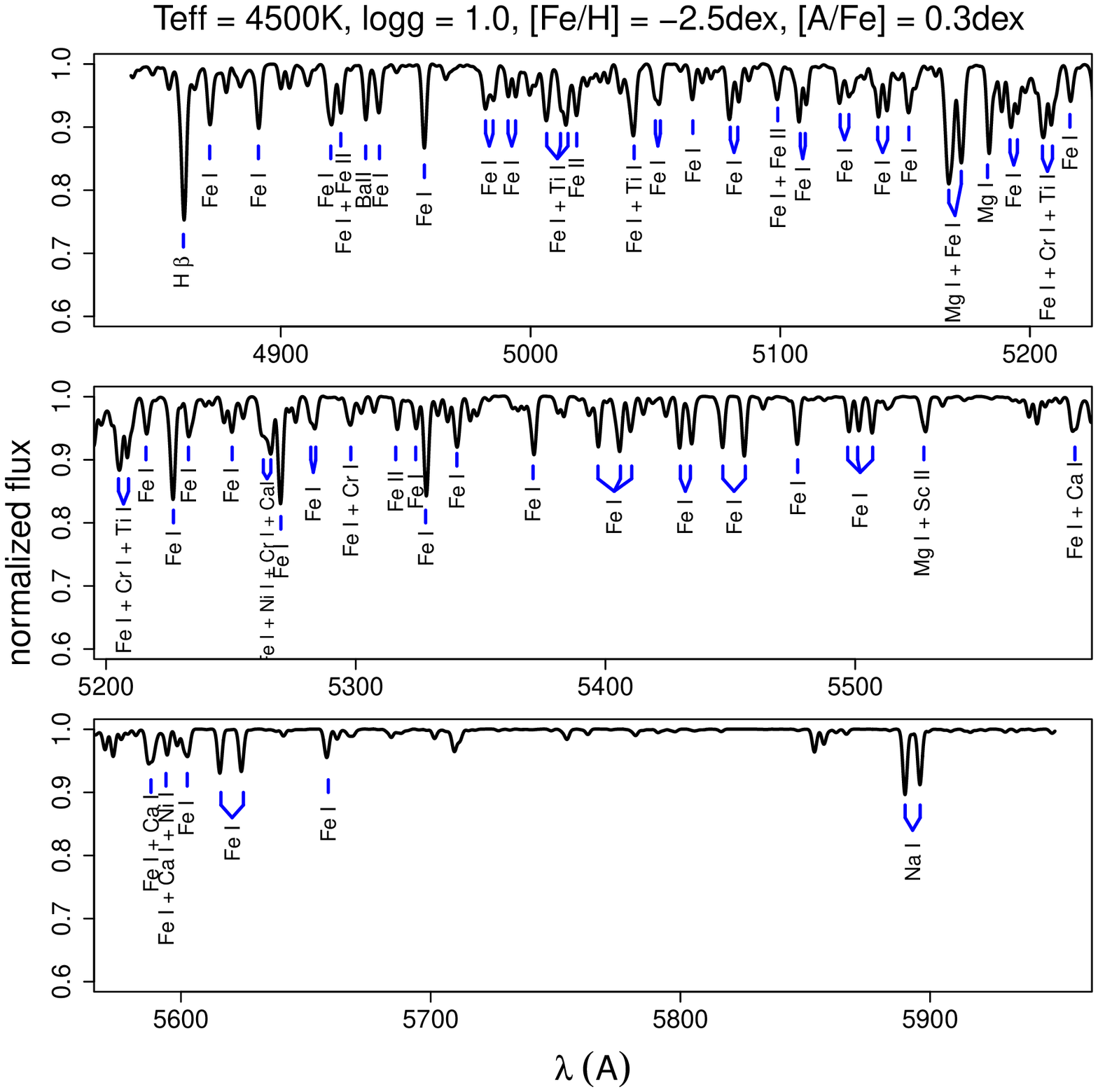}
\includegraphics[width=\columnwidth]{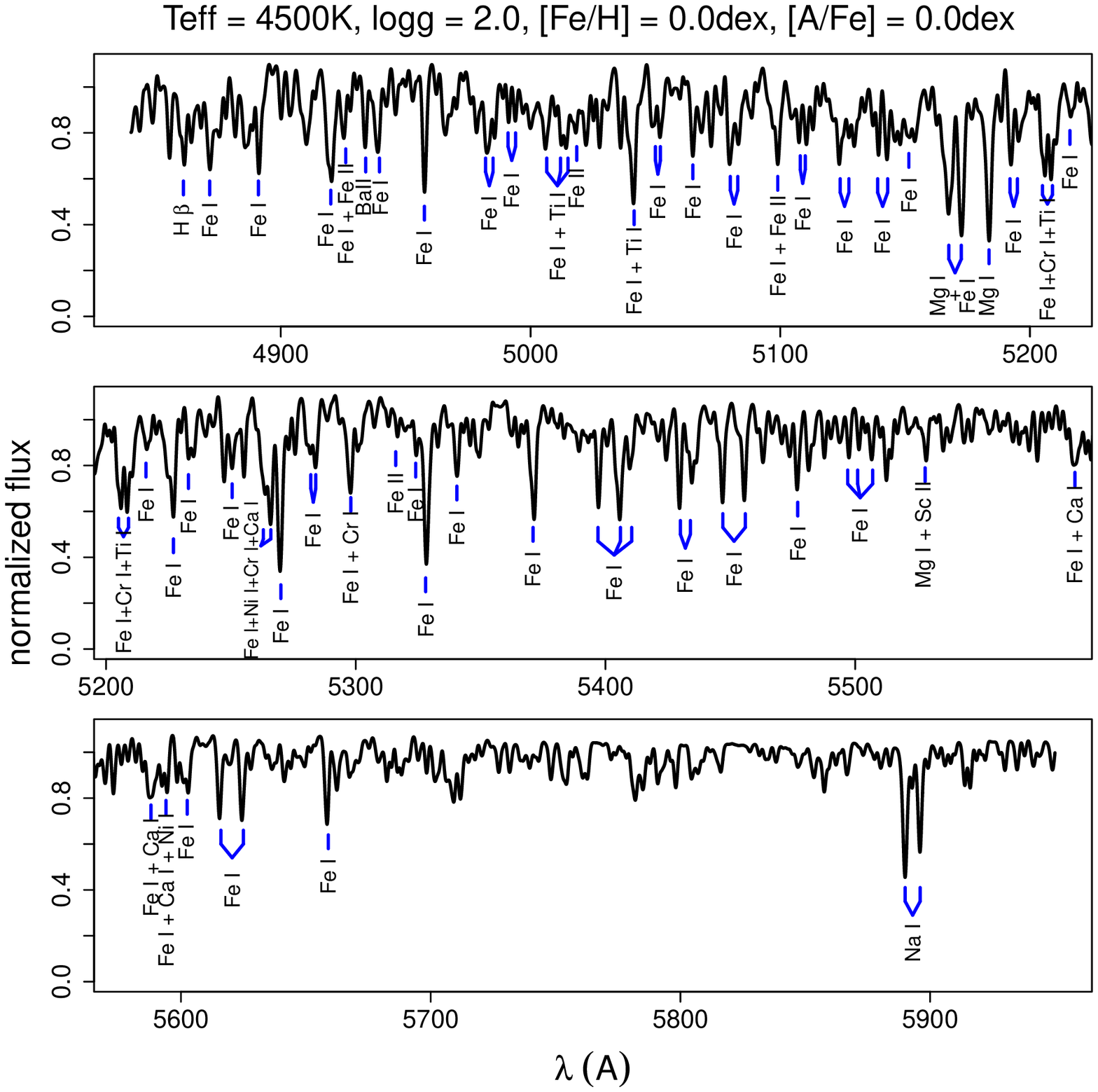}
\caption{Identification of most important atomic lines in each strong
feature of spectra with FORS2 resolution (R$\sim$2,000). The most
important molecular band in this region is MgH around $\lambda$ =
5165~${\rm \AA}$, on top of the \ion{Mg}{I} triplet lines. Left panel shows a
metal-poor red giant star from \cite{coelho+05} library, and right panel
shows a metal-rich one. The left panels were zoomed in the y-axis
direction for better visualization.}
\label{idlines}
\end{figure*}

The spectra were taken using the grism 1400V, centred at 5200~{\rm \AA},
covering the range 4560 - 5860~{\rm \AA}, with a resolution of 
R$\sim$2,000. Figure \ref{idlines} illustrates
the spectra of a metal-poor and a metal-rich red giant
 star, where many of the strongest lines are indicated. 

The spectra were reduced using esorex/FORS2
pipeline\footnote{http://www.eso.org/sci/software/pipelines/} with
default parameters for bias and flatfield correction, spectra
extraction, and wavelength calibration. The only modification relative to
default parameters, has been the introduction of a list of skylines, since
the default list had only one line. The wavelength calibration
proved to be satisfactory with such line list. 
A  last step in the reduction procedure
was a manual removal of cosmic rays.

%__________________________________________________________________
\section{Stellar parameters derivation}

\subsection{Radial velocities}

\begin{figure}[!htb]
\centering
\includegraphics[width=\columnwidth]{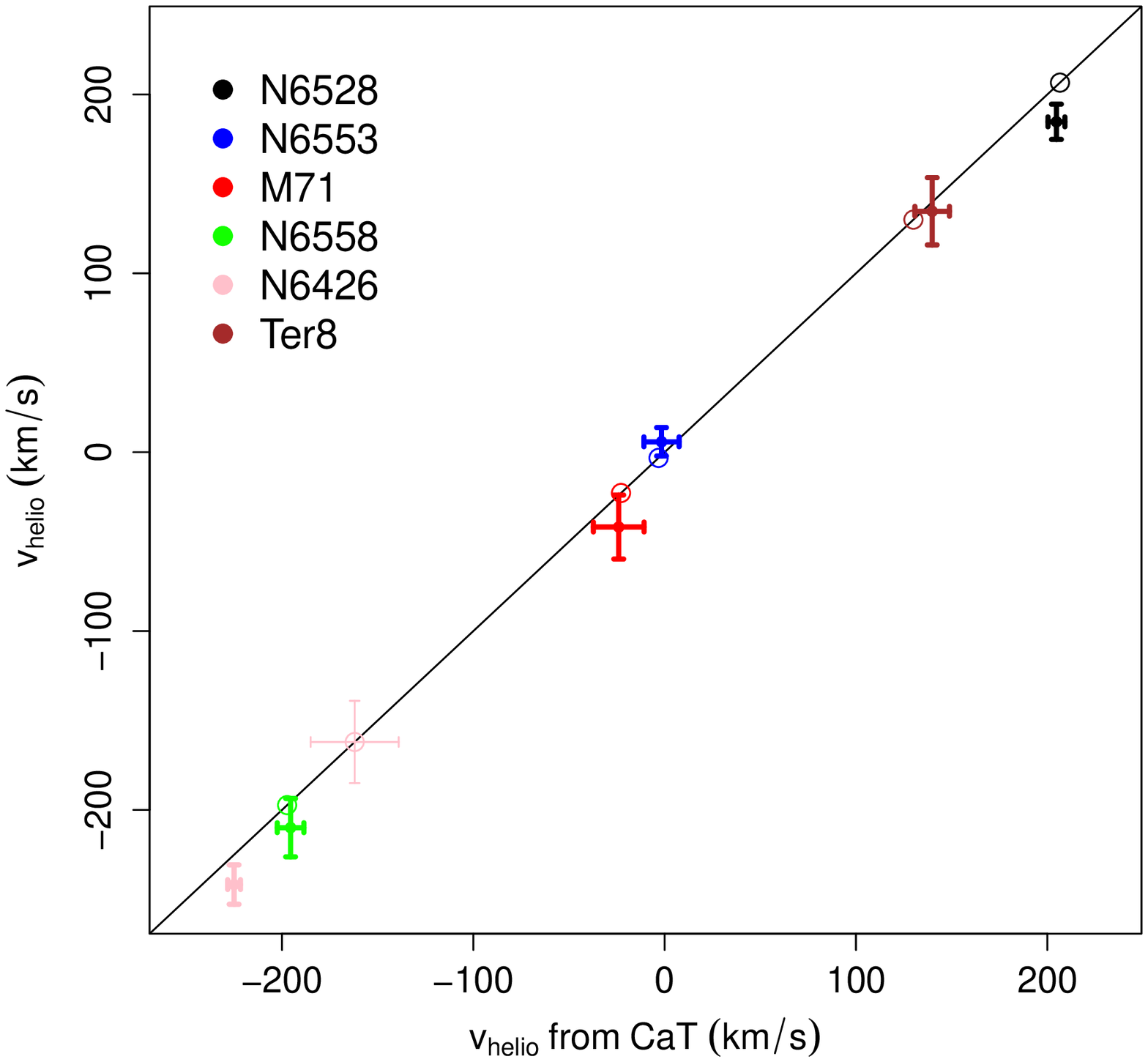}
\caption{Average of heliocentric radial velocities of member stars
  (see details in Section \ref{sec:membership} and values in Table
  \ref{starinfo}) for each globular cluster. Our results are plotted
  against those from CaT spectroscopy \citep{saviane+12} and
 Vasquez et al. (in prep.) showing
  good agreement. Error bars are the standard deviation of the
  average. One-to-one line is plotted for visual guidance, where
  radial velocities from \citet[2010 edition]{harris96} are
  overplotted as empty circles.}
\label{rvlit}
\end{figure}

Radial velocities were measured using the ETOILE code through cross
correlation with a template spectrum from the chosen library. Tests were done in
order to check the results, by measuring radial velocities using
fxcor@IRAF (cross correlation), and rvidlines@IRAF (using
wavelength of MgI triplet lines as a reference). The derived velocities
are consistent, therefore we used ETOILE also to determine radial
velocities. 
A mean FWHM of arc lines of 2.36$\pm$0.04{\rm \AA} (125~km/s) was
  measured. This leads to a radial velocity uncertainty of $\sim$13~km/s.
 Heliocentric radial velocities
 for each star can be found in Table \ref{starinfo},
where the last column refers to the values measured from the
 CaII triplet (CaT) lines in the near
infrared by \cite{saviane+12} for member stars for NGC~6528, NGC~6553,
M~71 and NGC~6558, and by Vasquez et al. (2014 in prep.) for
NGC~6426 and Terzan~8.
There is good agreement between the present
radial velocity values and those from
the CaT line region. 
A few exceptions are stars \#8 of NGC 6558, \#2, \#10 of M~71,
  among others. A possible explanation for this could be due to a not
  perfectly centred source in the slit in some cases, as suggested by
  \cite{katz+11} in using CFHT-MOS.
 Average values for member
stars in each cluster are presented in Figure \ref{rvlit}, where our
results are compared to CaT results (\citealp{saviane+12} and Vasquez
et al., in prep.), and with
 \citet[2010 edition]{harris96} catalogue. Error bars from the literature are
smaller than the empty circles that represent literature v$_{\rm
  helio}$, except for NGC~6426, as can be seen in 
Figure \ref{rvlit}. Our results are in good agreement with both
references. In particular, the radial velocity measured for NGC~6426  
is in agreement between this work and CaT results based on individual
member stars, but it is only compatible with the literature value
within 3$\sigma$. The explanation is that the only work that measured
radial velocities for this cluster was based on integrated light from
photographic plates \citep{hesser+86}. 
Therefore, the present radial velocity derivation for NGC~6426 is more 
reliable.

\subsection{Atmospheric parameters}
\label{atmosparam}

Full spectrum fitting with minimum distance method is employed, using
the ETOILE code described in \cite{katz+11} and \cite{katz01}.
We apply the calculations to the wavelength region 4600-5600 {\rm \AA},
similarly to the procedure described in  \cite{katz+11}.

Automated derivation of the 
atmospheric parameters (T$_{eff}$, log($g$), [Fe/H], [$\alpha$/Fe]) of
a stellar spectrum is carried out by comparing the target spectrum
with each library spectrum, thus covering a large range of
 atmospheric parameters.
In each comparison, ETOILE fits the template spectrum
to the observed spectrum. Mathematically, ETOILE solves,
by least squares, for the polynomial by which to multiply the
template spectrum to minimize the differences with the
observed spectrum 
(see Equations \ref{simil-def} and \ref{simil-min}).
The aim of these operations is to compensate for the differences
between the template and observed spectra which are not
from stellar origin: e.g. flux level/normalisation, instrumental
profile, interstellar reddening. In particular, concerning this
last point, no explicit reddening is applied to the template.
The differential reddening correction is included in the
fitting of the template to the observed spectrum.

\begin{equation}
S = \sqrt{ \sum\limits_{i = 0}^{n}{\bigg\{ F_{\rm obs}(i) - \Big[
  \sum\limits_{j=0}^{m}{ u_j \cdot \big( \lambda(i) - \lambda_{\rm
      central} \big)^j  } \Big] \cdot F_{\rm templ}(i) \bigg\}^2} } 
\label{simil-def}
\end{equation}

\noindent where $n$ is the number of pixels in the spectrum
being analysed, $F_{\rm obs}(i)$ and $F_{\rm templ}(i)$ are the fluxes
of the analysed and the template spectra respectively pixel by pixel
(i.e. lambda by lambda), $m$ is the order of the polynomial that
multiplies $F_{\rm templ}(i)$ and $u_j$ are the coefficients,
    $\lambda_{\rm central} = [\lambda(0) + \lambda(n)]/2$. Equation
\ref{simil-def} is minimized to find the multiplicative polynomial
that minimizes the differences in flux between observed and template
spectra solving the $m+1$ Equations below. In this work we
  adopted $m=4$.

\begin{equation}
\frac{\partial S}{\partial u_j} = 0 {\rm \ , \ where \ } j \in \{0,
..., m\} 
\label{simil-min}
\end{equation}

After determination of the polynomial that minimizes the
  difference between each template and the observed spectrum, as
  defined by Equation \ref{simil-def}, templates are ranked in order of increasing
  $S$ and the parameters of the top N templates are averaged out to
  produce the final results.  Determination of the optimal value of N
  is discussed in Section \ref{sec:validation}.
This is called the similarity method introduced by
\cite{katz+98}. For a more detailed explanation see \cite{katz01}.

Before running the code, two important steps are needed: to convolve
all the library spectra to the same resolution of the target spectra,
and to correct for radial velocities  $v_r$. Convolution calculations were
performed for the library spectra using the task GAUSS in IRAF. The
code ETOILE measures the radial velocities by comparison with template
spectra from the library, a reliable way to measure $v_r$ in each observed
spectrum and correct them. 

\begin{figure*}[!htb]
\centering
\includegraphics[width=\columnwidth]{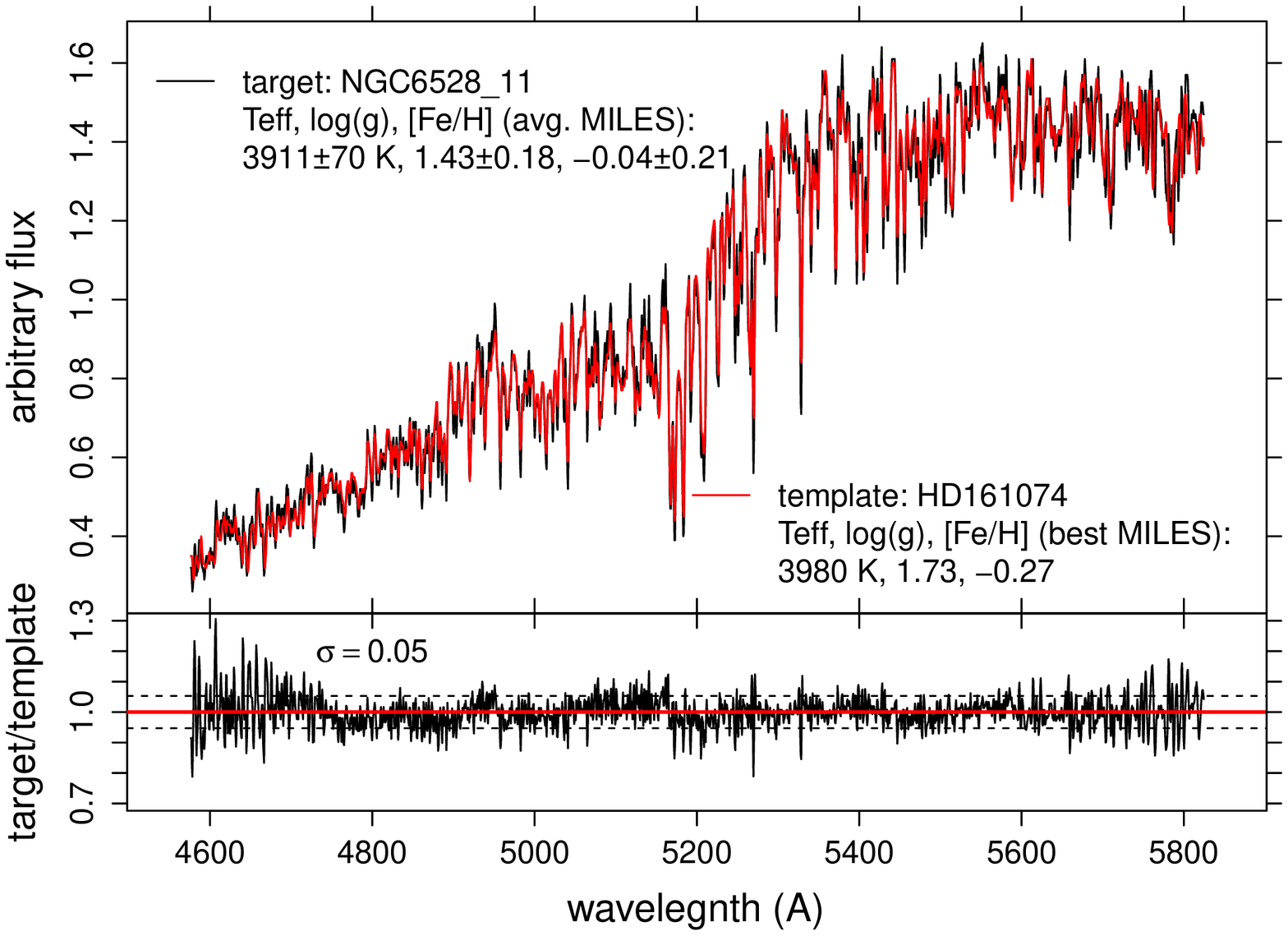}
\includegraphics[width=\columnwidth]{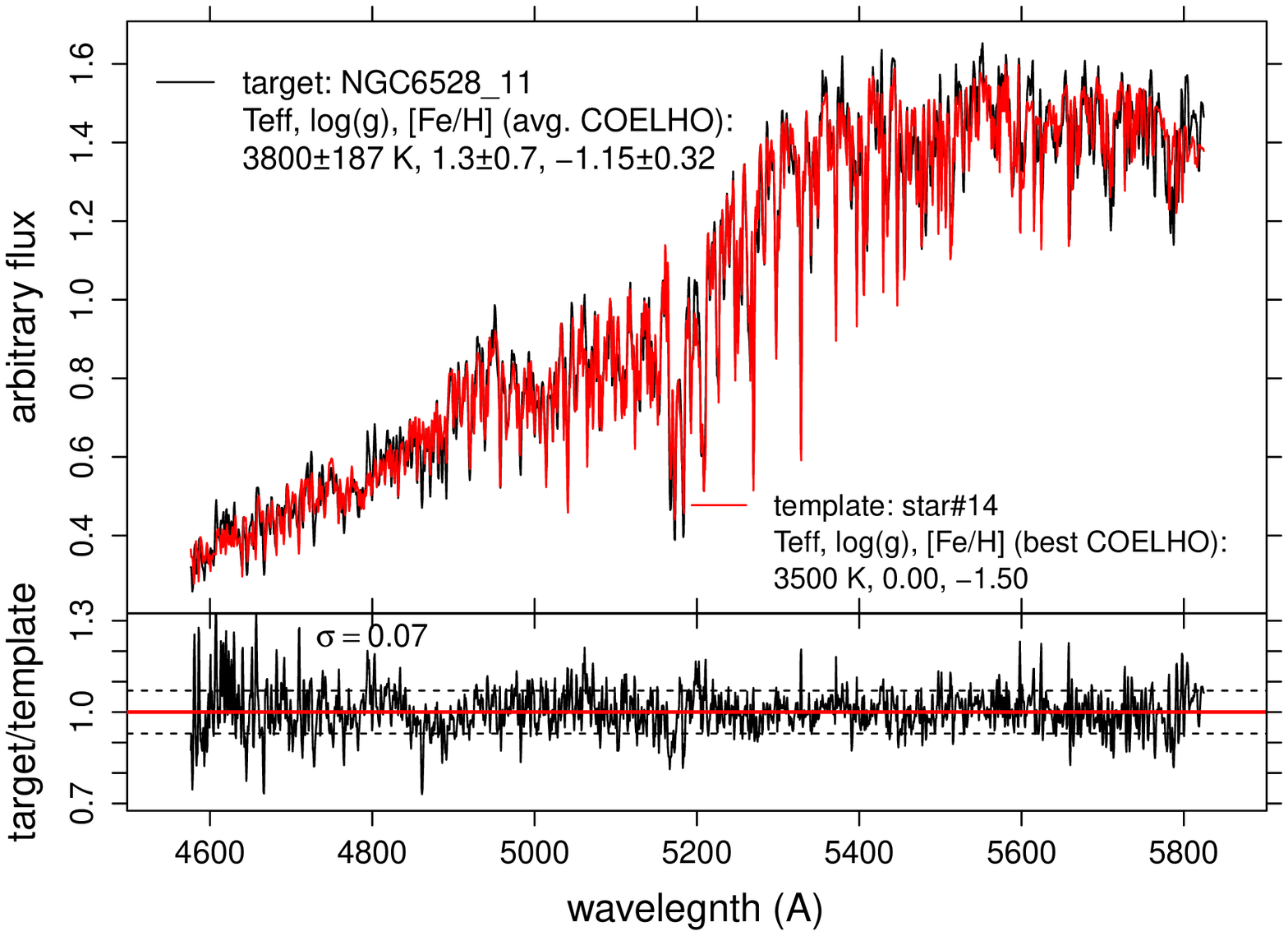}
\includegraphics[width=\columnwidth]{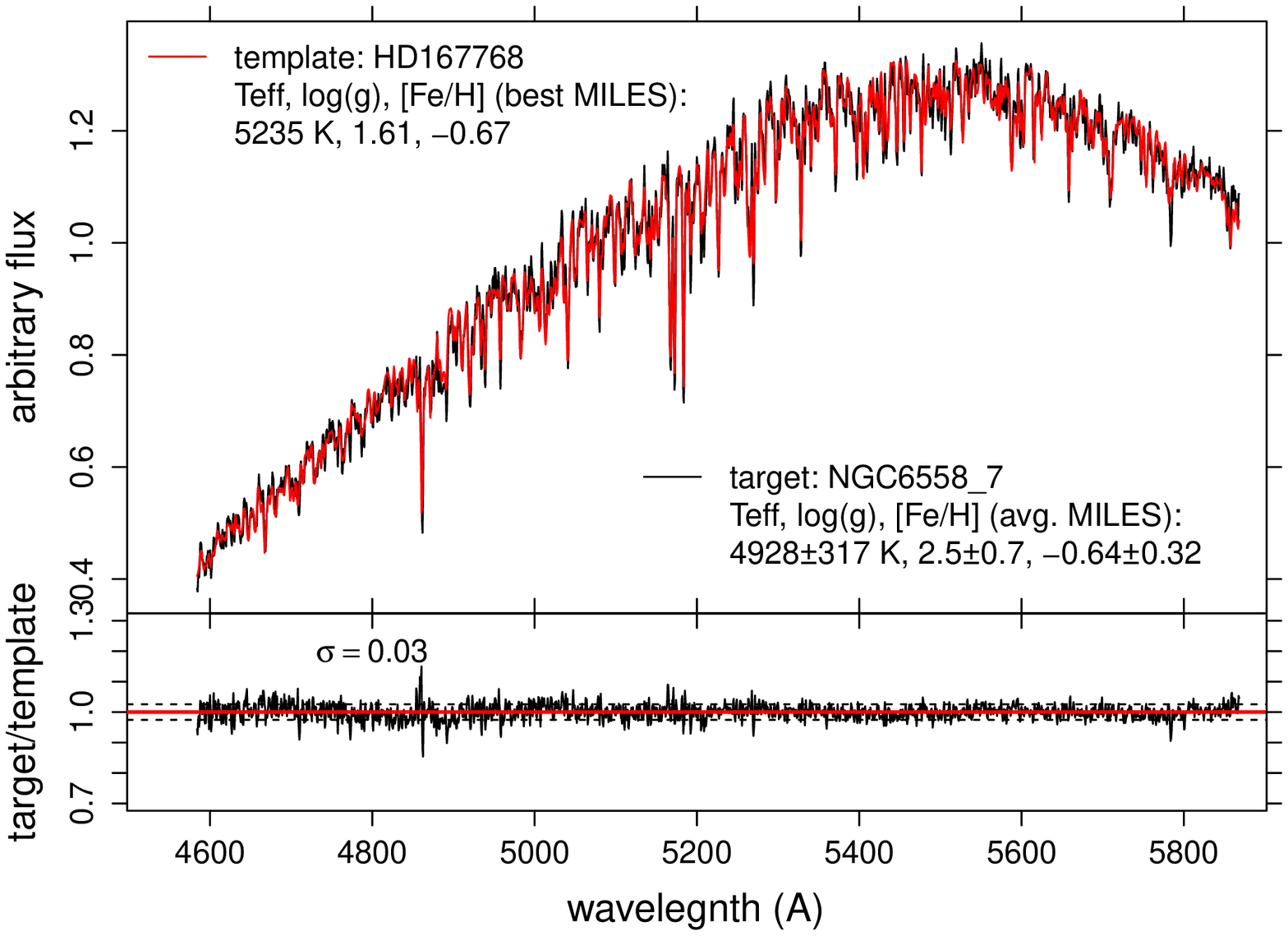}
\includegraphics[width=\columnwidth]{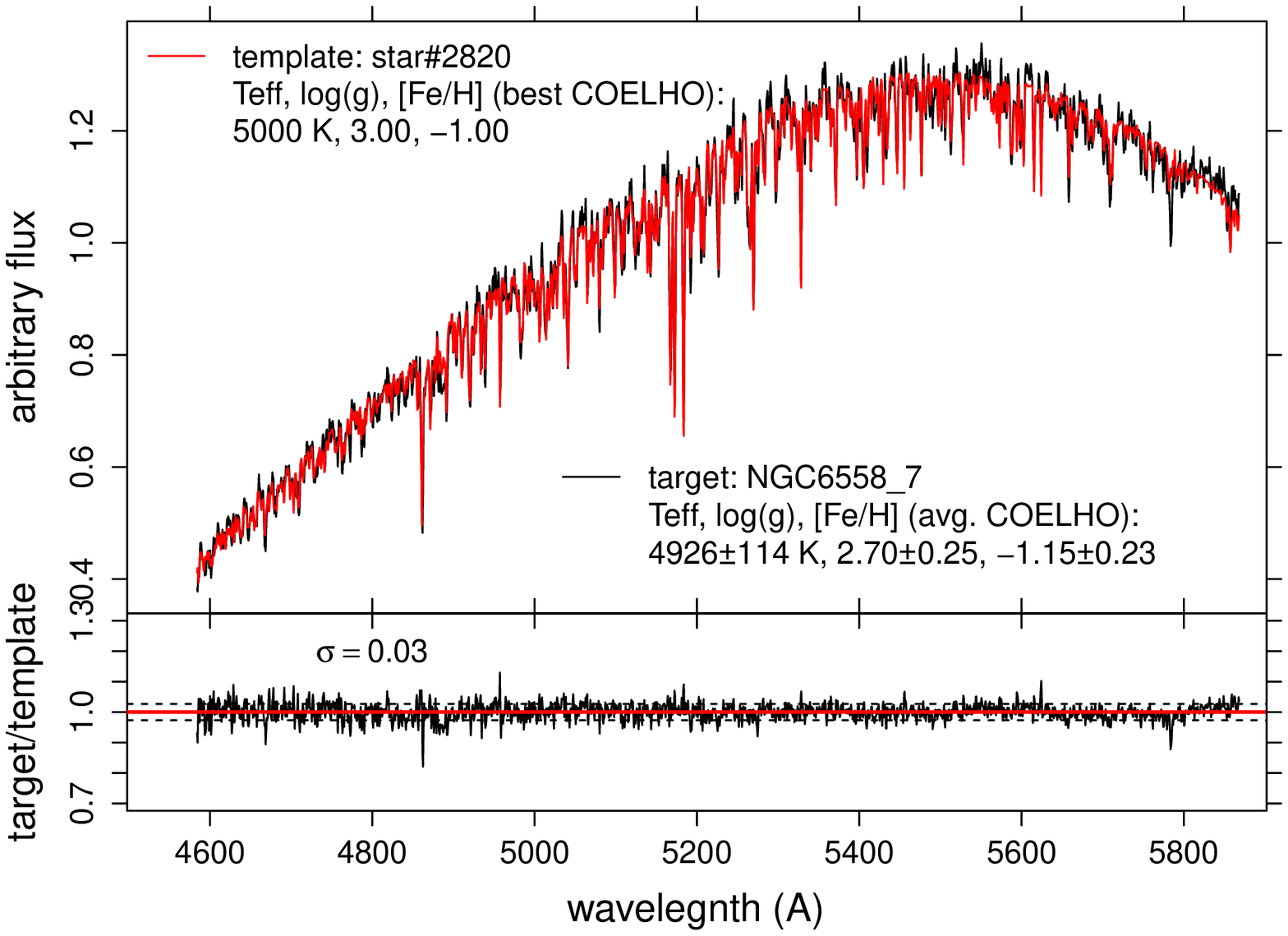}
\includegraphics[width=\columnwidth]{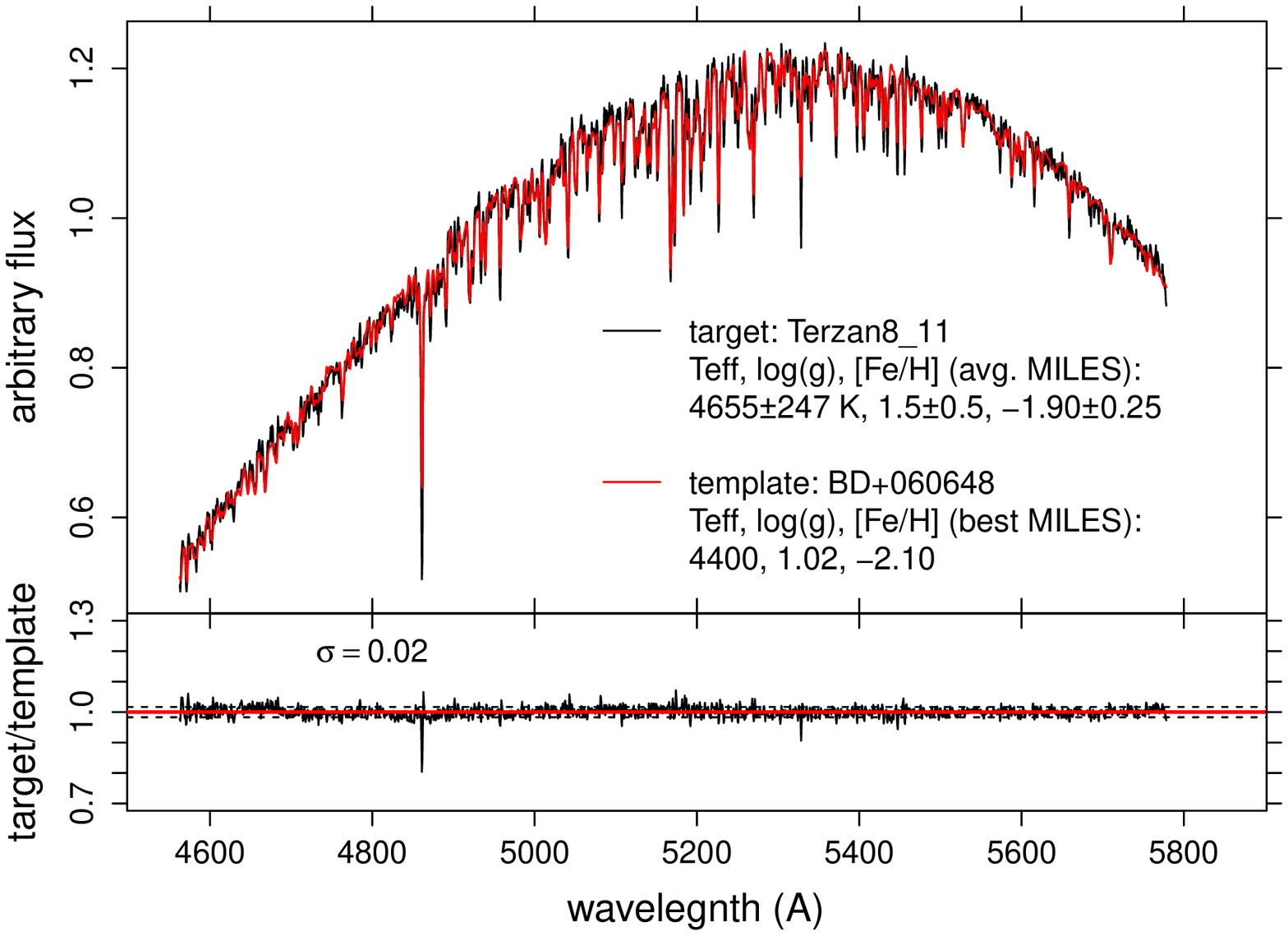}
\includegraphics[width=\columnwidth]{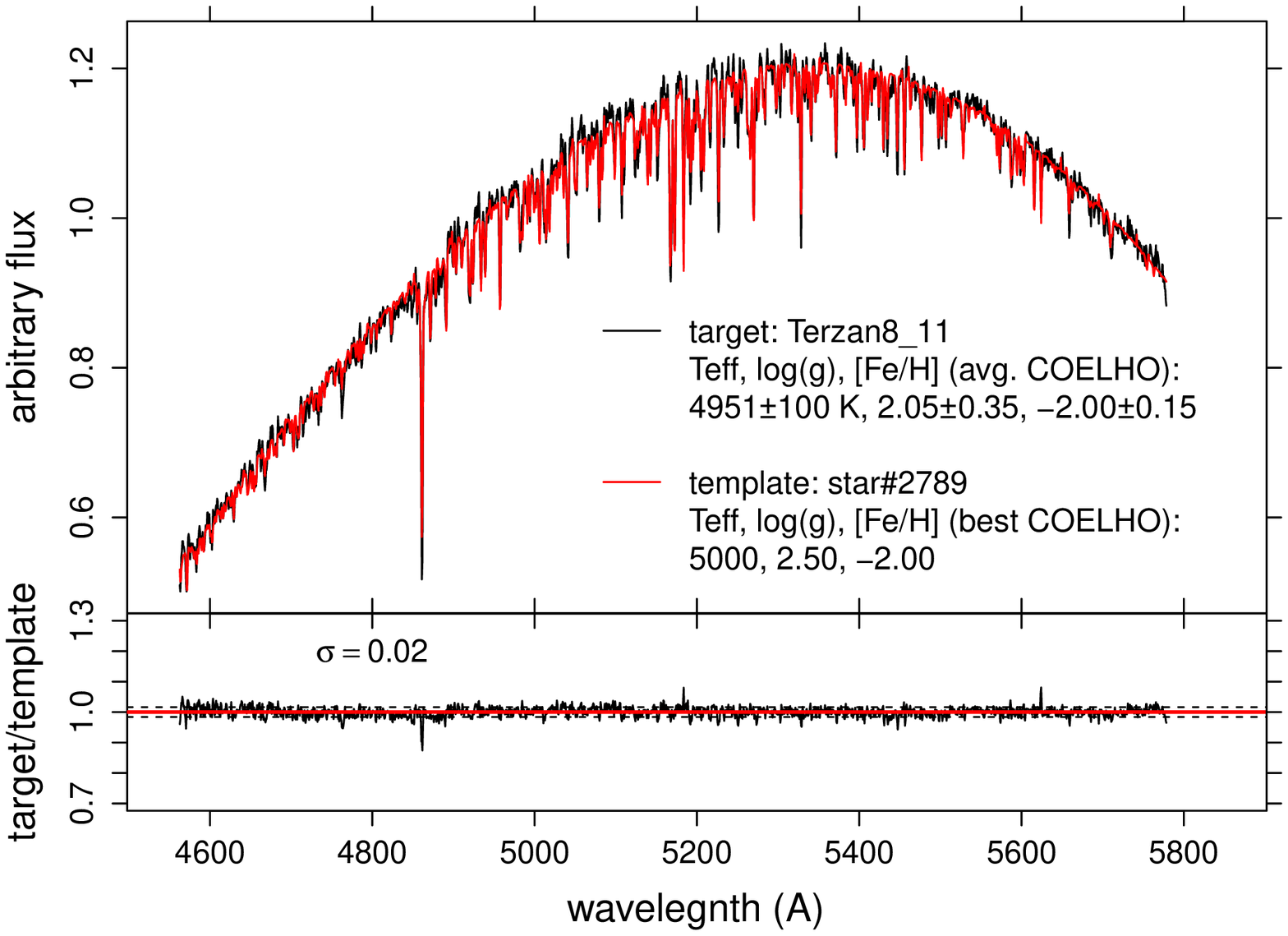}
\caption{Examples of spectral fitting carried out with ETOILE for a
metal-rich (NGC6528\_11), an intermediate metallicity (NGC6558\_7) and
a metal-poor (Terzan8\_11) star in the upper, middle and bottom
panels, respectively. Left panels represent the best fits using MILES
template spectra, and in the right panel the best fits using COELHO
spectra are shown.
For each star, its spectrum (black line) is overplotted by the template spectrum
  (red line) that best fits it.
  Below each stellar spectrum the
  residuals of each fit are presented. The match between the spectra
  is done following the procedures explained in Section
  \ref{atmosparam}. The fit appears very satisfactory for the 
  whole wavelength interval for all cases. In each comparison we give
  the parameters of the template spectrum and of the average
  parameters using only MILES or only COELHO spectra for each star, as
presented in Table \ref{finalparam}.}
\label{specfitting}
\end{figure*}

 Figure \ref{specfitting} shows six examples of spectral
 fitting, for a metal-poor (Terzan8\_11), a metal-rich (NGC6528\_11)
and an intermediate metallicity star (NGC6558\_7) using COELHO and MILES
   templates. The template stars that best fit these cluster 
  stars among the available spectra from MILES library are BD+060648,
  HD161074 and HD167768, respectively. For COELHO the best
    templates are the ones with the following parameters: (T$_{\rm eff}$,
    log(g), [Fe/H]) = (5000K, 2.5, -2.0), (3500K, 0.0, -1.5) and (5000K,
    3.0, -1.0), respectively.
 The residuals shown at the bottom of each panel indicate that the
  metal-poor target spectrum is similar to the template spectrum
  within 2\% for both libraries, except for a few strong features.
  The residuals for the
  metal-rich star shows a similarity between target and template
  spectra of 5\% for MILES and of 7\% for COELHO, except for the
  boundaries $\lambda \gtrsim 5700$~{\rm \AA} 
  and $\lesssim 4700$~{\rm \AA}, and for a few strong features.
For the intermediate metallicity star the residuals present a
  sigma of 3\% for both libraries with few stronger features varying
  more than 3\%.
These differences between the spectra are reflected in the atmospheric
parameters, and they are compensated by taking the average of
parameters of the most similar spectra. For Terzan8\_11, there are 8
MILES spectra close enough which were averaged, for NGC6528\_11,
21 MILES spectra were considered, and for NGC6558\_7 it were found 8
  stars. For all cases with COELHO library 10 templates were
  considered in the average. Details on the criterion to select
the number of template spectra are discussed in Section
\ref{sec:validation}.

%table with coords, S/N, phot, rv
%label="starinfo"
%\addtocounter{table}{2}

\subsubsection{Stellar libraries}
\label{sec:libraries}

The core of the atmospheric parameters derivation in this work is the
choice of a stellar library. There are two classes of stellar
libraries: based on observed or synthetic spectra. The real spectra
are more reliable, but the drawback is that they have abundances
typical of nearby stellar populations. The synthetic libraries have no
noise, and a large and uniform coverage of the atmospheric parameters
space, however there are still limitations on the completeness of
atomic and molecular line lists, plus uncertainties on oscillator
strengths, and assumptions on atmospheric models, such as 1-D and 
local thermodynamical equilibrium. For these reasons,  it is useful
to use both observational and synthetic libraries. In the present
work, we use two libraries, one observed and one synthetic, as
described below:

The MILES library \citep{sanchez-blazquez+06} has 985 stellar spectra
with resolution R$\sim$2\,080@5\,200~{\rm \AA}, and mean signal-to-noise ratio of 150 per
pixel for field and open cluster stars, and 50 for globular cluster
stars. Atmospheric parameters coverage is
\citep{cenarro+07,milone+11}:

\[\begin{array}{rcccl}
352.5{\rm nm} & < & \lambda & < & 750{\rm nm} \\
2\,747{\rm K} & < & T_{\rm eff} & < & 36\,000{\rm K} \\
-0.20 & < & {\rm log}(g) & < & 5.50 \\
-2.86 & < & {\rm [Fe/H]} & < & +1.65 \\
-0.54 & < & {\rm [Mg/Fe]} & < & +0.74 \\
\end{array}\]

The COELHO library \citep{coelho+05} has 6367 synthetic stellar
spectra\footnote{Interpolation on the original library was carried out
  to produce spectra with [$\alpha$/Fe] = 0.1, 0.2, 0.3~dex from the
  provided 0.0 and 0.4~dex spectra.} with wavelength steps of 0.02{\rm
  \AA} (resolution R=130\,000@5\,200~{\rm \AA}). Atmospheric parameters coverage is:

\[\begin{array}{rcccl}
300{\rm nm} & < & \lambda & < & 1800{\rm nm} \\
3\,500{\rm K} & < & T_{\rm eff} & < & 7\,000{\rm K} \\
0.0 & < & {\rm log}(g) & < & 5.0 \\
-2.5 & < & {\rm [Fe/H]} & < & +0.5 \\
0.0 & < & {\rm [\alpha/Fe]} & < & +0.4 \\
\end{array}\]

\noindent where $\alpha$-elements considered in this library are: O,
 Mg, Si, S, Ca and Ti.

\begin{figure}[!htb]
\centering
\includegraphics[width=\columnwidth]{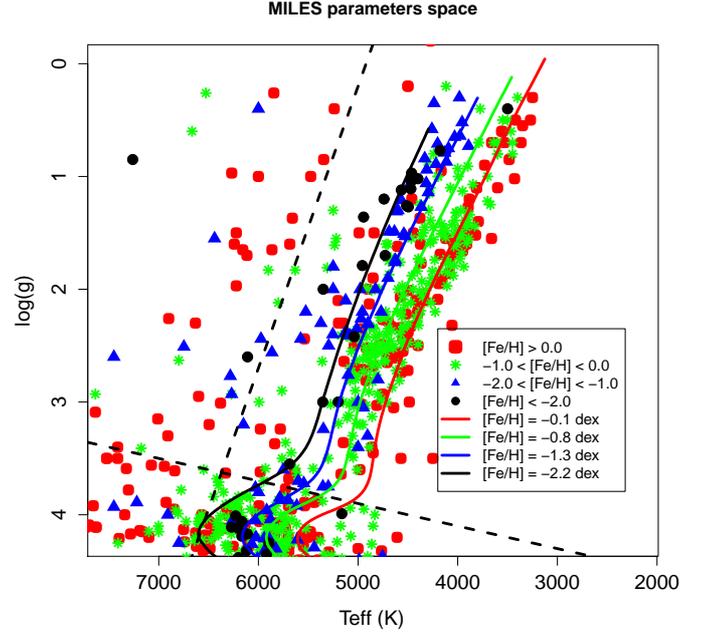}
\caption{HR diagram showing the parameter space available from the MILES
  library. Dartmouth isochrones \citep{dotter+08} are overplotted for
  metallicities close to those of the six analysed globular clusters:
  [Fe/H] = -0.1, -0.8, -1.3, -2.2 dex, with [$\alpha$/Fe] = 0.2, 0.4,
  0.4, 0.4, 0.4 dex, respectively, and ages = 13 Gyr for all
  cases. Colours of the dots indicate the metallicity range closer
  to the isochrones. Dashed black lines are the adopted limit to
  select RGB stars from MILES used as reference for the
  fits.}
\label{mileshrd}
\end{figure}

Given that all cluster stars are located in the red giant branch, as shown
in Figure \ref{cmds}, we selected only stars in this region in the
parameters space of the libraries (see Figure \ref{mileshrd}) to avoid
non-physical results.

\subsubsection{Average results and errors: validation with
  well-known stars}
\label{sec:validation}

 We define different criteria for MILES and COELHO libraries 
 for taking the average of
  stellar parameters from reference spectra, as mentioned in Section
  \ref{atmosparam}. For MILES the average results are based on different numbers
  of templates depending on the sampling as shown in Figure
  \ref{mileshrd}. For the synthetic library COELHO the sampling is
  homogeneous, therefore a constant number of templates is
  adopted. We found that 10 templates for COELHO cover satisfactorily the variations
  in the four stellar parameters (T$_{\rm eff}$, log($g$), [Fe/H] and
  [$\alpha$/Fe]). The COELHO library was built by varying all
  alpha-elements (O, Mg, Si, S, Ca, Ti) together, therefore [$\alpha$/Fe]
  is an average of the effect from enhancement of these element
  abundances. 

The criterion to define average results from the MILES library is more
complex, as follows.
The code provides a list of the closest reference spectra from the
library, ranked by the similarity parameter ($S$, as defined
in Equations \ref{simil-def} and \ref{simil-min}).
The final parameters T$_{\rm eff}$, log($g$), [Fe/H] and [Mg/Fe] are
the average of the parameters of first N reference stars from the ETOILE
output, where N depends on the sampling of the library for each
combination of parameters. 
The average is weighted by $1/S^2$ as shown
in the equation below for T$_{\rm eff}$ (the same is valid for the
other three parameters):

\begin{equation}
\centering
{\rm T}_{\rm eff}\ (N) = \frac{ \sum\limits_{i=1}^{N}{{\rm T}_{\rm eff, i}
    \times \frac{1}{S_i^2}} }{\sum\limits_{i=1}^{N}{\frac{1}{S_i^2}}}
\label{eq:average}
\end{equation}

The errors are defined as the average of the squared residuals,
weighted by $1/S^2$, as shown in the equation below for T$_{\rm eff}$
(the same is valid for the other three parameters). For N=1, we
adopted the same error of N=2.

\begin{equation}
\centering
\sigma_{{\rm T}_{\rm eff}\ (N)} = \sqrt{ \frac{ \sum\limits_{i=1}^{N}{
      ({\rm T}_{\rm eff, i} - {\rm T}_{\rm eff})^2 \times
      \frac{1}{S_i^2}} }{\sum\limits_{i=1}^{N}{\frac{1}{S_i^2}}} }
\label{eq:error}
\end{equation}

To estimate the number of reference stars to be averaged in each case,
we proceeded with some tests using 59 spectra of 49 well-known stars, listed in Table 
\ref{tab:etoilevalidation}. These stars were selected among red giant
stars (same log($g$) and T$_{\rm eff}$ intervals defined in Figure
\ref{mileshrd}) presented in the ELODIE 
library\footnote{http://www.obs.u-bordeaux1.fr/m2a/soubiran/elodie\_library.html}
\citep{prugniel+07}. Stellar spectra were taken from ELODIE
  library and convolved to the FORS2 resolution.
 Reference atmospheric parameters were
averaged from the PASTEL catalogue
(\citealp{soubiran+10}), and the quality filter 
was determined by a threshold in the standard deviation:
$\sigma_{T_{\rm eff}} < 200$K, $\sigma_{\rm log(g)} < 0.5$, $\sigma_{\rm [Fe/H]} <
0.2$.
We calculated the average parameters and respective errors for different
N and compared the results with the average values of T$_{\rm eff}$,
log($g$), [Fe/H] from the PASTEL
catalogue \citep{soubiran+10}.
We minimize the equation below to find the best N
that will give the final parameters and respective errors. This
equation considers the distance between the average for a given N and
literature average; in this way all the three parameters are minimized
at the same time. \cite{milone+11} have measured [Mg/Fe] for MILES
spectra, therefore it is possible to take averages for this
parameter as a function of N, and use [Mg/Fe] for the best N as an
estimation of the $\alpha$-enrichment for each star.

%% Table \ref{tab:etoilevalidation}.
%% 59 spectra of 49 well-known stars from ELODIE
\addtocounter{table}{1}

\begin{equation}
\begin{array}{ll}
{\rm RR}_{\rm tot}\ (N) = & \sqrt{ {\rm RR_{\rm N}}({\rm T}_{\rm eff})^2 + {\rm
    RR_{\rm N}}({\rm log}(g))^2  + {\rm RR_{\rm N}}({\rm[Fe/H]})^2 }\\
\end{array}
\label{eq:rrtot}
\end{equation}

\noindent where RR$_{\rm N}$(T$_{\rm eff}$) is given by the equation below (the
same is valid for the other three parameters):

\begin{equation}
{\rm RR_{\rm N}}({\rm T}_{\rm eff}) = \frac{ {\rm T}_{\rm eff}\ (N) - {\rm T}_{\rm
    eff}^{(lit)}\ (N) }{ {\rm T}_{\rm eff}^{(lit)}\ (N) }
\label{eq:rrteff}
\end{equation}

\begin{figure}[!htb]
\centering
\includegraphics[width=\columnwidth]{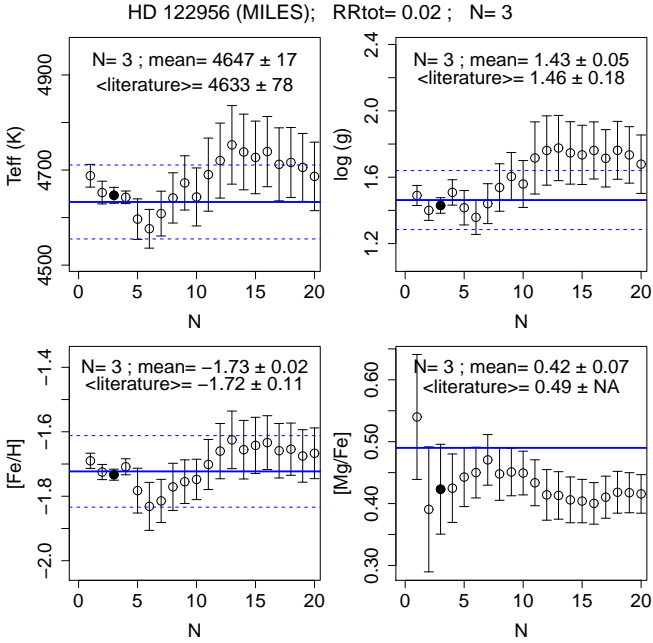}
\caption{Finding procedure of N for the case of star HD122956, based
  on the minimization of the residuals of T$_{\rm eff}$, log($g$) and
  [Fe/H] (equation \ref{eq:rrtot}). Circles represent the averages of
  the parameters of N best reference stars. Filled black circle
  indicates the closest values to the references. Blue solid lines and
  blue dashed lines are average from PASTEL catalogue and standard deviation
  (Table \ref{tab:etoilevalidation}). For this star, \cite{fulbright+00}
  published [Mg/Fe] and we compare also with the
  averages as a function of N. All the four parameters for the best
  N=3 are compatible with literature.}
\label{fig:etoilevalidation}
\end{figure}

Figure \ref{fig:etoilevalidation} illustrates the finding procedure of N
for the case of star HD122956, showing that ETOILE could recover all
the four parameters accurately. 
The resulting parameters, RR$_{\rm tot}$, N, and literature
values are presented in Table \ref{tab:etoilevalidation}. 
Different stars need different number N of templates to find the best
result. Moreover the ratio $S$(N)/$S$(1) for the best N is roughly
constant for all ETOILE template spectra, with an average value of
1.1$\pm$0.1. The best number N and the respective ratio $S$(N)/$S$(1)
are related to the library sampling, for example, for a given star
with best N=1 it means that there is only one reference star with
$S$(N)/$S$(1)~$\lesssim$~1.1, and there are two possible explanations:
either the target star matches perfectly some reference star, or the
library has no other reference spectra similar enough to that star to
be considered. In the cases with N=15, for instance, the library has
15 reference spectra very similar ($S$(N)/$S$(1)~$\lesssim$~1.1) to the
target spectra, and their parameters must be averaged in order to get
the parameters for the target star.

\begin{figure}[!htb]
\centering
\includegraphics[width=\columnwidth]{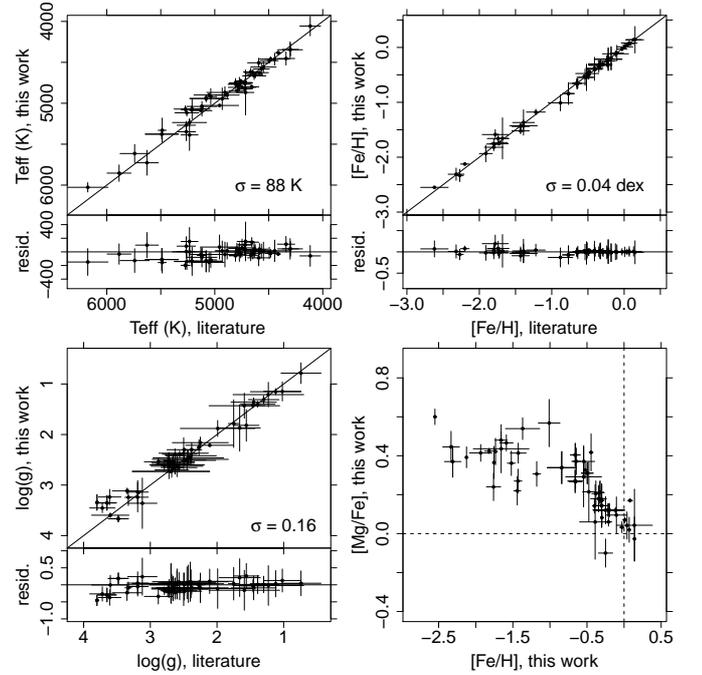}
\caption{Comparison of the parameters determined in this work with
   PASTEL catalogue average values for the well-known stars presented in
  Table \ref{tab:etoilevalidation}. These plots endorse the usage of
  ETOILE code for atmospheric parameters determination for red giant
  stars in the optical spectral region. Only stars with
  $\sigma_{T_{\rm eff}} < 200$K, $\sigma_{\rm log(g)} < 0.5$,
  $\sigma_{\rm [Fe/H]} < 0.2$ from PASTEL catalogue were
  selected as good quality candidates for validation of the
  method. Below the plots of T$_{\rm eff}$, log($g$) and [Fe/H]
    there is a residuals plot and the dispersion of the residuals is
    displayed in the respective plots.}
\label{elodie-lit}
\end{figure}

All results are plotted in Figure
\ref{elodie-lit} showing the good agreement of ETOILE results and
 PASTEL catalogue average for T$_{\rm eff}$, log($g$), [Fe/H] in the whole
range for RGB stars analysed in this work. 
The behaviour of the derived
values of [Mg/Fe] vs. [Fe/H] has a similar behaviour to field stars
(see e.g. Figure 6 of \citealp{alves-brito+10}).

After these tests we can consider that ETOILE code together with the MILES library
works well for low-resolution spectra of red giant stars in the
optical region. Additionally we define the criterion to consider a
reference spectrum similar enough to be considered in the average of
the parameters as $S$(N)/$S$(1)~$\leq$~1.1.

%______________________________________________________________

\section{Results}

  The derived T$_{\rm eff}$, log($g$), [Fe/H], [Mg/Fe] or
  [$\alpha$/Fe] are presented in Table \ref{finalparam}.
  In order to discuss these results, we proceed as follows:  
  in Sect. \ref{sec:isochrones} we
  plot T$_{\rm eff}$ and log($g$) for stars in each cluster together
  with isochrones of age and metallicity given in Table
  \ref{clusterparam}. Section \ref{sec:feshcat} compares
  [Fe/H] with CaT results from \cite{saviane+12} and Vasquez et al. (in
  prep.). Subsequently
  all checked parameters are used to select member stars for each
  cluster (Section \ref{sec:membership}). Finally, all parameters for
  member stars are compared individually with high-resolution
  analysis, when available in the literature. M~71 and NGC~6558 have three stars in
  common with \cite{cohen+01}, and \cite{barbuy+07} respectively,
 and Terzan~8 has four stars in common with
  \cite{carretta+14}, as described in
  Sections \ref{sec:validm71}, \ref{sec:validn6558}
  and \ref{sec:validter8}, respectively. For NGC~6528, NGC~6553 and
  NGC~6426 we did not find any star in common with high-resolution
  spectroscopic studies.

\subsection{T$_{\rm eff}$, log($g$) against isochrones}
\label{sec:isochrones}

In high-resolution spectroscopy studies, usually T$_{\rm eff}$
 is estimated from
photometry and log($g$) from theoretical equations\footnote{log($g$) =
  4.44 + 4log$\frac{\rm T}{\rm T_{\odot}}$ + 0.4(M$_{\rm bol}$ - 4.75)
+ log$\frac{M}{M_{\odot}}$, see for example \cite{barbuy+09}}. 
These parameters are employed as initial guesses to derive [Fe/H], 
 which is applied to redetermine  T$_{\rm eff}$ and log($g$) iteratively, 
until reaching a convergence of the three parameters.
In this work we fit all the three parameters at the same
time (Section \ref{atmosparam}), and a check on these
 parameters is carried out as explained below.

\begin{figure*}[!htb]
\centering
\includegraphics[width=0.7\textwidth]{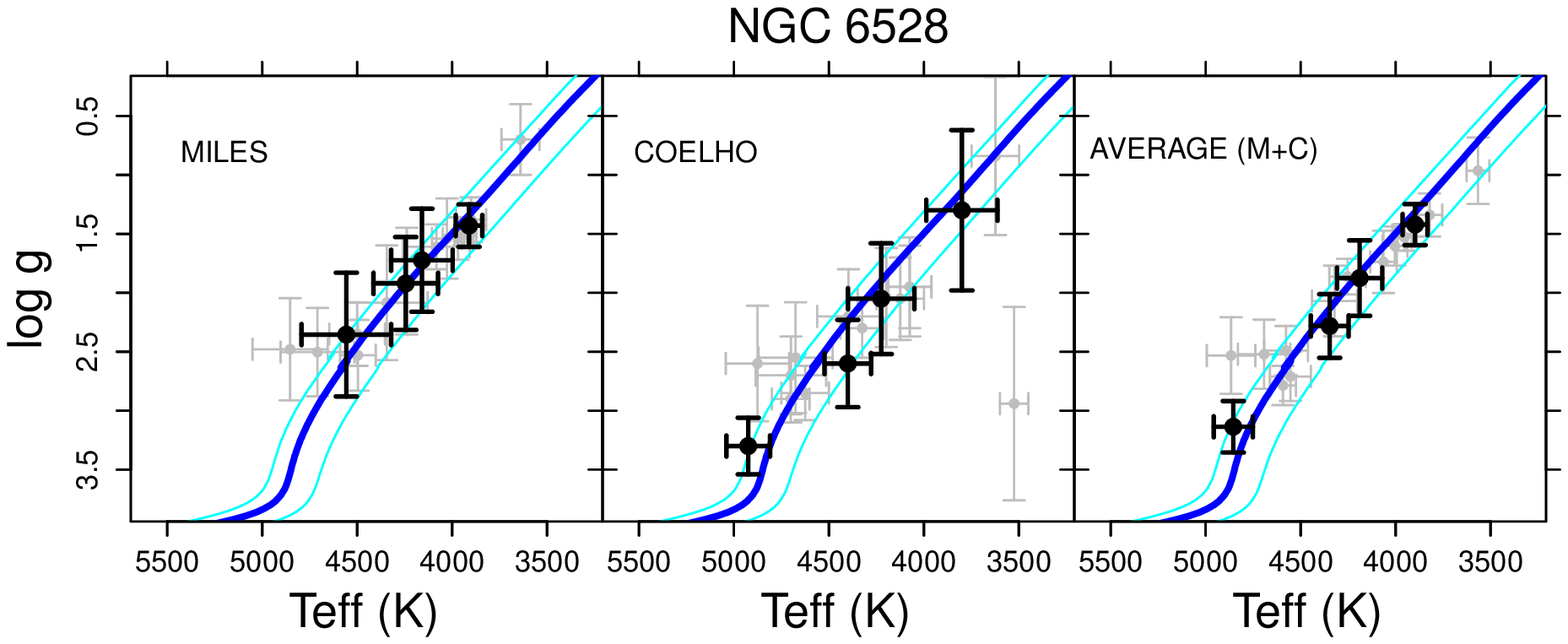}
\includegraphics[width=0.7\textwidth]{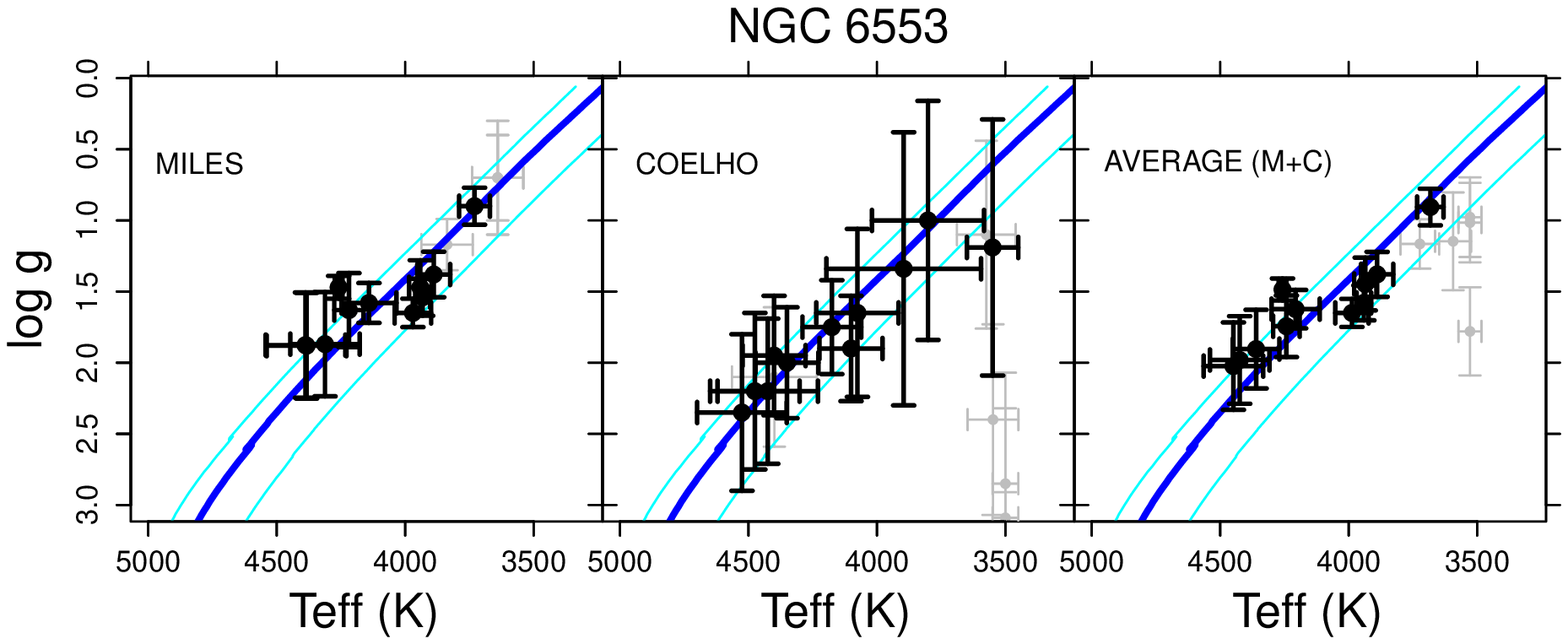}
\caption{Comparison of T$_{\rm eff}$ and log($g$) of stars in each
cluster with Dartmouth isochrones \citep{dotter+08} for the metal-rich
clusters NGC\,6528 and NGC\,6553. For each cluster we
show the results based on MILES and COELHO libraries and a third panel with
the weighted average of the results from both libraries. The parameters
of age, [Fe/H] and [$\alpha$/Fe] for the blue thick isochrones were
taken from Table \ref{clusterparam}. Cyan thin
isochrones have same age and [Fe/H] as blue ones, but with the limits
[$\alpha$/Fe] = -0.2dex and +0.8dex. Black dots represent member stars
of each cluster, and grey dots are not members.}
\label{isochrones}
\end{figure*}

Figure \ref{isochrones} displays the results of all stars in the six
clusters in a Hertzprung-Russell diagram form. Left, middle and right
panels show the results using MILES library, COELHO library and the
average of both results, respectively. Black dots represent member
stars of each cluster, and grey dots are not members, based on the
selection described in the next Section \ref{sec:membership}.
Dartmouth isochrones \citep{dotter+08} with age, [Fe/H] and
[$\alpha$/Fe] from Table \ref{clusterparam} are
overplotted in the diagrams of Figure \ref{isochrones} in blue. Cyan
lines have the same age and [Fe/H] as the respective blue lines, but
with the extreme values of [$\alpha$/Fe] = -0.2 and +0.8, available
from the models.

\begin{figure*}[!htb]
\centering
\includegraphics[width=0.7\textwidth]{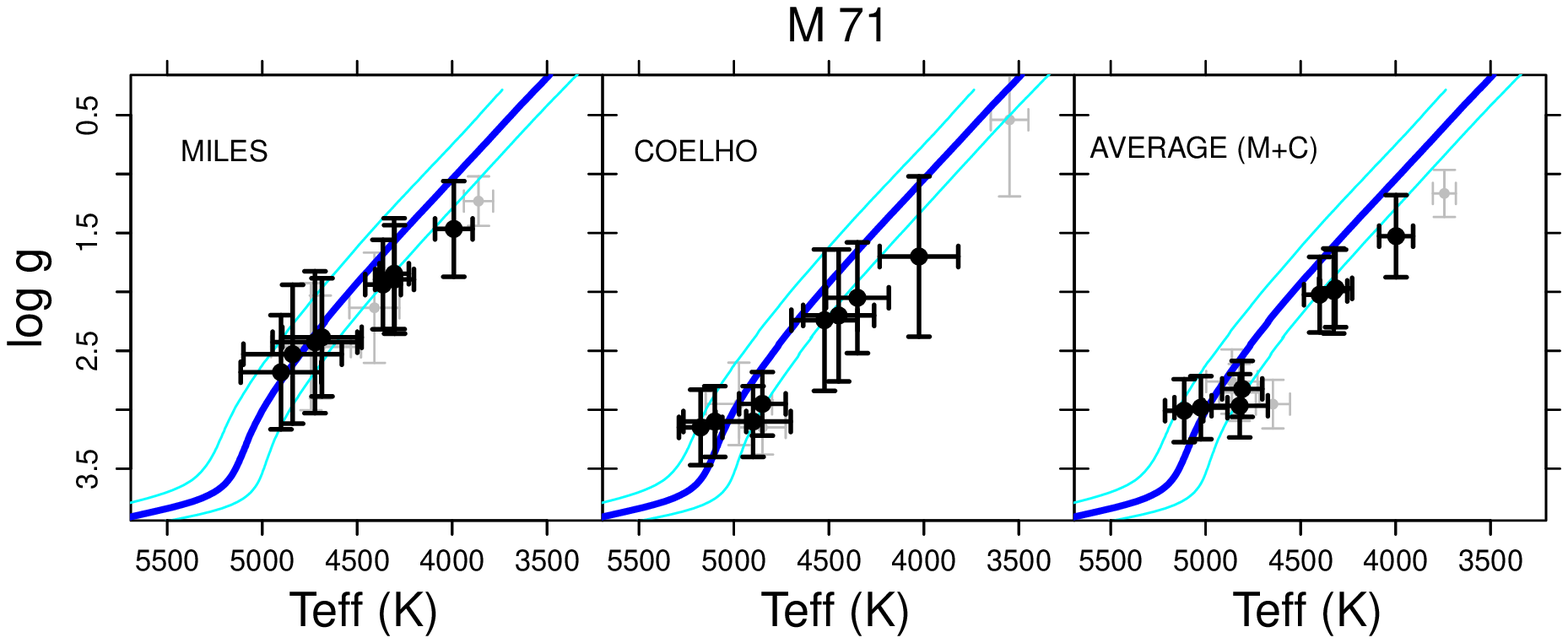}
\includegraphics[width=0.7\textwidth]{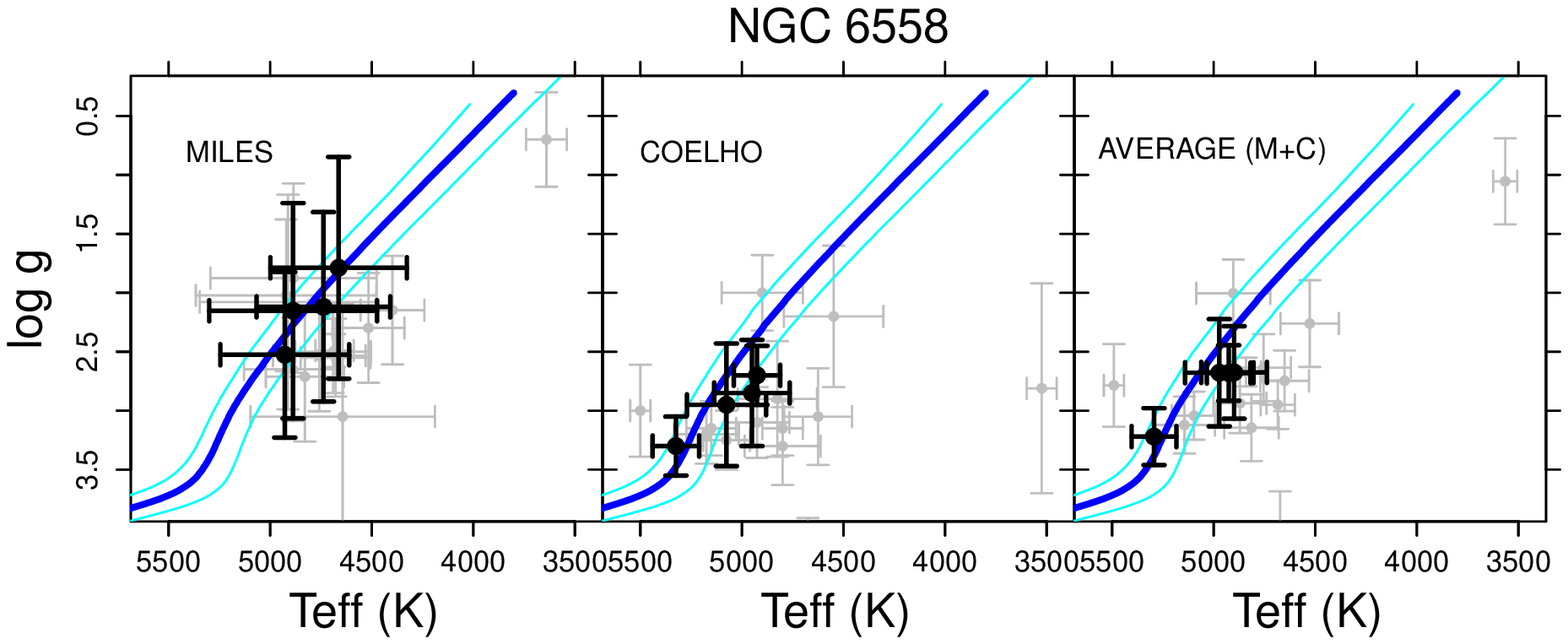}
\caption{Same as Figure \ref{isochrones} for the two clusters with
  intermediate metallicity M\,71 and NGC\,6558.}
\label{isochronesB}
\end{figure*} 

\begin{figure*}[!htb]
\centering
\includegraphics[width=0.7\textwidth]{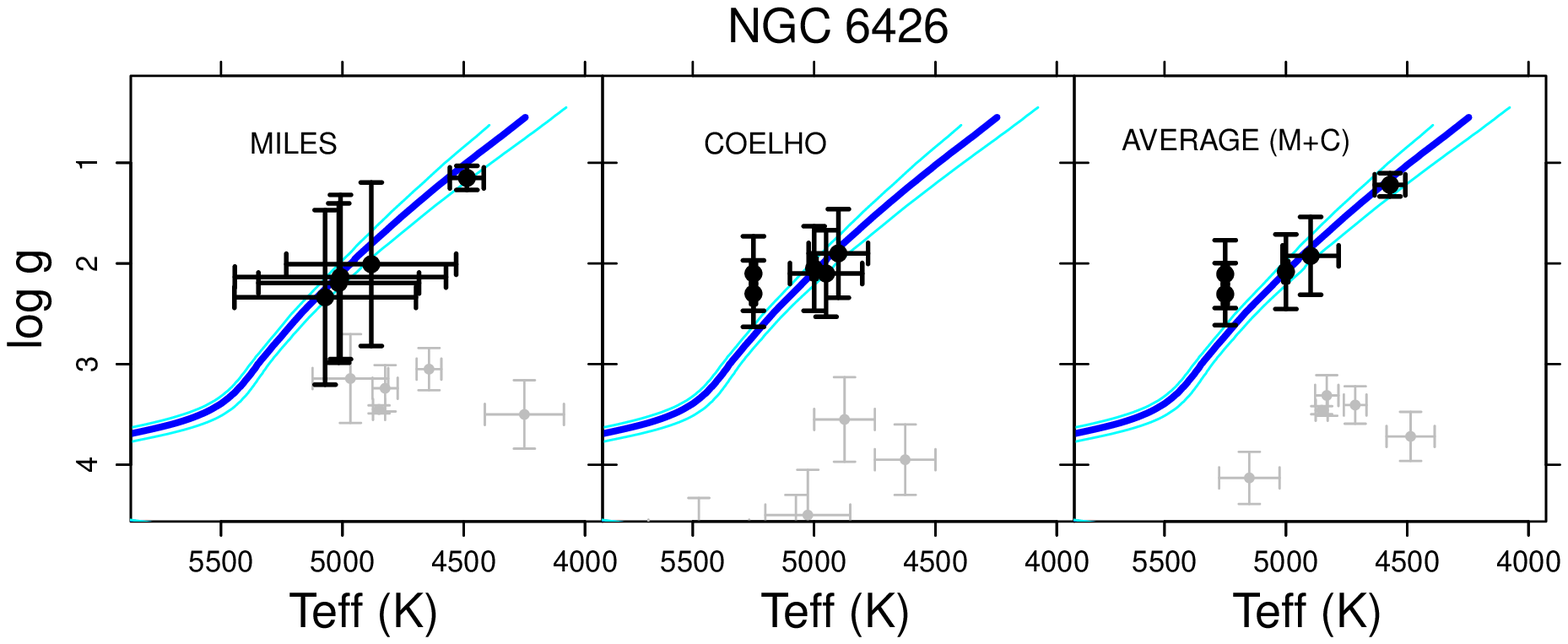}
\includegraphics[width=0.7\textwidth]{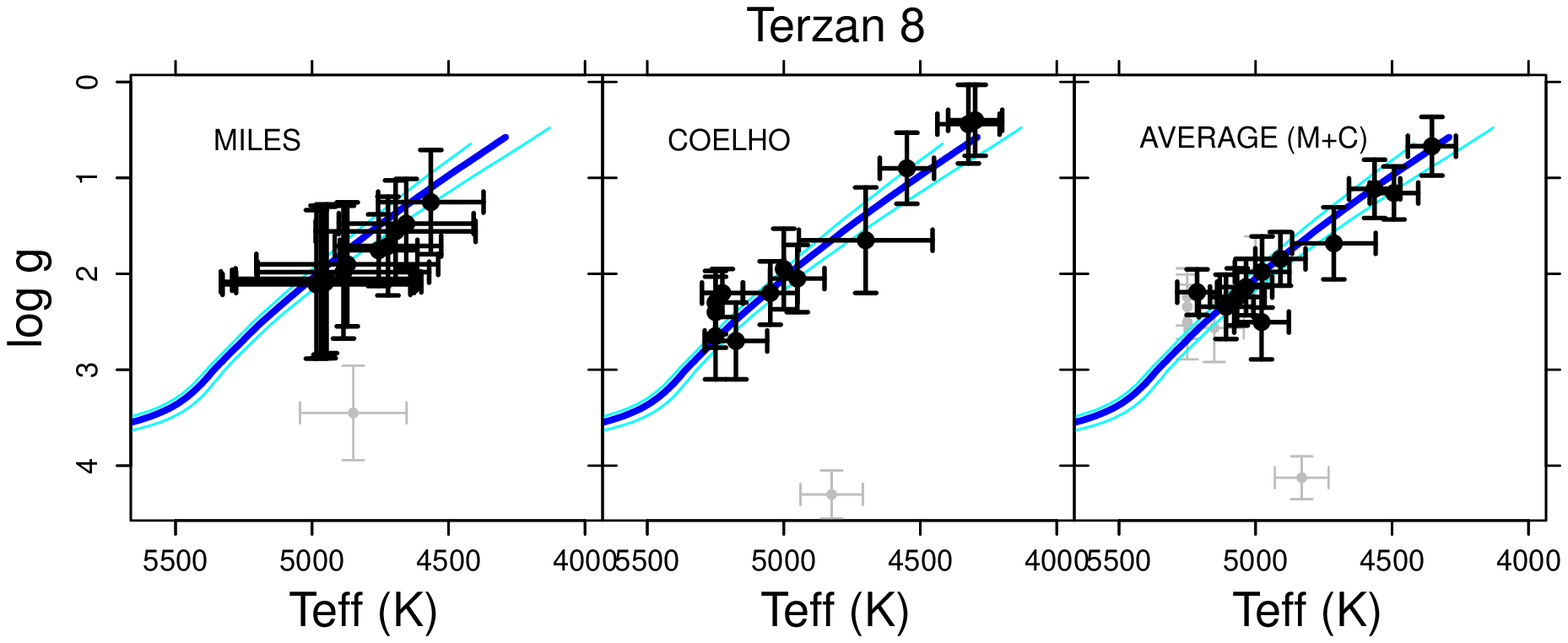}
\caption{Same as Figure \ref{isochrones} for the two metal-poor
  clusters NGC\,6426 and Terzan\,8.}
\label{isochronesC}
\end{figure*} 

 The results on T$_{\rm eff}$ and log($g$) from MILES and COELHO
  are in good agreement with the isochrones. We also computed an average
  of the results weighted by their uncertainties that are displayed 
in the right panels of Figure
  \ref{isochrones}.
For reasons explained in Section \ref{sec:feshcat}, we adopted
as the final results in this work the 
weighted average of MILES and COELHO results.

\subsection{Comparison of [Fe/H] with CaT results}
\label{sec:feshcat}

In the comparisons of results from CaT and the optical spectra,
it is important to keep in mind the facts that:
The synthetic spectra in the optical reproduce
less well the metal-rich stars, given the missing opacity due to
millions of very weak lines, not taken into account in the
calculations; this blanketing effect lowers the continuum in real
stars, and the measurable lines are shallower than in the present
synthetic spectra calculations by Coelho et al. which makes
  metal-rich stars more similar to synthetic spectra slightly more
  metal-poor. On the other hand, CaT-based abundances also suffer
    from significant uncertainties.  The modelling of the CaT region
    is affected by contamination by TiO lines and NLTE effects.
 Moreover, measuring a CaT index 
is very difficult, in particular for more metal-rich and luminous
  stars with the blanketing effect mentioned above, which complicates
  the definition of the continuum for equivalent widths (EW)
  measurements. 
The conversion of the EW to [Fe/H]
  has larger uncertainties which could recover even higher [Fe/H] for
  metal-rich stars. Another difficulty to measure
  EW for metal-rich ([Fe/H] $\gtrsim$ -0.7, 47 Tuc) is to choose the
  best function to fit the line profile: Gaussian,
  Gaussian+Lorentzian or Moffat, while for lower metallicities only
  a Gaussian function works well. This further step could introduce
  uncertainties in [Fe/H] from CaT in metal-rich regime.
A further issue is that the ratio between [Ca/H] vs. [Fe/H] is
not solar, i.e., since Ca is an alpha-element, it is enhanced
in old stars, even if not as enhanced as O and Mg. Detailed
  discussion about CaT metallicities can be found in
  \cite{saviane+12}. On the other hand,
there are some advantages to compare our results with CaT:
a) all selected stars from photometry were observed both
 in the near-infrared (CaT,
  \citealp{saviane+12} and Vasquez et al. in prep.) and in the optical
  spectral region which is very good for comparisons of the whole
    sample at once; b) The CaT-based
    metallicities were calibrated with the \cite{carretta+09}
 metallicity scale which makes CaT metallicities valid at least
   up to [Fe/H] $<$ -0.43 (the most metal-rich cluster observed by
   \citealp{carretta+09}, NGC~6441), with caution for metallicities
   higher than that.
  Finally, the optical region studied here is suitable to provide robust 
   values of [Fe/H] for each cluster, to be compared with the CaT value,
  and to converge ultimately to the average [Fe/H] for each cluster.

\begin{figure}[!htb]
\centering
\includegraphics[width=\columnwidth]{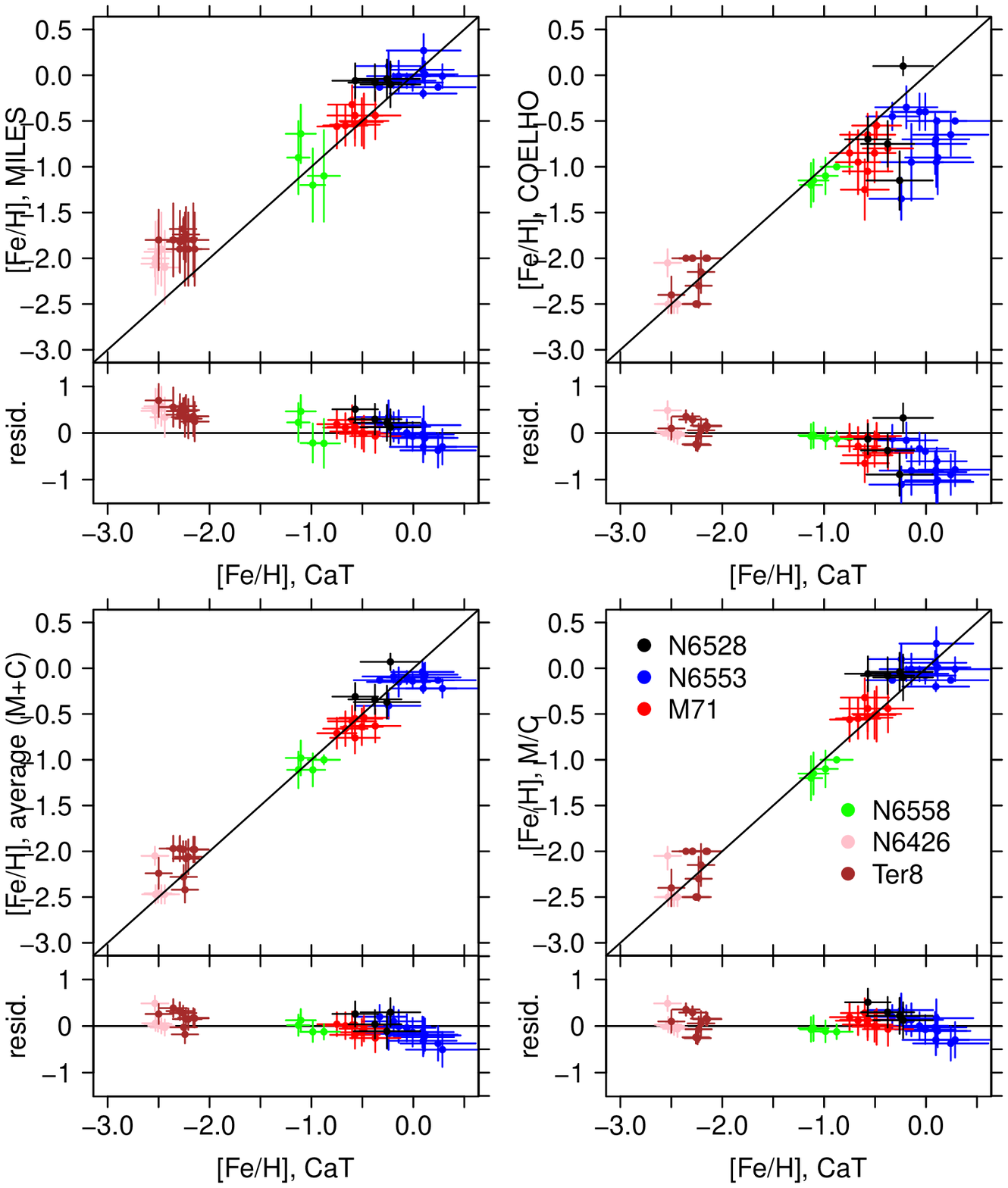}
\caption{Comparison of [Fe/H] from this work (mean of MILES and
      COELHO, see Table \ref{finalparam}) with those from equivalent widths of Ca II triplet
  for the same stars with the same instrument by \cite{saviane+12} and
Vasquez et al. (in prep). Upper panels compare CaT metallicities with
those obtained with MILES and COELHO libraries. Bottom panels are two
types of combination of the results: average of each star on the left,
and assuming MILES results for more metal-rich and COELHO results
for more metal-poor stars, on the right. Below the plots
    there is a residuals plot.}
\label{cat-bd}
\end{figure}

Figure \ref{cat-bd} shows the comparisons of the metallicity values
presented in Table \ref{finalparam} with those from CaT
analysis. Upper left panel shows that [Fe/H] using MILES library is in
good agreement with CaT results for the three most metal-rich clusters
NGC~6528, NGC~6553 and M~71. This is in agreement with the sampling
of metal-rich stars for all combinations of T$_{\rm eff}$ and log($g$)
as displayed in Figure \ref{mileshrd} in red and green. For NGC~6558
with [Fe/H]$\sim$-1.0 the dispersion on the parameters is larger which is
explained by the smaller number of stars available in the library with
such metallicity. 
MILES is based on the solar neighbourhood showing therefore only a
few stars with [Fe/H]$\sim$-1.0. For the metal-poor clusters NGC~6426
and Terzan~8 the library sampling is even more sparse, as becomes
evident in Figure \ref{mileshrd}. In this case, the average of
parameters from the library takes into account some more metal-rich
reference stars which results in higher values of [Fe/H] for NGC~6426
and Terzan~8 stars.

Metallicities using COELHO library are compared with CaT results in
the upper right panel of Figure \ref{cat-bd}.  The synthetic spectra reproduce
less well the metal-rich stars, given the missing opacity 
as mentioned above.
Because of this
effect, stars of NGC~6528, NGC~6553 and M~71 are more metal-poor than
CaT results. On the other hand, COELHO library is suitable to reproduce
the stars of the three more metal-poor clusters of this sample,
NGC~6558, NGC~6426 and Terzan~8.

To summarize, for the three more metal-rich clusters, MILES results
are better, and for the other three, COELHO results are
preferable. The bottom right panel of Figure \ref{cat-bd} shows the
concatenation of this conclusion, i.e., it displays MILES results for
NGC~6528, NGC~6553 and M~71, and COELHO results for NGC~6558, NGC~6426
and Terzan~8. An alternative combination of results from MILES and
COELHO is to take the average of the results weighted by their
uncertainties. This average combines the best of both libraries and
gives a good correlation with CaT results, as shown in the bottom left
panel of Figure \ref{cat-bd}. Both criteria to combine MILES and
COELHO (two bottom panels) are in good agreement with CaT results, and
we adopted [Fe/H] from the average results represented in the bottom
left panel.

We adopted as final parameters the mean of MILES and COELHO
  results, because they show better compatibility with the isochrones
  for T$_{\rm eff}$ and log($g$), and with the CaT results
  for metallicities. We recall that CaT-based metallicities were
  calibrated in the \cite{carretta+09} scale.

\subsection{Membership selection}
\label{sec:membership}

\begin{figure}[!htb]
\centering
\includegraphics[width=\columnwidth]{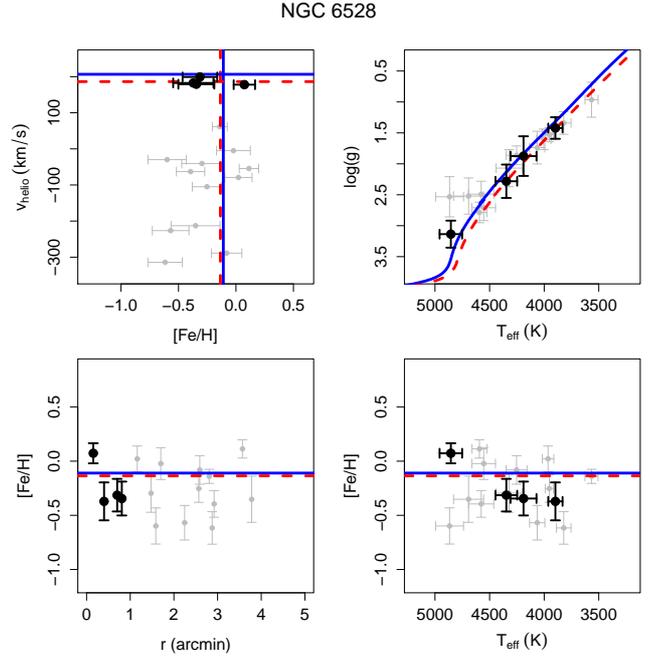}
\caption{Step-by-step of selection of member stars for
  NGC\,6528. Black dots are selected member stars, grey dots are 
  non-members and green circles shows stars considered as cluster
  member by \cite{saviane+12} but non-members in the present
  work. Blue solid lines are drawn based on values of Table
  \ref{clusterparam}, which were also applied to the ischrones from
  \cite{dotter+08}. Red dashed lines refer to the weighted average
  of the member stars parameters.}
\label{memberselection}
\end{figure} 

\begin{figure}[!htb]
\centering
\includegraphics[width=\columnwidth]{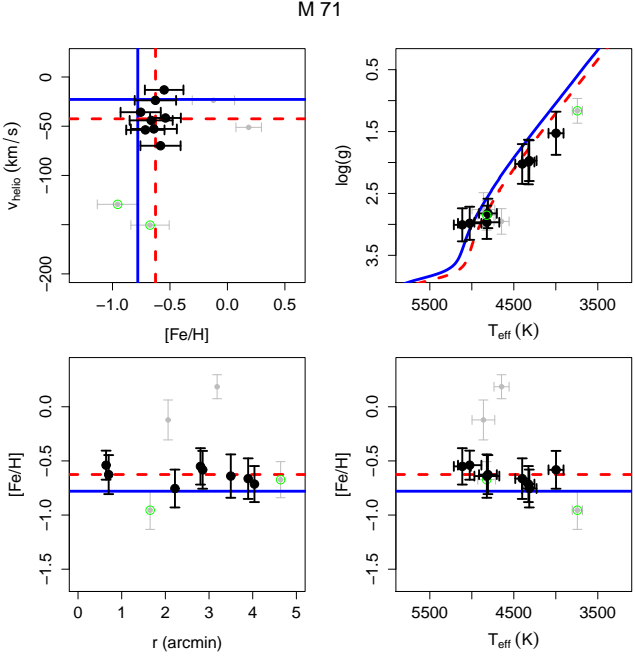}
\caption{Same as Figure \ref{memberselection} for M\,71.}
\label{memberselectionB}
\end{figure} 

\begin{figure}[!htb]
\centering
\includegraphics[width=\columnwidth]{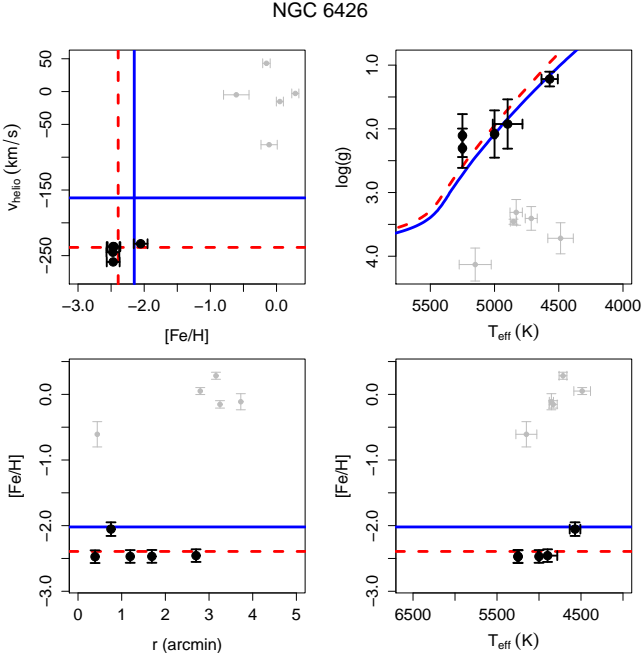}
\caption{Same as Figure \ref{memberselection} for NGC\,6426.}
\label{memberselectionC}
\end{figure} 

\begin{figure}[!htb]
\centering
\includegraphics[width=\columnwidth]{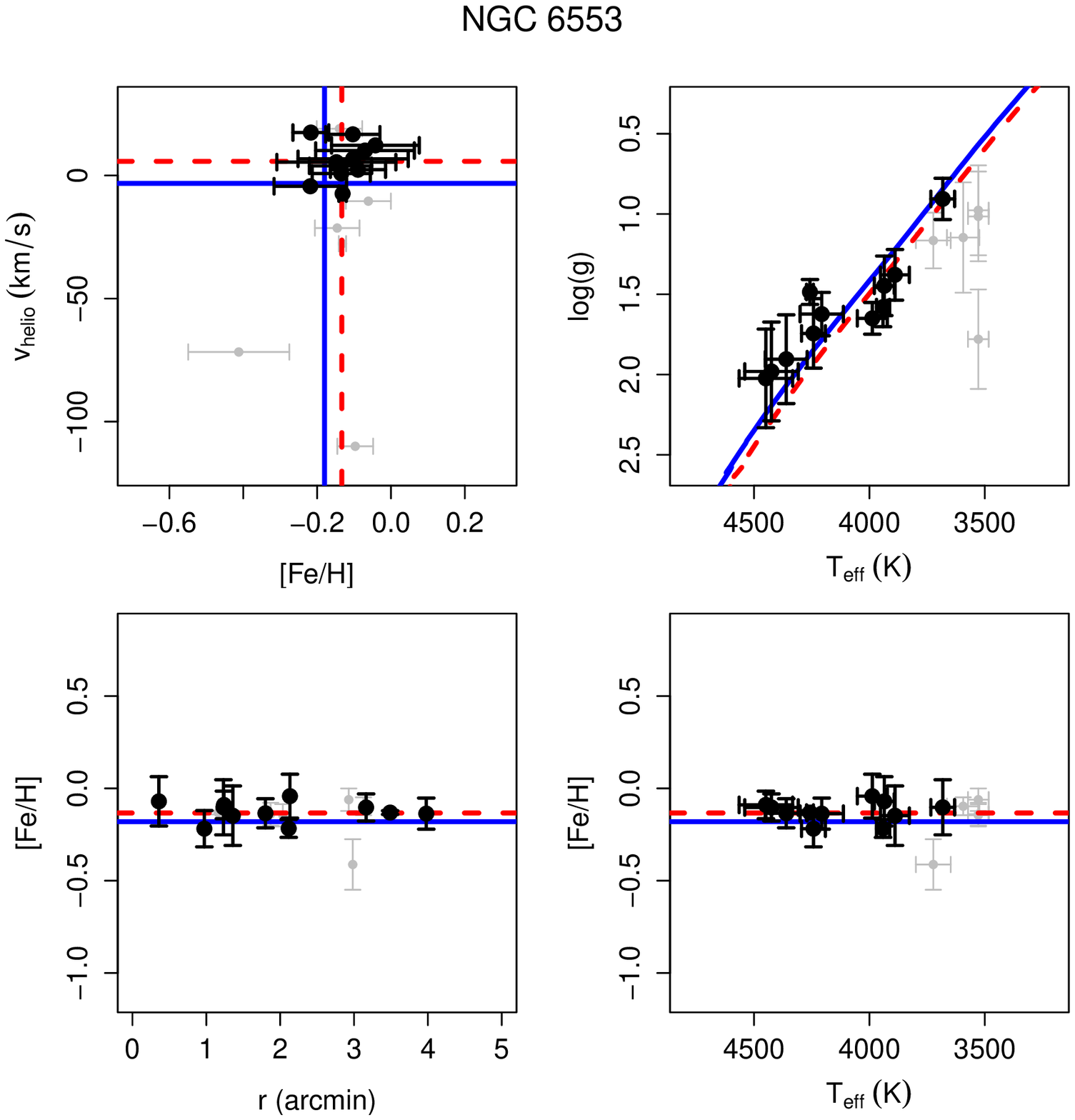}
\caption{Same as Figure \ref{memberselection} for NGC\,6553.}
\label{memberselectionD}
\end{figure} 

\begin{figure}[!htb]
\centering
\includegraphics[width=\columnwidth]{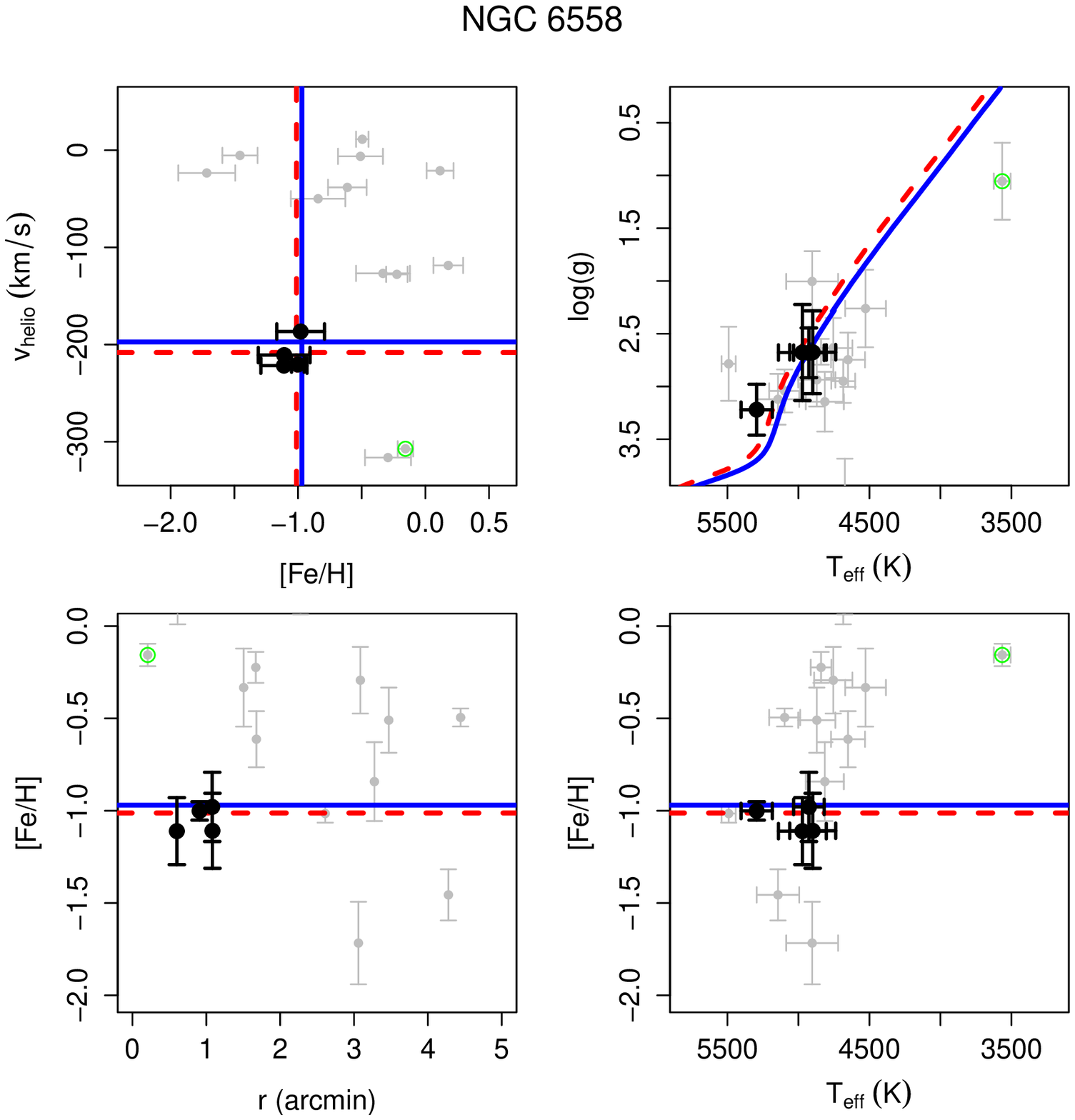}
\caption{Same as Figure \ref{memberselection} for NGC\,6558.}
\label{memberselectionE}
\end{figure} 

\begin{figure}[!htb]
\centering
\includegraphics[width=\columnwidth]{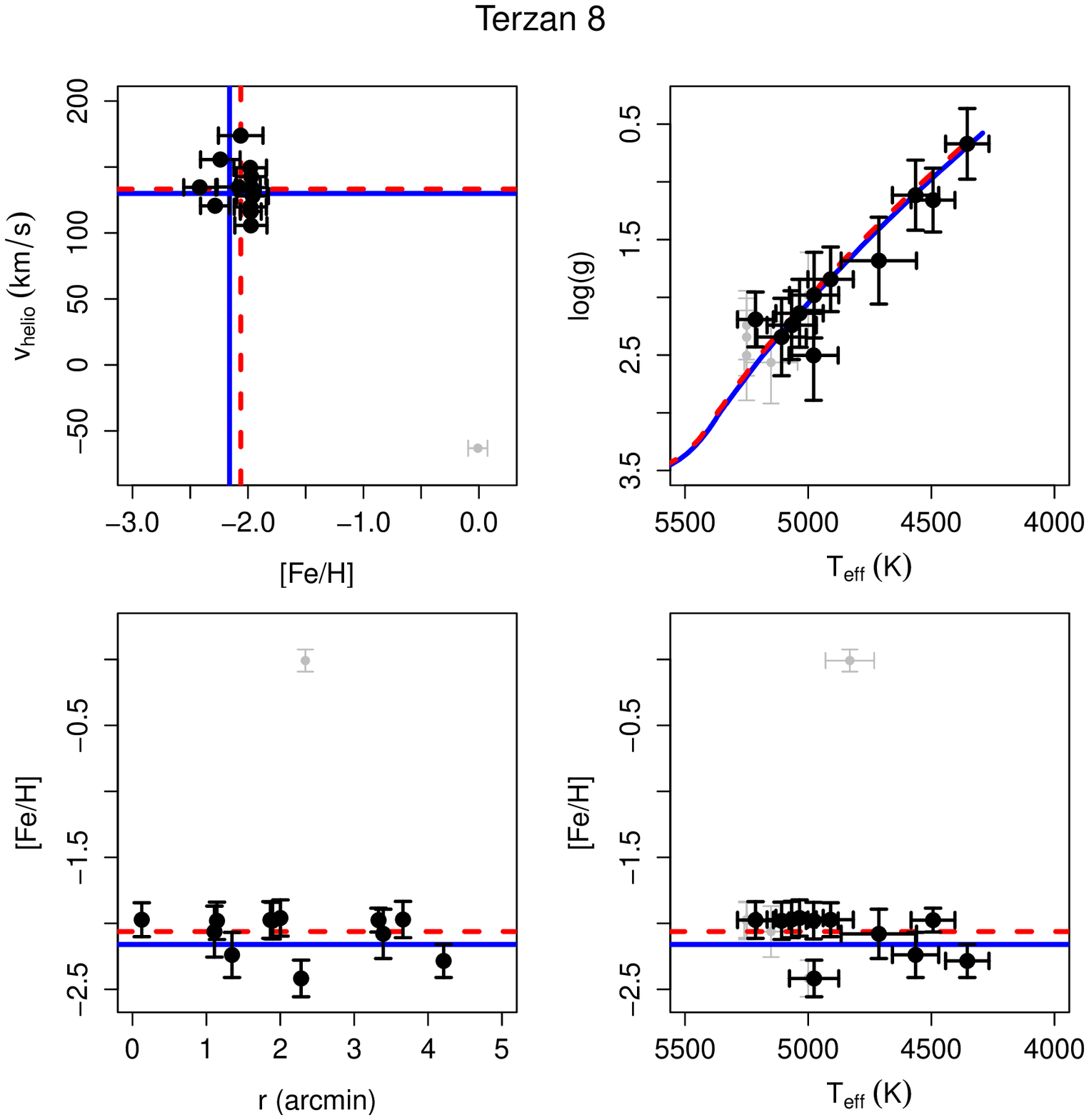}
\caption{Same as Figure \ref{memberselection} for Terzan\,8.}
\label{memberselectionF}
\end{figure}

  In Figure \ref{memberselection} we show four panels for each globular
  cluster. In upper left panels radial velocities against
  metallicities are displayed
(this is the classical plot for membership selection for globular
clusters, e.g. \citealp{zoccali+08}).
Blue solid lines are parameters from 
Table \ref{clusterparam} and red dashed lines are the average of
parameters for cluster members only (for isochrones we use age,
  [Fe/H] and [$\alpha$/Fe] information). For all clusters there is a clear
concentration of stars around literature values, and we
  considered as members stars with v$_{\rm helio} < \pm 2\sigma$
($\sigma$ is given in Sect. 3.1) of
  the literature value, and with [Fe/H] $< \pm 0.3$~dex.
  Upper right panels show T${\rm eff}$  vs. log($g$) compared with Dartmouth isochrones
\citep{dotter+08} as done in Sect. \ref{sec:isochrones}. 
Blue solid lines are isochrones with parameters from Table
\ref{clusterparam}, and red dashed lines are based on the average
(Table \ref{clusterparamavg}) of the parameters for member stars
(Table \ref{finalparam}). All
member stars are close to the isochrones, 
confirming the membership selection. This
  extra criterion led to the exclusion of a few more stars from the
  v${\rm helio}$-[Fe/H] selection. Also excluded in some cases are
  stars cooler than T$_{\rm eff}$ $<$ 4000~K that give a poor fit to
  template spectra due to TiO bands.
Bottom left panels
show [Fe/H] against distance to the cluster centre, and no trends
were found for any sample cluster. Bottom right panels
 show no  correlation between T$_{\rm eff}$ and [Fe/H] for any of 
the six clusters analysed in this work.

  In conclusion all T${\rm eff}$, log($g$) and [Fe/H] values are
  found to sit in well defined sequences following the isochrones.
  Bottom-right panels show no correlation between
[Fe/H] and T${\rm eff}$, suggesting that the [Fe/H]/T${\rm eff}$ degeneracy does not
affect our fitting procedure. Further evidence of this is provided by
the comparisons of our results with [Fe/H] measures from high-resolution analysis available
in the literature, presented in Sections \ref{sec:validation}, \ref{sec:validm71}, \ref{sec:validn6558}
  and \ref{sec:validter8}. 

\begin{table*}[!htb]
  \caption{Final average parameters for member stars in each globular
    cluster, and respective internal errors.}
\label{clusterparamavg}
%\scriptsize
  \centering
  \begin{tabular}{l|cccccc}
\hline \hline \noalign{\smallskip}
Cluster &  $<$v$_{\rm helio}>$ (km/s) & { $<$[Fe/H]$>^{(a)}$}  &   { $<$[Fe/H]$>^{(b)}$ } &    { $<$[Fe/H]$>^{(avg)}$ } &  $<$[Mg/Fe]$>^{(a)}$  &    $<$[$\alpha$/Fe]$>^{(b)}$   \\
\noalign{\smallskip}\hline \noalign{\smallskip}
NGC 6528   &  185$\pm$10    &  { -0.07$\pm$0.10  }    &   {-0.18$\pm$0.08  }   &   { -0.13$\pm$0.05    }   &   0.05$\pm$0.09     &    0.26$\pm$0.05  \\
NGC 6553   &    6$\pm$8     & { -0.125$\pm$0.009  }   &  { -0.55$\pm$0.07  }   &   { -0.133$\pm$0.017  }  &  0.107$\pm$0.009    &   0.302$\pm$0.025 \\
\noalign{\smallskip}
M 71       &  -42$\pm$18    &  { -0.48$\pm$0.08  }    &  { -0.77$\pm$0.08   }  &   { -0.63$\pm$0.15    }   &   0.25$\pm$0.07     &   0.293$\pm$0.032 \\
NGC 6558   & -210$\pm$16    &  { -0.88$\pm$0.20  }    &  { -1.02$\pm$0.05  }   &   { -1.012$\pm$0.013  }    &   0.26$\pm$0.06     &    0.23$\pm$0.06  \\
\noalign{\smallskip}
NGC 6426   & -242$\pm$11    &   { -2.03$\pm$0.11  }    &  { -2.46$\pm$0.05 }    &   { -2.39$\pm$0.11  }     &   0.38$\pm$0.06     &    0.24$\pm$0.05  \\
Terzan 8   &  135$\pm$19    &  { -1.76$\pm$0.07  }    &  { -2.18$\pm$0.05  }   &    { -2.06$\pm$0.17  }     &   0.41$\pm$0.04     &    0.21$\pm$0.04  \\
\noalign{\smallskip}\hline
  \end{tabular}  
\tablefoot{ 
\tablefoottext{a}{ MILES library}
\tablefoottext{b}{ COELHO library}
\tablefoottext{avg}{ Average of MILES and COELHO results}
}
\end{table*}

%%Table \ref{finalparam}
%%parameters for all stars analysed in this work
\addtocounter{table}{1}

\subsection{Validation with high-resolution spectroscopy}
We found stars in common with literature high-resolution spectroscopy
  for three clusters: M~71, NGC~6558 and Terzan~8.
In Sect. \ref{sec:membership} we were able to identify ten member
stars of M~71, five member stars of
NGC~6558, and twelve member stars in Terzan~8,
these being the same selected by \cite{saviane+12}, and
 Vasquez et al. (in prep.).
The derived stellar parameters are reported in Table \ref{finalparam}
for member and non-member stars.
We were able to find detailed analyses in the
literature for 3 member stars in M~71, 3 in NGC~6558 and 4 in Terzan~8,
as reported below.

\subsubsection{M 71}
\label{sec:validm71}

\cite{cohen+01} observed 25 member red giant stars of M~71 using
HIRES@Keck (R$\sim$34,000), and derived their T$_{\rm eff}$ and
log($g$). In two subsequent papers, they derived [Fe/H]
\citep{ramirez+01} and [Mg/Fe] \citep{ramirez+02} for them. We have
three stars in common that are presented in Table \ref{m71-cohen}.
Temperature and gravity values are compatible within 0.5 to
2-$\sigma$, [Fe/H] and [Mg/Fe] are compatible within 0.1 to
1.5-$\sigma$.

\begin{table}[!htb]
  \caption{ Final atmospheric parameters for the three stars of M~71 in
    common with \cite{cohen+01}, and also their 
    determinations for the respective parameters.}
\label{m71-cohen}
\scriptsize
  \centering
  \begin{tabular}{l|l|l|l|l}
\hline \hline \noalign{\smallskip}
Star & T$_{\rm eff}$ (K)  & log($g$) & [Fe/H]  &  [Mg/Fe]   \\
       & T$_{\rm eff}$-C01 (K) & log($g$)-C01 & [Fe/H]-C01  &  [Mg/Fe]-C01   \\
\noalign{\smallskip}\hline \noalign{\smallskip}
 M71\_7  & 3997$\pm$89  &      1.53$\pm$0.35   &      -0.58$\pm$0.17   &   0.15$\pm$0.18         \\
  1-45      & 3950                   &      0.9                        &      -0.60$\pm$0.03   &  0.43$\pm$0.09          \\
\noalign{\smallskip}
 M71\_9   &  4316$\pm$87 &      1.97$\pm$0.33    &     -0.76$\pm$0.17    &   0.27$\pm$0.21      \\
   1-64     &  4200                  &     1.35                        &     -0.61$\pm$0.03   &    0.43$\pm$0.09      \\
\noalign{\smallskip}
 M71\_13             &  4808$\pm$106 &   2.82$\pm$0.24  &     -0.63$\pm$0.18      &   0.23$\pm$0.20      \\
 G53476\_4543  &    4900                    &   2.65                      &    -0.61$\pm$0.03      &   0.36$\pm$0.06      \\
\noalign{\smallskip}\hline
  \end{tabular}  
\end{table}

\subsubsection{NGC 6558}
\label{sec:validn6558}

\cite{barbuy+07} observed six RGB stars using the high-resolution
(R$\sim$22,000) spectrograph FLAMES+GIRAFFE@VLT/ESO, and derived
T$_{\rm eff}$, log($g$), [Fe/H], and [Mg/Fe] for each of them. We 
have three stars in common with their sample: \#6, \#8, \#9,
corresponding to their identification as B11, F42, F97, respectively
(see Table \ref{n6558-barbuy}). 
For stars \#6 and \#9, full spectrum fitting recovers all
parameters within 1-$\sigma$.
Star \#8 is a more complicated case because it is a very cool star (T$_{\rm
  eff} < 4000$~K) and molecular bands of TiO are important. They
change a lot the continuum which is not fitted perfectly. In fact, the
derived parameters for this star led us to select it as non-member.
Although temperature agrees with that from \cite{barbuy+07}, the
gravity is much lower than their results.

These results show that full spectrum fitting method is reliable,
consistent among all libraries, and present reasonable
errors for RGB stars hotter than $\sim$4000~K. Stars cooler than that
must be analysed with a better suited reference library, 
  cointaining a sufficient number of cool stars at all metallicities.

\begin{table}[!htb]
  \caption{Final atmospheric parameters for the three stars of NGC
    6558 in common with \cite{barbuy+07}, and also their
    determinations for the respective parameters.}
\label{n6558-barbuy}
\scriptsize
  \centering
  \begin{tabular}{l|l|l|l|l}
\hline \hline \noalign{\smallskip}
Star & T$_{\rm eff}$ (K)  & log($g$) & [Fe/H]  &  [Mg/Fe]   \\
       & T$_{\rm eff}$-B07 (K) & log($g$)-B07 & [Fe/H]-B07  &  [Mg/Fe]-B07   \\
\noalign{\smallskip}\hline \noalign{\smallskip}
 6558\_6  & 4899$\pm$162  &      2.68$\pm$0.39   &      -1.11$\pm$0.20   &   0.22$\pm$0.07         \\
B11       &   4650                 &        2.2                     &         -1.04                 &       0.20                        \\
\noalign{\smallskip}
 6558\_8   &  3565$\pm$ 59 &      1.05$\pm$0.36    &     -0.16$\pm$0.06    &   0.23$\pm$0.00      \\
F42        &  3800                  &       0.5                      &       -1.01                    &      0.30                    \\
\noalign{\smallskip}
 6558\_9   &  4972$\pm$168 &   2.68$\pm$0.46      &     -1.11$\pm$0.18       &   0.41$\pm$0.16      \\
F97       &    4820                &      2.3                       &      -0.97                        &      0.23               \\
\noalign{\smallskip}\hline
  \end{tabular}  
\end{table}

\subsubsection{Terzan 8}
\label{sec:validter8}

\cite{carretta+14} observed six stars with UVES@VLT/ESO (R$\sim$45,000)
 and 14 with GIRAFFE@VLT/ESO (R$\sim$22,500 - 24,200), 
with among them four stars in common with our
FORS2@VLT/ESO sample. Their parameters for these stars are presented
in Table \ref{terzan8-carretta}. For
temperature and gravity the compatibility is within 1 to 3-$\sigma$,
except for T$_{\rm eff}$ of star Ter8\_8 which is in the limit of
3.9-$\sigma$ of distance. For [Fe/H] all stars have compatible values
with \cite{carretta+14} within 1-$\sigma$, except for star Ter8\_1
which is in the limit of 3.9-$\sigma$ of distance. [Mg/Fe] is
compatible within 1-$\sigma$. 

\begin{table}[!htb]
  \caption{Final atmospheric parameters for the four stars of Terzan~8 in 
common with \cite{carretta+14}, together with their determinations for the 
respective parameters.}
\label{terzan8-carretta}
\scriptsize
  \centering
  \begin{tabular}{l|l|l|l|l}
\hline \hline \noalign{\smallskip}
Star & T$_{\rm eff}$ (K)  & log($g$) & [Fe/H]  &  [Mg/Fe]   \\
       & T$_{\rm eff}$-C14 (K) & log($g$)-C14 & [Fe/H]-C14  &  [Mg/Fe]-C14   \\
\noalign{\smallskip}\hline \noalign{\smallskip}
 Ter8\_1    & 5067$\pm$314  &      2.24$\pm$0.30     &      -1.97$\pm$0.14   &   0.40$\pm$0.13         \\
2913      &   4628                 &        1.49                     &     -2.52$\pm$0.07   &       0.58                        \\
\noalign{\smallskip}
 Ter8\_4   &  4354$\pm$88   &      0.67$\pm$0.31    &     -2.28$\pm$0.13    &   0.40$\pm$0.14      \\
2357     &  4188                  &       0.66                      &   -2.29$\pm$0.10   &      0.48$\pm$0.14     \\
\noalign{\smallskip}
 Ter8\_8   &  5151$\pm$108   &      2.56$\pm$0.36    &     -2.06$\pm$0.19    &   0.40$\pm$0.18      \\
2124     &  4730                   &       1.67                      &   -2.28$\pm$0.26   &      0.56                       \\
\noalign{\smallskip}
 Ter8\_9   &  4564$\pm$94   &      1.12$\pm$0.30    &     -2.24$\pm$0.17    &   0.42$\pm$0.13      \\
1658      &  4264                  &       0.80                      &   -2.40$\pm$0.07   &      0.51$\pm$0.02     \\
\noalign{\smallskip}\hline
  \end{tabular}  
\end{table}

\begin{figure}[!htb]
\centering
\includegraphics[width=\columnwidth]{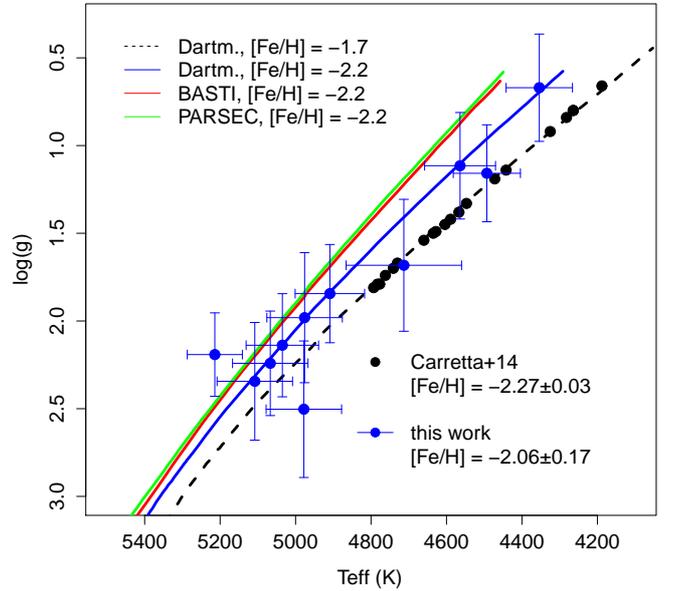}
\caption{Diagram log($g$) vs. T${\rm eff}$ for Terzan 8 showing
  present (blue points with error bars) and 
  \citet[][black filled circles]{carretta+14} results, overplotted with
  Dartmouth, PARSEC and BASTI isochrones for ages = 13~Gyr, and
  metallicities indicated on the Figure.}
\label{carr14hrd}
\end{figure}

 Differently from M~71 and NGC~6558, the comparison between our
results and \cite{carretta+14} for Terzan~8 give
 all three parameters T$_{\rm eff}$, log($g$) and [Fe/H] 
systematically larger. For this reason we inspected the T$_{\rm
  eff}$-log($g$) diagram of both sets of data, compared with the
Dartmouth \citep{dotter+08}, PARSEC \citep{bressan+12} and BASTI
\citep{pietrinferni+04} isochrones, as shown in Figure
\ref{carr14hrd}. The \cite{carretta+14} results are compatible with an
isochrone of [Fe/H] = -1.7, [$\alpha$/Fe]=+0.4 and 13~Gyr, whereas the
present results fit better with an isochrone [Fe/H] = -2.2,
[$\alpha$/Fe]=+0.4 and 13~Gyr.  Except for star Terzan8\_4
showing very similar gravities, for the other stars 
they are different. Given that are results 
are consistent with isochrones, we suggest that in the high-resolution
analysis of metal-poor stars, the effect
ofver-ionization at low temperature
atmospheres, may have led to lower gravities.

\section{Discussion}

\begin{figure}[!htb]
\centering
\includegraphics[width=\columnwidth]{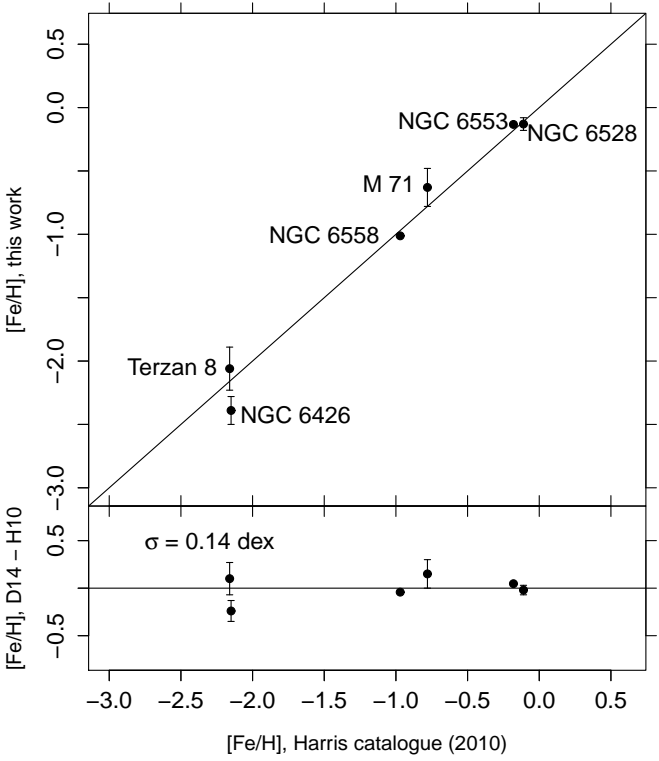}
\caption{Comparison of [Fe/H] from this work (Table
  \ref{clusterparamavg}) with those from literature, as revised in
  Table \ref{clusterparam}. Error bars are the weighted average of the
  [Fe/H] of member stars in each cluster, as presented in Table
  \ref{finalparam}. For NGC~6553 and NGC~6558 the error bars are not
  visible because they are too small (see Table
  \ref{clusterparamavg}). In the lower panel the residuals are
    shown.} 
\label{met-lit}
\end{figure}

\begin{figure}[!htb]
\centering
\includegraphics[width=\columnwidth]{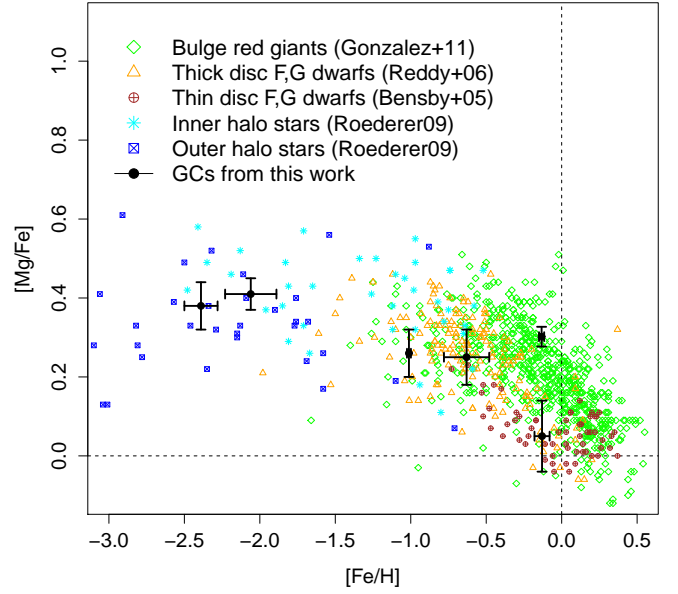}
\caption{Chemical evolution of the Milky Way based on bulge field stars
\citep{gonzalez+11}, thin disc \citep{bensby+05}, thick disc
\citep{reddy+06}, inner and outer halo \citep{roederer+09}. Results from 
this work for globular clusters are overplotted. For NGC~6553 the
point corresponds to [$\alpha$/Fe] instead of [Mg/Fe], as discussed in
Section \ref{sec:rich}}
\label{mgfeXfesh}
\end{figure}

Results for individual stars in each cluster (Table \ref{finalparam})
and average results (Table \ref{clusterparamavg}) are discussed below
and compared with literature results. 
Figure \ref{met-lit} displays the comparison of our
[Fe/H] results for each cluster with reference values, showing
good agreement for the whole range of metallicities from
[Fe/H] = -2.5 to solar. Figure \ref{mgfeXfesh} gives the comparison of
the average results of [Fe/H] and [Mg/Fe] with abundances
for field stars of the different Galactic components: bulge,
thin/thick disc and inner/outer halo. We discuss case by case below.

\subsection{Metal-rich clusters NGC 6528, NGC 6553}
NGC~6528 and NGC~6553 have similar CMDs (\citealp{ortolani+95}),
  as shown in Figure \ref{cmds}. Their metallicities and element
  abundance ratios are also similar for most elements, as reported in
  Table \ref{6553-abund}. As given in Table \ref{clusterparam}
they have reddening 
E(B-V) = 0.54~mag and 0.63~mag, [Fe/H] = -0.11 and -0.18, [Mg/Fe] =
0.24 and 0.26, respectively. 
They are located in the Milky Way bulge, distant 0.6~kpc
and 2.2~kpc from the Galactic centre, and
in opposite ``southern legs'' of the X-shaped bulge (see Figure 3 of
\citealp{saito+11}).
Figure \ref{isochrones} shows that ETOILE recovers
parameters for member stars coherent with a simple stellar population
represented by the isochrones. Although all the results are
compatible between the libraries, error bars for MILES results are lower than from COELHO results.
This synthetic
library, on the other hand, gives better alpha-enhancement values
compatible with average values from  high-resolution studies
([$\alpha$/Fe] = 0.26$\pm$0.05 and 0.302$\pm$0.025, respectively, from
Table \ref{clusterparamavg}). In particular these values
are compatible with NGC~6553 (\citealp{alves-brito+06},
\citealp{cohen+99}), whereas for NGC~6528 the 
alpha-enhancement is lower (\citealp{zoccali+04} ).
In this respect, MILES cannot give an alpha-enhancement, since
their metal-rich stars are basically solar neighbourhood stars, that have no
alpha-enhancement for metal-rich stars (see Figure 2 in
\citealp{milone+11}). The results from MILES are [Mg/Fe] =
0.05$\pm$0.09 and 0.107$\pm$0.009, respectively. This is a
particularity of the bulge, where stars are metal-rich and old.
As mentioned above, \cite{zoccali+04} find
 [Mg/Fe]=+0.07 from high-resolution
spectroscopy of three stars of NGC~6528, which is compatible with MILES
and not with COELHO. For N6553, \cite{cohen+99} finds [Mg/Fe]=+0.41
also from high-resolution spectroscopy of five stars, which is
closer to COELHO results.

Figure \ref{memberselection} compares the final results (average of
MILES and COELHO results) with isochrones with literature parameters
(same as Figure \ref{isochrones}) and an additional
isochrone considering the parameters derived from this work (Table
\ref{clusterparamavg}). The isochrones consider [$\alpha$/Fe] from COELHO
results as discussed above. We derived [Fe/H] = -0.13$\pm$0.06 and
-0.133$\pm$0.009 for NGC~6528 and NGC~6553, respectively, in agreement
with high spectral resolution analyses of \cite{carretta+01} and
  \cite{alves-brito+06}.

\begin{table*}
\caption{Literature abundances in NGC 6528 and NGC 6553.}
\label{6553-abund}
\centering
%\begin{flushleft}
%\tabcolsep 0.15cm
\begin{tabular}{cccccccccccc}
\noalign{\smallskip}
\hline
\noalign{\smallskip}
\hline
\noalign{\smallskip}
[Fe/H] & [O/Fe] & [Mg/Fe] & [Si/Fe] & [Ca/Fe] & [Ti/Fe] &
[Na/Fe] & [Eu/Fe] & [Ba/Fe] & [Mn/Fe] &[Sc/Fe] & {\rm ref.} \\
%\noalign{\vskip 0.2cm}
%\noalign{\hrule\vskip 0.2cm}
%\noalign{\vskip 0.2cm}
\noalign{\smallskip}
\hline
\noalign{\smallskip}
\multicolumn{12}{l}{NGC 6528} \\
\noalign{\smallskip}
\hline
\noalign{\smallskip}
+0.07   & +0.07 & +0.14 & +0.36 & +0.23 &+0.03  & +0.40 & --- & +0.14&$-$0.37&$-$0.05 & (1)\\
$-$0.11 & +0.10 & +0.05 & +0.05 & $-$0.40 & $-$0.25 & +0.60 & --- &---&---&--- & (2) \\
\noalign{\smallskip}
\hline
\noalign{\smallskip}
\multicolumn{12}{l}{NGC 6553} \\
\noalign{\smallskip}
\hline
\noalign{\smallskip}
$-$0.16 & +0.50 & +0.41 & +0.14 & +0.26 & +0.19 & --- & ---   &---&---&-0.12 & (3) \\
$-$0.20 & +0.20 & ---   & ---   & ---   & ---   & --- & ---   &--- &---&--- & (4)\\
$-$0.20 & ---   & +0.28 & +0.21 & +0.05 & $-$0.01 & +0.16 & +0.10 &$-$0.28 & ---&--- & (5) \\
\noalign{\smallskip} \hline 
\end{tabular}
%\end{flushleft}
\tablefoot{ 
\tablefoottext{1}{\cite{carretta+01};}
\tablefoottext{2}{\cite{zoccali+04};}
\tablefoottext{3}{\cite{cohen+99};}
\tablefoottext{4}{\cite{melendez+03};}
\tablefoottext{5}{\cite{alves-brito+06}.}
}
\end{table*}

\subsection{Moderately metal-rich clusters M~71, NGC~6558}
\label{sec:rich}

According to \citet[2010 edition]{harris96}, M~71 (or NGC~6838) is
located 6.7~kpc from the Galactic centre, with a perpendicular distance to
the Galactic plane of only 0.3~kpc towards the South Galactic Pole,
which means that this globular cluster is located in the Milky Way
disc. Because of that its reddening is high, although not as high as
for most bulge clusters, with E(B-V) = 0.25. NGC~6558 is located only 1.0~kpc
from the Galactic centre, in particular between the two ``southern
legs'' of the X-shaped bulge (see Figure 3 of
\citealp{saito+11}). It has a reddening of E(B-V) = 0.44. Although
they are located in different components of the Milky Way, these
clusters share similar metallicities, [Fe/H]$\sim$-0.8, and
[Fe/H]$\sim$-1.0, respectively. 

Figure \ref{cmds} shows a red horizontal branch (HB) for M~71 and a blue
HB for NGC~6558, and this difference is not due to
metallicity \citep{lee+94}. It is true that a few other parameters can
change the HB morphology at a fixed metallicity, as discussed by
\cite{catelan+01}, for instance, but in this case the age is
probably playing the
major role. Literature ages are 11~Gyr \citep{vandenberg+13} and
14~Gyr old \citep{barbuy+07}, respectively.
 M~71 is younger and moderately metal-rich,
therefore a red horizontal branch is expected. In particular, we
derived a slightly higher metallicity for this cluster in comparison
with the literature, [Fe/H] = -0.63$\pm$0.06. NGC~6558 has a high
metallicity for such a blue horizontal branch, and \cite{barbuy+07}
argue that if this is interpreted as pure factor of age, this cluster
should be one of the oldest objects in the Milky Way. We derived
[Fe/H] = -1.01$\pm$0.04, compatible with their findings ([Fe/H] =
-0.97$\pm$0.15), and more metal-rich than \citet[2010
edition]{harris96} value of [Fe/H] = -1.32. 
\cite{saviane+12} also found [Fe/H] = -1.03$\pm$0.14 for NGC~6558 from
their Ca II triplet spectroscopy, with error being dominated by the
external calibration uncertainty.

Comparing the error bars of T$_{\rm eff}$ and log($g$) in Figure
\ref{isochrones} between the two libraries, for M~71 they are of the
same order, but for NGC~6558 MILES results present larger error
bars. The main reason for this is that MILES library is based on solar
neighbouhood stars, and only a few of them have metallicities 
[Fe/H] $\approx$ -1.0 (see Figure 2 of \citealp{sanchez-blazquez+06}).
Synthetic libraries such as that of \cite{coelho+05} have spectra for
any combination of atmospheric parameters evenly distributed in the
parameter space. Therefore COELHO library is more suitable for
the analysis of these moderately metallicity clusters.

For this range of metallicity and for expected values of [Mg/Fe] from the
literature (0.19 and 0.24, respectively), MILES and COELHO results are
compatible, as shown in Table \ref{clusterparamavg}, [Mg/Fe] = 0.25$\pm$0.07
and 0.26$\pm$0.06, [$\alpha$/Fe] = 0.293$\pm$0.032 and 0.23$\pm$0.06,
respectively. \cite{ramirez+02} give an average [Mg/Fe]=+0.37 from
24 stars observed with high-resolution in M71, which is compatible with our findings.
The number of stars in MILES with these metallicities is
low, but it is sampled enough for determinations of [Fe/H] and [Mg/Fe]
for Milky Way stars. COELHO spectra have [$\alpha$/Fe] varying from
0.0 to 0.4, which is also enough for this kind of objects.

\subsection{Metal-poor clusters NGC~6426, Terzan~8}

NGC~6426 and Terzan~8
present similar CMDs with a same literature age and metallicity
of 13~Gyr \citep{dotter+11,vandenberg+13} and [Fe/H] = -2.15
\citep[2010 edition]{harris96}.
NGC~6426 is located in the northern halo of the Milky Way, 14.4~kpc
from the Galactic centre and 5.8~kpc above the Galactic
plane. Its height is much larger than the height scale of the thick
disc (0.75~kpc, \citealp{dejong+10}), but it has a non-negligible
reddening of E(B-V) = 0.36, at a galactic
latitude of b=16.23$^{\circ}$.
The best CMD available for this cluster was observed with
the ACS imager onboard the Hubble Space
Telescope by \cite{dotter+11}, and they derived an age of
13.0$\pm$1.5~Gyr from isochrone fitting. Our pre-image photometry based
on observations with the Very Large Telescope/ESO produced a rather 
well-defined CMD for this
cluster which is compatible with a 13~Gyr isochrone (see Figure
\ref{cmds}).

Terzan~8 is located in the southern halo of the Milky Way, 19.4~kpc
from the Galactic centre and 10.9~kpc below the Galactic plane. It has
the lowest reddening of all clusters analysed in this work, E(B-V) =
0.12. This is one of the four Milky Way globular clusters believed 
to be captured
from the Sagittarius dwarf galaxy \citep{ibata+94}, the other three being
M~54, Terzan~7 and Arp~2 \citep{dacosta+95}. \cite{carretta+14} did
not find a strong evidence for Na-O anticorrelation, typical for
globular clusters which may indicate that these clusters may have
simple stellar populations.

Atmospheric parameters derived in this work for both clusters are in
good agreement with literature values, when compared to the isochrones in
Figure \ref{isochrones}. As for the moderately metal-rich clusters,
MILES results present larger error bars than COELHO results for
T$_{\rm eff}$ and log($g$), and the reason is the same as mentioned
above, i.e., the sampling
of MILES library is poorer for this metallicity range (see Figure 2 of
\citealp{sanchez-blazquez+06}). Derived metallicities for these
clusters are [Fe/H] = -2.39$\pm$0.04 and -2.06$\pm$0.04,
respectively. For NGC~6426 our determination is 0.24 more metal-poor
than given in Harris's catalogue ([Fe/H] = -2.15). However the only
derivation of the metallicity for this cluster was done by
\cite{zinn+84} based on integrated light ([Fe/H] = -2.20$\pm$0.17)
which is compatible with our findings. The value from Harris catalogue
was obtained by applying the \cite{carretta+09} metallicity scale. We
present for the first time a direct measurement of metallicity of individual
red giant stars finding a more metal-poor value than previously
attributed to this cluster. Terzan~8 is compatible with the
Harris (2010 edition) catalogue of [Fe/H] = -2.16. There are three
papers with metallicities for this cluster: \cite{mottini+08} and
\cite{carretta+14}, based on high-resolution spectra of individual
star and average metallicity [Fe/H] = -2.35$\pm$0.04 and
-2.27$\pm$0.08, respectively. The third paper is \cite{dacosta+95} who
derived [Fe/H] = -1.99$\pm$0.08 based on CaII triplet
spectroscopy. Our result is compatible with the more metal-rich
results.

For alpha-enhancement in this metallicity range, MILES spectra goes up
to [Mg/Fe] = 0.74, while COELHO is limited to the models of
[$\alpha$/Fe] = 0.4. The results based on COELHO
([$\alpha$/Fe] = 0.24$\pm$0.05 and 0.21$\pm$0.04, respectively) are
less enhanced than MILES ([Mg/Fe] = 0.38$\pm$0.06 and 0.41$\pm$0.04,
respectively), the latter being closer to literature abundance ratios.

%______________________________________________________________

\section{Summary and conclusions}

We present a method of full spectrum fitting, based on the ETOILE code, to
derive v$_{\rm helio}$, T$_{\rm eff}$, log($g$), [Fe/H], [Mg/Fe] and
[$\alpha$/Fe] for red giant stars in Milky Way globular clusters. The
observations were carried out with FORS2@VLT/ESO with resolution R$\sim$2,000.

We validated the method using well known red giant stars covering the 
parameter space of 4000 K $<$ T$_{\rm eff}$ $<$ 6000 K, 0.0 $<$ log($g$) $<$
4.0, -2.5 $<$ [Fe/H] $<$ +0.3 and -0.2 $<$ [Mg/Fe] $<$ +0.6. The spectra
of these reference stars are from the ELODIE library and the parameters 
from the PASTEL catalogue. The parameters were recovered for 
the whole range of parameters. We applied the method to red giant
stars, and the code ETOILE has been applied and validated also for
dwarf stars by \cite{katz+11}.

In order to establish the methodology to be adopted for a larger
 sample of clusters, we chose two metal-rich (NGC~6528, NGC~6553), two moderately
metal-rich (M~71, NGC~6558) and two metal-poor (NGC~6426 and
Terzan~8) clusters. NGC~6528, NGC~6553 and NGC~6558 are located in the bulge,
M~71 in the disc, NGC~6426 and Terzan~8 in the halo.
For all clusters the effective temperatures and gravities
are well determined using the MILES and Coelho et al. (2005)
 libraries of spectra. Metallicities and alpha-element enhancement are
also derived with the caveats that for alpha-enhanced bulge clusters
with [Fe/H] $>$ -0.5, MILES is not suitable, since it has only solar
neighbourhood stars, therefore in this case we use COELHO results because
synthetic libraries have all combinations of parameters. For [Fe/H]
$\approx$ -1.0 MILES has few stars due to a lack of such stars in the
solar vicinity. For metal-poor clusters, with high [$\alpha$/Fe],
MILES may be more suitable than COELHO because the latter is limited to
0 $<$ [$\alpha$/Fe] $<$ 0.4dex, if such high Mg enhancements are confirmed. 

The present results are in agreement with literature parameters
 available for five of the
six template clusters. NGC~6426 is analysed 
for the first time using spectroscopy of individual stars.
Therefore we provide a more precise radial velocity of
-242$\pm$11~km/s, a metallicity
[Fe/H] = -2.39$\pm$0.04, and [Mg/Fe] = 0.38$\pm$0.06. 
The comparison of
our results of [Fe/H] and [Mg/Fe] to those from field stars from all
Galactic components shows that the globular clusters follows the same
chemical enrichment pattern as the field stars. 

In conclusion, full spectrum fitting technique using ETOILE code together
with MILES and COELHO libraries appears to be
suitable to derive chemical abundances for Milky Way globular
clusters, from low/medium-resolution spectra  of red giant branch stars. 
Depending on the stellar population studied, the choice of
 library with parameter space covering the expected values
for the clusters is a crucial ingredient, observed library being
better for more metal-rich stars and synthetic library being
preferable for the more metal-poor ones.
This method will be applied to the other Milky Way
globular clusters from this survey.
It is also promising for extragalactic stars, 
that can be more easily observed with similar resolutions of R$\sim$2,000,
for studies of galaxy formation and evolution.

%______________________________________________________________

\begin{acknowledgements}
We are grateful to Paula Coelho for useful discussions.
BD acknowledges financial support from CNPq and ESO.
BB acknowledges partial financial support from CNPq and Fapesp.
\end{acknowledgements}

%______________________________________________________________
\bibliographystyle{aa}
\bibliography{bibliography}

\begin{thebibliography}{66}
\expandafter\ifx\csname natexlab\endcsname\relax\def\natexlab#1{#1}\fi

\bibitem[{{Allende Prieto} {et~al.}(2008){Allende Prieto}, {Sivarani}, {Beers},
  {Lee}, {Koesterke}, {Shetrone}, {Sneden}, {Lambert}, {Wilhelm}, {Rockosi},
  {Lai}, {Yanny}, {Ivans}, {Johnson}, {Aoki}, {Bailer-Jones}, \& {Re
  Fiorentin}}]{allende+08}
{Allende Prieto}, C., {Sivarani}, T., {Beers}, T.~C., {et~al.} 2008, \aj, 136,
  2070

\bibitem[{{Alves-Brito} {et~al.}(2006){Alves-Brito}, {Barbuy}, {Zoccali},
  {Minniti}, {Ortolani}, {Hill}, {Renzini}, {Pasquini}, {Bica}, {Rich},
  {Mel{\'e}ndez}, \& {Momany}}]{alves-brito+06}
{Alves-Brito}, A., {Barbuy}, B., {Zoccali}, M., {et~al.} 2006, \aap, 460, 269

\bibitem[{{Alves-Brito} {et~al.}(2010){Alves-Brito}, {Mel{\'e}ndez}, {Asplund},
  {Ram{\'{\i}}rez}, \& {Yong}}]{alves-brito+10}
{Alves-Brito}, A., {Mel{\'e}ndez}, J., {Asplund}, M., {Ram{\'{\i}}rez}, I., \&
  {Yong}, D. 2010, \aap, 513, A35

\bibitem[{{Appenzeller} {et~al.}(1998){Appenzeller}, {Fricke}, {F{\"u}rtig},
  {G{\"a}ssler}, {H{\"a}fner}, {Harke}, {Hess}, {Hummel}, {J{\"u}rgens},
  {Kudritzki}, {Mantel}, {Meisl}, {Muschielok}, {Nicklas}, {Rupprecht},
  {Seifert}, {Stahl}, {Szeifert}, \& {Tarantik}}]{appenzeller+98}
{Appenzeller}, I., {Fricke}, K., {F{\"u}rtig}, W., {et~al.} 1998, The
  Messenger, 94, 1

\bibitem[{{Barbuy} {et~al.}(2009){Barbuy}, {Zoccali}, {Ortolani}, {Hill},
  {Minniti}, {Bica}, {Renzini}, \& {G{\'o}mez}}]{barbuy+09}
{Barbuy}, B., {Zoccali}, M., {Ortolani}, S., {et~al.} 2009, \aap, 507, 405

\bibitem[{{Barbuy} {et~al.}(2007){Barbuy}, {Zoccali}, {Ortolani}, {Minniti},
  {Hill}, {Renzini}, {Bica}, \& {G{\'o}mez}}]{barbuy+07}
{Barbuy}, B., {Zoccali}, M., {Ortolani}, S., {et~al.} 2007, \aj, 134, 1613

\bibitem[{{Bensby} {et~al.}(2005){Bensby}, {Feltzing}, {Lundstr{\"o}m}, \&
  {Ilyin}}]{bensby+05}
{Bensby}, T., {Feltzing}, S., {Lundstr{\"o}m}, I., \& {Ilyin}, I. 2005, \aap,
  433, 185

\bibitem[{{Bressan} {et~al.}(2012){Bressan}, {Marigo}, {Girardi}, {Salasnich},
  {Dal Cero}, {Rubele}, \& {Nanni}}]{bressan+12}
{Bressan}, A., {Marigo}, P., {Girardi}, L., {et~al.} 2012, \mnras, 427, 127

\bibitem[{{Carretta} {et~al.}(2009){Carretta}, {Bragaglia}, {Gratton},
  {D'Orazi}, \& {Lucatello}}]{carretta+09}
{Carretta}, E., {Bragaglia}, A., {Gratton}, R., {D'Orazi}, V., \& {Lucatello},
  S. 2009, \aap, 508, 695

\bibitem[{{Carretta} {et~al.}(2014){Carretta}, {Bragaglia}, {Gratton},
  {D'Orazi}, {Lucatello}, \& {Sollima}}]{carretta+14}
{Carretta}, E., {Bragaglia}, A., {Gratton}, R.~G., {et~al.} 2014, \aap, 561,
  A87

\bibitem[{{Carretta} {et~al.}(2010){Carretta}, {Bragaglia}, {Gratton},
  {Recio-Blanco}, {Lucatello}, {D'Orazi}, \& {Cassisi}}]{carretta+10}
{Carretta}, E., {Bragaglia}, A., {Gratton}, R.~G., {et~al.} 2010, \aap, 516,
  A55

\bibitem[{{Carretta} {et~al.}(2001){Carretta}, {Cohen}, {Gratton}, \&
  {Behr}}]{carretta+01}
{Carretta}, E., {Cohen}, J.~G., {Gratton}, R.~G., \& {Behr}, B.~B. 2001, \aj,
  122, 1469

\bibitem[{{Catelan} {et~al.}(2001){Catelan}, {Ferraro}, \& {Rood}}]{catelan+01}
{Catelan}, M., {Ferraro}, F.~R., \& {Rood}, R.~T. 2001, \apj, 560, 970

\bibitem[{{Cayrel}(1988)}]{cayrel88}
{Cayrel}, R. 1988, in IAU Symposium, Vol. 132, The Impact of Very High S/N
  Spectroscopy on Stellar Physics, ed. G.~{Cayrel de Strobel} \& M.~{Spite},
  345

\bibitem[{{Cayrel} {et~al.}(1991){Cayrel}, {Perrin}, {Barbuy}, \&
  {Buser}}]{cayrel+91}
{Cayrel}, R., {Perrin}, M.-N., {Barbuy}, B., \& {Buser}, R. 1991, \aap, 247,
  108

\bibitem[{{Cenarro} {et~al.}(2007){Cenarro}, {Peletier},
  {S{\'a}nchez-Bl{\'a}zquez}, {Selam}, {Toloba}, {Cardiel},
  {Falc{\'o}n-Barroso}, {Gorgas}, {Jim{\'e}nez-Vicente}, \&
  {Vazdekis}}]{cenarro+07}
{Cenarro}, A.~J., {Peletier}, R.~F., {S{\'a}nchez-Bl{\'a}zquez}, P., {et~al.}
  2007, \mnras, 374, 664

\bibitem[{{Coelho} {et~al.}(2005){Coelho}, {Barbuy}, {Mel{\'e}ndez},
  {Schiavon}, \& {Castilho}}]{coelho+05}
{Coelho}, P., {Barbuy}, B., {Mel{\'e}ndez}, J., {Schiavon}, R.~P., \&
  {Castilho}, B.~V. 2005, \aap, 443, 735

\bibitem[{{Cohen} {et~al.}(2001){Cohen}, {Behr}, \& {Briley}}]{cohen+01}
{Cohen}, J.~G., {Behr}, B.~B., \& {Briley}, M.~M. 2001, \aj, 122, 1420

\bibitem[{{Cohen} {et~al.}(1999){Cohen}, {Gratton}, {Behr}, \&
  {Carretta}}]{cohen+99}
{Cohen}, J.~G., {Gratton}, R.~G., {Behr}, B.~B., \& {Carretta}, E. 1999, \apj,
  523, 739

\bibitem[{{Da Costa} \& {Armandroff}(1995)}]{dacosta+95}
{Da Costa}, G.~S. \& {Armandroff}, T.~E. 1995, \aj, 109, 2533

\bibitem[{{Da Costa} {et~al.}(2009){Da Costa}, {Held}, {Saviane}, \&
  {Gullieuszik}}]{dacosta+09}
{Da Costa}, G.~S., {Held}, E.~V., {Saviane}, I., \& {Gullieuszik}, M. 2009,
  \apj, 705, 1481

\bibitem[{{de Jong} {et~al.}(2010){de Jong}, {Yanny}, {Rix}, {Dolphin},
  {Martin}, \& {Beers}}]{dejong+10}
{de Jong}, J.~T.~A., {Yanny}, B., {Rix}, H.-W., {et~al.} 2010, \apj, 714, 663

\bibitem[{{Dotter} {et~al.}(2008){Dotter}, {Chaboyer}, {Jevremovi{\'c}},
  {Kostov}, {Baron}, \& {Ferguson}}]{dotter+08}
{Dotter}, A., {Chaboyer}, B., {Jevremovi{\'c}}, D., {et~al.} 2008, \apjs, 178,
  89

\bibitem[{{Dotter} {et~al.}(2011){Dotter}, {Sarajedini}, \&
  {Anderson}}]{dotter+11}
{Dotter}, A., {Sarajedini}, A., \& {Anderson}, J. 2011, \apj, 738, 74

\bibitem[{{Faber} {et~al.}(1985){Faber}, {Friel}, {Burstein}, \&
  {Gaskell}}]{faber+85}
{Faber}, S.~M., {Friel}, E.~D., {Burstein}, D., \& {Gaskell}, C.~M. 1985,
  \apjs, 57, 711

\bibitem[{{Fulbright}(2000)}]{fulbright+00}
{Fulbright}, J.~P. 2000, \aj, 120, 1841

\bibitem[{{Gilmore} {et~al.}(2012){Gilmore}, {Randich}, {Asplund}, {Binney},
  {Bonifacio}, {Drew}, {Feltzing}, {Ferguson}, {Jeffries}, {Micela},
  {Negueruela}, {Prusti}, {Rix}, {Vallenari}, {Alfaro}, {Allende-Prieto},
  {Babusiaux}, {Bensby}, {Blomme}, {Bragaglia}, {Flaccomio}, {Fran{\c c}ois},
  {Irwin}, {Koposov}, {Korn}, {Lanzafame}, {Pancino}, {Paunzen},
  {Recio-Blanco}, {Sacco}, {Smiljanic}, {Van Eck}, \& {Walton}}]{gilmore+12}
{Gilmore}, G., {Randich}, S., {Asplund}, M., {et~al.} 2012, The Messenger, 147,
  25

\bibitem[{{Gonzalez} {et~al.}(2011){Gonzalez}, {Rejkuba}, {Zoccali}, {Hill},
  {Battaglia}, {Babusiaux}, {Minniti}, {Barbuy}, {Alves-Brito}, {Renzini},
  {Gomez}, \& {Ortolani}}]{gonzalez+11}
{Gonzalez}, O.~A., {Rejkuba}, M., {Zoccali}, M., {et~al.} 2011, \aap, 530, A54

\bibitem[{{Harris}(1996)}]{harris96}
{Harris}, W.~E. 1996, \aj, 112, 1487

\bibitem[{{Hesser} {et~al.}(1986){Hesser}, {Shawl}, \& {Meyer}}]{hesser+86}
{Hesser}, J.~E., {Shawl}, S.~J., \& {Meyer}, J.~E. 1986, \pasp, 98, 403

\bibitem[{{Ibata} {et~al.}(1994){Ibata}, {Gilmore}, \& {Irwin}}]{ibata+94}
{Ibata}, R.~A., {Gilmore}, G., \& {Irwin}, M.~J. 1994, \nat, 370, 194

\bibitem[{{Katz}(2001)}]{katz01}
{Katz}, D. 2001, Journal of Astronomical Data, 7, 8

\bibitem[{{Katz} {et~al.}(1998){Katz}, {Soubiran}, {Cayrel}, {Adda}, \&
  {Cautain}}]{katz+98}
{Katz}, D., {Soubiran}, C., {Cayrel}, R., {Adda}, M., \& {Cautain}, R. 1998,
  \aap, 338, 151

\bibitem[{{Katz} {et~al.}(2011){Katz}, {Soubiran}, {Cayrel}, {Barbuy}, {Friel},
  {Bienaym{\'e}}, \& {Perrin}}]{katz+11}
{Katz}, D., {Soubiran}, C., {Cayrel}, R., {et~al.} 2011, \aap, 525, A90+

\bibitem[{{Kirby} {et~al.}(2009){Kirby}, {Guhathakurta}, {Bolte}, {Sneden}, \&
  {Geha}}]{kirby+09}
{Kirby}, E.~N., {Guhathakurta}, P., {Bolte}, M., {Sneden}, C., \& {Geha}, M.~C.
  2009, \apj, 705, 328

\bibitem[{{Koleva} {et~al.}(2009){Koleva}, {Prugniel}, {Bouchard}, \&
  {Wu}}]{koleva+09}
{Koleva}, M., {Prugniel}, P., {Bouchard}, A., \& {Wu}, Y. 2009, \aap, 501, 1269

\bibitem[{{Lee} {et~al.}(2008){Lee}, {Beers}, {Sivarani}, {Allende Prieto},
  {Koesterke}, {Wilhelm}, {Re Fiorentin}, {Bailer-Jones}, {Norris}, {Rockosi},
  {Yanny}, {Newberg}, {Covey}, {Zhang}, \& {Luo}}]{lee+08}
{Lee}, Y.~S., {Beers}, T.~C., {Sivarani}, T., {et~al.} 2008, \aj, 136, 2022

\bibitem[{{Lee} {et~al.}(1994){Lee}, {Demarque}, \& {Zinn}}]{lee+94}
{Lee}, Y.-W., {Demarque}, P., \& {Zinn}, R. 1994, \apj, 423, 248

\bibitem[{{Mel{\'e}ndez} {et~al.}(2003){Mel{\'e}ndez}, {Barbuy}, {Bica},
  {Zoccali}, {Ortolani}, {Renzini}, \& {Hill}}]{melendez+03}
{Mel{\'e}ndez}, J., {Barbuy}, B., {Bica}, E., {et~al.} 2003, \aap, 411, 417

\bibitem[{{Mel{\'e}ndez} \& {Cohen}(2009)}]{melendez+09}
{Mel{\'e}ndez}, J. \& {Cohen}, J.~G. 2009, \apj, 699, 2017

\bibitem[{{M{\'e}sz{\'a}ros} {et~al.}(2013){M{\'e}sz{\'a}ros}, {Holtzman},
  {Garc{\'{\i}}a P{\'e}rez}, {Allende Prieto}, {Schiavon}, \&
  {Basu}}]{meszaros+13}
{M{\'e}sz{\'a}ros}, S., {Holtzman}, J., {Garc{\'{\i}}a P{\'e}rez}, A.~E.,
  {et~al.} 2013, \aj, 146, 133

\bibitem[{{Milone} {et~al.}(2011){Milone}, {Sansom}, \&
  {S{\'a}nchez-Bl{\'a}zquez}}]{milone+11}
{Milone}, A.~D.~C., {Sansom}, A.~E., \& {S{\'a}nchez-Bl{\'a}zquez}, P. 2011,
  \mnras, 414, 1227

\bibitem[{{Mottini} {et~al.}(2008){Mottini}, {Wallerstein}, \&
  {McWilliam}}]{mottini+08}
{Mottini}, M., {Wallerstein}, G., \& {McWilliam}, A. 2008, \aj, 136, 614

\bibitem[{{Ortolani} {et~al.}(1995){Ortolani}, {Renzini}, {Gilmozzi},
  {Marconi}, {Barbuy}, {Bica}, \& {Rich}}]{ortolani+95}
{Ortolani}, S., {Renzini}, A., {Gilmozzi}, R., {et~al.} 1995, \nat, 377, 701

\bibitem[{{Perryman} {et~al.}(2001){Perryman}, {de Boer}, {Gilmore}, {H{\o}g},
  {Lattanzi}, {Lindegren}, {Luri}, {Mignard}, {Pace}, \& {de
  Zeeuw}}]{perryman+01}
{Perryman}, M.~A.~C., {de Boer}, K.~S., {Gilmore}, G., {et~al.} 2001, \aap,
  369, 339

\bibitem[{{Pietrinferni} {et~al.}(2004){Pietrinferni}, {Cassisi}, {Salaris}, \&
  {Castelli}}]{pietrinferni+04}
{Pietrinferni}, A., {Cassisi}, S., {Salaris}, M., \& {Castelli}, F. 2004, \apj,
  612, 168

\bibitem[{{Prugniel} {et~al.}(2007){Prugniel}, {Soubiran}, {Koleva}, \& {Le
  Borgne}}]{prugniel+07}
{Prugniel}, P., {Soubiran}, C., {Koleva}, M., \& {Le Borgne}, D. 2007,
  arXiv:astro-ph/0703658

\bibitem[{{Ram{\'{\i}}rez} \& {Cohen}(2002)}]{ramirez+02}
{Ram{\'{\i}}rez}, S.~V. \& {Cohen}, J.~G. 2002, \aj, 123, 3277

\bibitem[{{Ram{\'{\i}}rez} {et~al.}(2001){Ram{\'{\i}}rez}, {Cohen}, {Buss}, \&
  {Briley}}]{ramirez+01}
{Ram{\'{\i}}rez}, S.~V., {Cohen}, J.~G., {Buss}, J., \& {Briley}, M.~M. 2001,
  \aj, 122, 1429

\bibitem[{{Reddy} {et~al.}(2006){Reddy}, {Lambert}, \& {Allende
  Prieto}}]{reddy+06}
{Reddy}, B.~E., {Lambert}, D.~L., \& {Allende Prieto}, C. 2006, \mnras, 367,
  1329

\bibitem[{{Roederer}(2009)}]{roederer+09}
{Roederer}, I.~U. 2009, \aj, 137, 272

\bibitem[{{Saito} {et~al.}(2011){Saito}, {Zoccali}, {McWilliam}, {Minniti},
  {Gonzalez}, \& {Hill}}]{saito+11}
{Saito}, R.~K., {Zoccali}, M., {McWilliam}, A., {et~al.} 2011, \aj, 142, 76

\bibitem[{{S{\'a}nchez Almeida} \& {Allende Prieto}(2013)}]{sanchez-almeida+13}
{S{\'a}nchez Almeida}, J. \& {Allende Prieto}, C. 2013, \apj, 763, 50

\bibitem[{{S{\'a}nchez-Bl{\'a}zquez} {et~al.}(2006){S{\'a}nchez-Bl{\'a}zquez},
  {Peletier}, {Jim{\'e}nez-Vicente}, {Cardiel}, {Cenarro},
  {Falc{\'o}n-Barroso}, {Gorgas}, {Selam}, \& {Vazdekis}}]{sanchez-blazquez+06}
{S{\'a}nchez-Bl{\'a}zquez}, P., {Peletier}, R.~F., {Jim{\'e}nez-Vicente}, J.,
  {et~al.} 2006, \mnras, 371, 703

\bibitem[{{Saviane} {et~al.}(2012){Saviane}, {da Costa}, {Held}, {Sommariva},
  {Gullieuszik}, {Barbuy}, \& {Ortolani}}]{saviane+12}
{Saviane}, I., {da Costa}, G.~S., {Held}, E.~V., {et~al.} 2012, \aap, 540, A27

\bibitem[{{Soubiran} {et~al.}(2010){Soubiran}, {Le Campion}, {Cayrel de
  Strobel}, \& {Caillo}}]{soubiran+10}
{Soubiran}, C., {Le Campion}, J.-F., {Cayrel de Strobel}, G., \& {Caillo}, A.
  2010, \aap, 515, A111

\bibitem[{{Steinmetz} {et~al.}(2006){Steinmetz}, {Zwitter}, {Siebert},
  {Watson}, {Freeman}, {Munari}, {Campbell}, \& {Williams}}]{steinmetz+06}
{Steinmetz}, M., {Zwitter}, T., {Siebert}, A., {et~al.} 2006, \aj, 132, 1645

\bibitem[{{VandenBerg} {et~al.}(2013){VandenBerg}, {Brogaard}, {Leaman}, \&
  {Casagrande}}]{vandenberg+13}
{VandenBerg}, D.~A., {Brogaard}, K., {Leaman}, R., \& {Casagrande}, L. 2013,
  \apj, 775, 134

\bibitem[{{Worthey} {et~al.}(1994){Worthey}, {Faber}, {Gonzalez}, \&
  {Burstein}}]{worthey+94}
{Worthey}, G., {Faber}, S.~M., {Gonzalez}, J.~J., \& {Burstein}, D. 1994,
  \apjs, 94, 687

\bibitem[{{Wu} {et~al.}(2011){Wu}, {Luo}, {Li}, {Shi}, {Prugniel}, {Liang},
  {Zhao}, {Zhang}, {Bai}, {Wei}, {Dong}, {Zhang}, \& {Chen}}]{wu+11}
{Wu}, Y., {Luo}, A.-L., {Li}, H.-N., {et~al.} 2011, Research in Astronomy and
  Astrophysics, 11, 924

\bibitem[{{Wylie-de Boer} \& {Freeman}(2010)}]{wylie+10}
{Wylie-de Boer}, E. \& {Freeman}, K. 2010, in IAU Symposium, Vol. 262, IAU
  Symposium, ed. G.~R. {Bruzual} \& S.~{Charlot}, 448--449

\bibitem[{{York} {et~al.}(2000){York}, {Adelman}, {Anderson}, {Anderson},
  {Annis}, {Bahcall}, {Bakken}, \& {SDSS Collaboration}}]{york+00}
{York}, D.~G., {Adelman}, J., {Anderson}, Jr., J.~E., {et~al.} 2000, \aj, 120,
  1579

\bibitem[{{Zinn} \& {West}(1984)}]{zinn+84}
{Zinn}, R. \& {West}, M.~J. 1984, \apjs, 55, 45

\bibitem[{{Zoccali} {et~al.}(2004){Zoccali}, {Barbuy}, {Hill}, {Ortolani},
  {Renzini}, {Bica}, {Momany}, {Pasquini}, {Minniti}, \& {Rich}}]{zoccali+04}
{Zoccali}, M., {Barbuy}, B., {Hill}, V., {et~al.} 2004, \aap, 423, 507

\bibitem[{{Zoccali} {et~al.}(2008){Zoccali}, {Hill}, {Lecureur}, {Barbuy},
  {Renzini}, {Minniti}, {G{\'o}mez}, \& {Ortolani}}]{zoccali+08}
{Zoccali}, M., {Hill}, V., {Lecureur}, A., {et~al.} 2008, \aap, 486, 177

\bibitem[{{Zoccali} {et~al.}(2001){Zoccali}, {Renzini}, {Ortolani}, {Bica}, \&
  {Barbuy}}]{zoccali+01}
{Zoccali}, M., {Renzini}, A., {Ortolani}, S., {Bica}, E., \& {Barbuy}, B. 2001,
  \aj, 121, 2638

\end{thebibliography}

\longtab{3}{
\tiny
\begin{longtable}{lllcccrrrc}
\caption{\label{starinfo} Star by star coordinates,
  magnitude, colour, heliocentric radial
  velocity. Velocities from CaT were taken from from \cite{saviane+12}
  for NGC~6528, NGC~6553, M~71 and NGC~6558, and from Vasquez et
  al. (2014, in prep.) for NGC~6426 and Terzan~8.}\\
\hline\hline
\noalign{\smallskip}
 {\rm Star ID} & {\rm RA (J2000)} & {\rm DEC (J2000)} & {V} & {V-I} & {\rm v$_{\rm helio}$} & {\rm
   v$_{\rm helio-CaT}$} & members\\
 & (deg)  & (deg) & (mag) & (mag) & {\rm (km/s)} & {\rm (km/s)} & \\
\noalign{\smallskip}
\hline
\endfirsthead
\caption{continued.}\\
\hline\hline
\noalign{\smallskip}
 {\rm Star ID} & {\rm RA (J2000)} & {\rm DEC (J2000)} & {V} & {V-I} & {\rm v$_{\rm helio}$} & {\rm
   v$_{\rm helio-CaT}$} & members\\
 & (deg)  & (deg) & (mag) & (mag) & {\rm (km/s)} & {\rm (km/s)} & \\
\noalign{\smallskip}
\hline
\endhead
\hline
\noalign{\smallskip}
\endfoot
\noalign{\smallskip}
NGC6528\_2   &  271.1807715771910   &   -30.0070234836130    &   15.987   &   1.982   &      -51.55 & --- &   \\  
NGC6528\_3   &  271.1933667526377   &   -30.0114575457940    &   15.647   &   2.836   &       61.28 & --- &   \\  
NGC6528\_4   &  271.1789935185810   &   -30.0172537974630    &   15.596   &   2.257   &     -314.24 & --- &   \\  
NGC6528\_5   &  271.1883528713980   &   -30.0237445060760    &   15.939   &   1.937   &     -226.44 & --- &   \\  
NGC6528\_6   &  271.1974416677920   &   -30.0295599638440    &   17.060   &   1.741   &       -4.93 & --- &   \\  
NGC6528\_7   &  271.2102328036860   &   -30.0372195159180    &   15.887   &   2.140   &      -79.24 & --- &   \\  
NGC6528\_8   &  271.2027175539120   &   -30.0434602130820    &   16.428   &   1.831   &      179.09 & 200 & M  \\  
NGC6528\_9   &  271.1969387171677   &   -30.0500739849720    &   16.501   &   1.774   &      199.18 & 208 & M  \\  
NGC6528\_10  &  271.2070341687930   &   -30.0537176170110    &   16.145   &   2.007   &      177.87 & 209 & M  \\  
NGC6528\_11  &  271.2013524700317   &   -30.0600904185180    &   15.511   &   2.255   &      182.67 & 202 & M  \\  
NGC6528\_13  &  271.1879324598030   &   -30.0747465095010    &   16.429   &   1.731   &      -29.84 & --- &   \\  
NGC6528\_14  &  271.2012738946260   &   -30.0802412070160    &   16.032   &   2.056   &      -40.56 & --- &   \\  
NGC6528\_15  &  271.1774205888063   &   -30.0879237445260    &   15.892   &   2.062   &     -288.86 & --- &   \\  
NGC6528\_16  &  271.1843122189650   &   -30.0926974725130    &   15.954   &   2.150   &     -104.91 & --- &   \\  
NGC6528\_17  &  271.1904270294383   &   -30.1021087576640    &   16.537   &   1.844   &      -62.96 & --- &   \\  
NGC6528\_18  &  271.1726747862950   &   -30.1092237523470    &   16.666   &   1.767   &     -212.87 & --- &   \\  
NGC6528\_19  &  271.1832358189100   &   -30.1108940656500    &   16.468   &   1.778   &      -54.47 & --- &   \\  
\noalign{\smallskip}
\hline
\noalign{\smallskip}
 NGC6553\_1   &  272.3536106669400   &  -25.8497112721910   &   15.816  &   2.109 &        3.90 &  -8  & M \\ 
 NGC6553\_3   &  272.3507085422689   &  -25.8636412811220   &   15.832  &   2.010 &       16.71 &  19  & M \\ 
 NGC6553\_4   &  272.2982167426244   &  -25.8658039783470   &   15.310  &   2.618 &      -71.71 & -39  & M \\ 
 NGC6553\_5   &  272.3267021788947   &  -25.8733214097360   &   15.775  &   2.522 &       12.23 & -12  & M \\ 
 NGC6553\_6   &  272.3161974470756   &  -25.8795540059450   &   16.237  &   2.177 &        0.81 &  -6  & M \\ 
 NGC6553\_7   &  272.3258216397713   &  -25.8883122417460   &   15.338  &   2.881 &        6.76 &   3  & M \\ 
 NGC6553\_8   &  272.3501542761960   &  -25.8928074210270   &   15.370  &   3.088 &       18.98 &  --- &  \\ 
 NGC6553\_9   &  272.3409674217040   &  -25.9007568498840   &   16.253  &   2.097 &      -27.72 & -32  & M \\ 
NGC6553\_10  &  272.3439213760840   &  -25.9071903257530   &   15.985  &   1.998  &        2.34 &   2  & M \\ 
NGC6553\_11  &  272.3287891424960   &  -25.9110912319810   &   15.441  &   2.443  &       10.12 &  -3  & M \\ 
NGC6553\_13  &  272.3235733284760   &  -25.9249134402580   &   15.906  &   2.075  &       -4.38 & -12  & M \\ 
NGC6553\_14  &  272.3157993633750   &  -25.9300477560180   &   15.187  &   2.382  &        5.40 &  -5  & M \\ 
NGC6553\_15  &  272.3047106358403   &  -25.9370597367310   &   14.980  &   2.996  &      -21.36 &  --- &  \\ 
NGC6553\_16  &  272.3280756666470   &  -25.9436286781210   &   15.708  &   2.332  &       17.42 &   8  & M \\ 
NGC6553\_17  &  272.2895515368990   &  -25.9484055025060   &   15.065  &   2.849  &     -110.00 &  --- &  \\ 
NGC6553\_18  &  272.3144982937293   &  -25.9566926017970   &   15.407  &   3.681  &      -10.44 &  --- &  \\ 
NGC6553\_19  &  272.3440933085433   &  -25.9630097530170   &   15.843  &   2.209  &       -7.38 &  -4  & M \\ 
\noalign{\smallskip}
\hline
\noalign{\smallskip}
M71\_2      &   298.4578984488130  &    18.8301452486084   &    14.606  &     1.269   &    -150.46 & -34  & M \\
M71\_4      &   298.4609554670447  &    18.8192204176758   &    13.030  &     1.503   &     -53.75 & -41  & M \\
M71\_5      &   298.4632902665389  &    18.8090499016740   &    14.391  &     1.282   &     -52.74 & -26  & M \\
M71\_6      &   298.4900822754500  &    18.7995866372160   &    13.534  &     1.367   &     -44.07 & -37  & M \\
M71\_7      &   298.4510869475747  &    18.8011235588184   &    12.376  &     1.719   &     -69.84 & -34  & M \\
M71\_8      &   298.4854008510480  &    18.7869940682184   &    14.421  &     1.280   &     -51.13 & ---  &  \\
M71\_9      &   298.4422411559920  &    18.7910962573810   &    13.146  &     1.530   &     -35.78 & -24  & M \\
M71\_10     &   298.4476878071520  &    18.7813713381542   &    12.140  &     2.042   &    -129.33 & -26  & M \\
M71\_13     &   298.4496765965810  &    18.7641970806387   &    14.281  &     1.303   &     -23.72 &  -5  & M \\ 
M71\_14     &   298.4902210311380  &    18.7496150618418   &    14.582  &     1.210   &     -13.11 & -17  & M \\
M71\_15     &   298.4521026881480  &    18.7474809643955   &    14.580  &     1.209   &     -41.62 &  -8  & M \\
M71\_16     &   298.4732927595957  &    18.7366583486676   &    14.727  &     1.231   &     -23.67 &  --- &  \\
\noalign{\smallskip}
\hline
\noalign{\smallskip}
 NGC6558\_3	&  272.6225545017980 & -31.7335169754310   &  16.862  &   1.434     &       -6.32   &  ---  &    \\
 NGC6558\_4	&  272.5685475821600 & -31.7363376028500   &  16.541  &   1.441     &      -38.29   &  ---  &    \\ 
 NGC6558\_5	&  272.6214120279947 & -31.7468917846890   &  16.349  &   1.360     &      -23.46   &  ---  &    \\  
 NGC6558\_6	&  272.5796695743913 & -31.7469946267620   &  15.982  &   1.524     &     -210.85   &  -196 &  M  \\ 
 NGC6558\_7	&  272.5899712149563 & -31.7570504471050   &  15.803  &   1.393     &     -186.59   &  -187 &  M  \\ 
 NGC6558\_8	&  272.5739816190383 & -31.7605125893880   &  13.651  &   2.044     &     -307.26   &  -210 &  M  \\ 
 NGC6558\_9	&  272.5637020636180 & -31.7666306345590   &  16.026  &   1.499     &     -221.66   &  -204 &  M  \\
NGC6558\_10	&  272.5771880085779 & -31.7733227718450   &  16.710  &   1.580     &      -21.13   &  ---  &    \\ 
NGC6558\_11	&  272.5757815235553 & -31.7788392433250   &  16.753  &   1.329     &     -220.56   &  -195 &  M  \\
NGC6558\_12	&  272.5805489015007 & -31.7879129654280   &  16.521  &   1.452     &     -126.79   &  ---  &    \\
NGC6558\_13	&  272.5717825299750 & -31.7916739864870   &  15.626  &   1.477     &     -127.74   &  ---  &    \\
NGC6558\_14	&  272.5809044425367 & -31.8010529051890   &  16.740  &   1.188     &     -118.59   &  ---  &    \\
NGC6558\_15	&  272.5652668675210 & -31.8066271885810   &  16.480  &   1.293     &      105.58   &  ---  &    \\ 
NGC6558\_16	&  272.5564861827613 & -31.8124812867920   &  16.366  &   1.450     &     -316.32   &  ---  &    \\
NGC6558\_17	&  272.5801632324050 & -31.8180552937480   &  16.964  &   1.348     &      -49.93   &  ---  &    \\
NGC6558\_18	&  272.6131914789087 & -31.8262981571080   &  16.911  &   1.288     &       11.27   &  ---  &    \\ 
NGC6558\_19	&  272.5939188278727 & -31.8321512784620   &  17.013  &   1.299     &       -5.38   &  ---  &    \\
\noalign{\smallskip}
\hline
\noalign{\smallskip}
 NGC6426\_1   &   266.2553357204670	    &    3.2257821720617	 	  &    18.018	  &     1.359        &     -80.97    &    -49.927       &    		     \\
 NGC6426\_2   &   266.2088188480099	    &    3.2192772570129	 	  &    17.597	  &     1.394        &      -2.85    &    11.325  	  &  		     \\
 NGC6426\_3   &   266.2200960712349	    &    3.2161573577027	 	  &    16.468	  &     1.586        &     -15.18    &    12.662  	  &  		     \\
 NGC6426\_4   &   266.2059346355020	    &    3.2095717676480	 	  &    16.072	  &     1.644        &    -236.29    &    -230.019	  &  	M	     \\
 NGC6426\_7   &   266.2472044411759	    &    3.1904905620518	 	  &    17.666	  &     1.459        &    -259.55    &    -225.413	  &  	M	     \\
 NGC6426\_9   &   266.2324246578110	    &    3.1746245634344	 	  &    17.170	  &     1.515        &    -244.78    &    -222.660	  &  	M	     \\
NGC6426\_10   &   266.2152297889280	    &    3.1685396261324	 	  &    15.544	  &     1.749        &    -231.87    &    -221.297	  &  	M	     \\
NGC6426\_11   &   266.2225277929239	    &    3.1649406775723	 	  &    17.892	  &     1.354        &      -4.92    &    12.270  	  &  		     \\
NGC6426\_13   &   266.2203523684530	    &    3.1516328071727	 	  &    16.745	  &     1.482        &    -236.15    &    -225.730	  &  	M	     \\
NGC6426\_18   &   266.2405437043161	    &    3.1175053813749	 	  &    16.970	  &     1.434        &      43.08    &    72.570  	  &  		     \\
\noalign{\smallskip}
\hline
\noalign{\smallskip}
Terzan8\_1	 &    295.4542164758450	 &    -33.9415478435970		 &    16.706   &   1.232 &        134.45 &   138.8238	 &  M  \\
Terzan8\_4	 &    295.4999075670249	 &    -33.9726804796670		 &    15.268   &   1.436 &        120.63 &   137.4998	 &  M  \\
Terzan8\_5	 &    295.4297741922069	 &    -33.9618073863440		 &    17.002   &   1.135 &        134.61 &   152.6001	 &  M  \\
Terzan8\_6	 &    295.4324851623650	 &    -33.9679803480730		 &    17.517   &   1.076 &        119.81 &   138.3819	 &  M  \\
Terzan8\_8	 &    295.4284871648290	 &    -33.9821965410250		 &    17.089   &   1.108 &        173.81 &   150.6620	 &  M  \\
Terzan8\_9	 &    295.4571652028810	 &    -33.9956864225960		 &    15.447   &   1.352 &        155.67 &   139.3673	 &  M  \\
Terzan8\_10	 &    295.4738836006140	 &    -34.0028288269560		 &    17.381   &   1.076 &        -63.09 &   -25.6758	 &    \\
Terzan8\_11	 &    295.4369091804230	 &    -34.0003943267700		 &    15.682   &   1.277 &        142.74 &   150.6086	 &  M  \\
Terzan8\_13	 &    295.4233824419659	 &    -34.0145360437400		 &    17.364   &   1.101 &        149.42 &   144.2031	 &  M  \\
Terzan8\_14	 &    295.4815071780459	 &    -34.0298070702590		 &    15.530   &   1.388 &        116.26 &   131.9558	 &  M  \\
Terzan8\_15	 &    295.4375324314209	 &    -34.0304794278020		 &    16.433   &   1.219 &        105.77 &   121.5507	 &  M  \\
Terzan8\_16	 &    295.4285209120099	 &    -34.0321898355090		 &    16.778   &   1.122 &        128.12 &   140.9906	 &  M  \\
Terzan8\_18	 &    295.4529047094940	 &    -34.0531331777060		 &    16.143   &   1.273 &        134.81 &   131.1500	 &  M  \\
\end{longtable}
}% End \longtab

\longtab{4}{
\tiny
\begin{longtable}{rl|crc|lcr|lcrr}
\caption{\label{tab:etoilevalidation} List of 49 well known stars selected from ELODIE spectral
  library. Literature parameters are average from PASTEL
  catalogue. Results from this work are using MILES library. See
  details in Section \ref{sec:validation}.}\\
\hline\hline
\noalign{\smallskip}
 {\rm ELODIE} & {\rm Star} & {\rm RR$_{\rm tot}$}  & {\rm N} & S$_{\rm lim}$
 & T$_{\rm eff}$ (K) & {\rm log($g$)} & {\rm [Fe/H]} & T$_{\rm eff}$ (K) & {\rm
   log$g$} & {\rm [Fe/H]} & {\rm [Mg/Fe]} \\
& &  & &  & & (literature) & & \multicolumn{4}{c}{(this work)} \\
\noalign{\smallskip}
\hline
\endfirsthead
\caption{continued.}\\
\hline\hline
\noalign{\smallskip}
 {\rm ELODIE} & {\rm Star} & {\rm RR$_{\rm tot}$} & {\rm N} & S$_{\rm lim}$
 & T$_{\rm eff}$ (K) & {\rm log($g$)} & {\rm [Fe/H]} & T$_{\rm eff}$ (K) & {\rm
   log$g$} & {\rm [Fe/H]} & {\rm [Mg/Fe]} \\
\noalign{\smallskip}
\hline
\endhead
\noalign{\smallskip}
\hline
\endfoot
\noalign{\smallskip}
   1 &   HD000245 & 0.107 &  2  & 1.042 & 5490$\pm$153 & 3.48$\pm$0.15 & -0.77$\pm$0.08  &  5378$\pm$60 & 3.67$\pm$0.06 & -0.84$\pm$0.20 &  0.34$\pm$0.08 \\ 
   9 &   HD002796 & 0.062 &  1  & 1.000 & 4931$\pm$60 & 1.45$\pm$0.34 & -2.32$\pm$0.11  &  4945$\pm$133 & 1.36$\pm$0.08 & -2.31$\pm$0.11 &  0.37$\pm$0.08 \\ 
  19 &   HD004395 & 0.042 &  5  & 1.116 & 5487$\pm$38 & 3.33$\pm$0.05 & -0.35$\pm$0.04  &  5330$\pm$149 & 3.24$\pm$0.16 & -0.34$\pm$0.11 &  0.18$\pm$0.07 \\ 
  31 &   HD006833 & 0.092 &  1  & 1.000 & 4426$\pm$95 & 1.28$\pm$0.32 & -0.91$\pm$0.14  &  4380$\pm$217 & 1.25$\pm$0.64 & -0.99$\pm$0.29 &  0.30$\pm$0.11 \\ 
  32 &   HD006920 & 0.035 &  8  & 1.319 & 5886$\pm$111 & 3.60$\pm$0.28 & -0.10$\pm$0.09  &  5854$\pm$100 & 3.60$\pm$0.05 & -0.10$\pm$0.14 &  0.10$\pm$0.09 \\ 
  33 &   HD008724 & 0.018 &  2  & 1.001 & 4586$\pm$84 & 1.39$\pm$0.26 & -1.73$\pm$0.13  &  4626$\pm$6 & 1.40$\pm$0.06 & -1.75$\pm$0.00 &  0.37$\pm$0.08 \\ 
  47 &   HD013530 & 0.013 &  6  & 1.036 & 4772$\pm$106 & 2.60$\pm$0.39 & -0.54$\pm$0.11  &  4769$\pm$87 & 2.63$\pm$0.21 & -0.54$\pm$0.15 &  0.37$\pm$0.08 \\ 
  66 &   HD015596 & 0.047 &  9  & 1.211 & 4808$\pm$59 & 2.66$\pm$0.32 & -0.65$\pm$0.06  &  4760$\pm$58 & 2.54$\pm$0.14 & -0.66$\pm$0.07 &  0.40$\pm$0.06 \\ 
  67 &   HD015596 & 0.037 & 10  & 1.298 & 4808$\pm$59 & 2.66$\pm$0.32 & -0.65$\pm$0.06  &  4751$\pm$52 & 2.57$\pm$0.12 & -0.64$\pm$0.06 &  0.37$\pm$0.07 \\ 
  68 &   HD015596 & 0.047 &  9  & 1.248 & 4808$\pm$59 & 2.66$\pm$0.32 & -0.65$\pm$0.06  &  4759$\pm$56 & 2.54$\pm$0.13 & -0.66$\pm$0.06 &  0.40$\pm$0.06 \\ 
  69 &   HD016458 & 0.118 &  2  & 1.115 & 4593$\pm$123 & 1.84$\pm$0.26 & -0.35$\pm$0.05  &  4992$\pm$513 & 1.97$\pm$0.95 & -0.34$\pm$0.25 &  0.33$\pm$0.23 \\ 
  88 &   HD020512 & 0.086 &  3  & 1.118 & 5212$\pm$63 & 3.65$\pm$0.14 & -0.22$\pm$0.19  &  5074$\pm$38 & 3.35$\pm$0.14 & -0.22$\pm$0.20 &  0.12$\pm$0.03 \\ 
  89 &   HD020512 & 0.094 &  3  & 1.101 & 5212$\pm$63 & 3.65$\pm$0.14 & -0.22$\pm$0.19  &  5074$\pm$39 & 3.35$\pm$0.15 & -0.21$\pm$0.21 &  0.12$\pm$0.03 \\ 
 117 &   HD026297 & 0.133 & 11  & 1.263 & 4445$\pm$140 & 1.02$\pm$0.28 & -1.74$\pm$0.15  &  4460$\pm$73 & 1.15$\pm$0.19 & -1.66$\pm$0.08 &  0.48$\pm$0.04 \\ 
 151 &   HD035369 & 0.032 &  6  & 1.062 & 4885$\pm$112 & 2.57$\pm$0.27 & -0.21$\pm$0.08  &  4900$\pm$45 & 2.65$\pm$0.08 & -0.21$\pm$0.05 &  0.06$\pm$0.02 \\ 
 227 &   HD045282 & 0.023 &  1  & 1.000 & 5264$\pm$86 & 3.19$\pm$0.16 & -1.43$\pm$0.12  &  5348$\pm$110 & 3.24$\pm$0.29 & -1.44$\pm$0.05 &  0.22$\pm$0.07 \\ 
 228 &   HD045282 & 0.065 &  3  & 1.042 & 5264$\pm$86 & 3.19$\pm$0.16 & -1.43$\pm$0.12  &  5268$\pm$48 & 3.14$\pm$0.23 & -1.52$\pm$0.05 &  0.36$\pm$0.06 \\ 
 253 &   HD046480 & 0.016 & 15  & 1.218 & 4785$\pm$26 & 2.63$\pm$0.12 & -0.49$\pm$0.01  &  4791$\pm$55 & 2.65$\pm$0.13 & -0.50$\pm$0.08 &  0.31$\pm$0.05 \\ 
 254 &   HD046480 & 0.016 & 15  & 1.214 & 4785$\pm$26 & 2.63$\pm$0.12 & -0.49$\pm$0.01  &  4791$\pm$55 & 2.65$\pm$0.13 & -0.50$\pm$0.08 &  0.31$\pm$0.05 \\ 
 314 &   HD063791 & 0.040 &  4  & 1.028 & 4715$\pm$78 & 1.75$\pm$0.08 & -1.68$\pm$0.08  &  4868$\pm$275 & 1.78$\pm$0.47 & -1.66$\pm$0.38 &  0.44$\pm$0.12 \\ 
 384 &   HD087140 & 0.033 &  2  & 1.038 & 5129$\pm$103 & 2.66$\pm$0.25 & -1.80$\pm$0.13  &  5090$\pm$5 & 2.58$\pm$0.10 & -1.82$\pm$0.06 &  0.42$\pm$0.01 \\ 
 425 &   HD108317 & 0.039 &  2  & 1.034 & 5259$\pm$111 & 2.68$\pm$0.25 & -2.27$\pm$0.05  &  5117$\pm$40 & 2.70$\pm$0.15 & -2.33$\pm$0.11 &  0.45$\pm$0.08 \\ 
 452 &   HD117876 & 0.075 &  3  & 1.091 & 4747$\pm$128 & 2.27$\pm$0.03 & -0.48$\pm$0.02  &  4806$\pm$18 & 2.25$\pm$0.15 & -0.44$\pm$0.03 &  0.42$\pm$0.10 \\ 
 454 &   HD122956 & 0.023 &  3  & 1.018 & 4633$\pm$78 & 1.46$\pm$0.18 & -1.72$\pm$0.11  &  4646$\pm$16 & 1.43$\pm$0.04 & -1.73$\pm$0.02 &  0.42$\pm$0.07 \\ 
 459 &   HD124897 & 0.136 &  5  & 1.023 & 4302$\pm$115 & 1.66$\pm$0.31 & -0.52$\pm$0.11  &  4346$\pm$98 & 1.87$\pm$0.46 & -0.55$\pm$0.24 &  0.29$\pm$0.15 \\ 
 470 &   HD135722 & 0.011 &  3  & 1.016 & 4795$\pm$76 & 2.60$\pm$0.41 & -0.40$\pm$0.10  &  4846$\pm$8 & 2.60$\pm$0.03 & -0.40$\pm$0.03 &  0.14$\pm$0.04 \\ 
 473 &   HD137759 & 0.142 &  2  & 1.014 & 4549$\pm$118 & 2.88$\pm$0.21 &  0.13$\pm$0.11  &  4558$\pm$64 & 2.54$\pm$0.07 &  0.14$\pm$0.02 & -0.03$\pm$0.11 \\ 
 509 &   HD159181 & 0.180 &  4  & 1.215 & 5234$\pm$158 & 1.56$\pm$0.19 &  0.15$\pm$0.12  &  5235$\pm$48 & 1.82$\pm$0.31 &  0.14$\pm$0.24 &  0.04$\pm$0.19 \\ 
 566 &   HD166161 & 0.059 &  2  & 1.040 & 5210$\pm$167 & 2.25$\pm$0.42 & -1.22$\pm$0.13  &  5071$\pm$122 & 2.16$\pm$0.11 & -1.18$\pm$0.05 &  0.31$\pm$0.06 \\ 
 568 &   HD166208 & 0.161 &  1  & 1.000 & 5037$\pm$56 & 2.71$\pm$0.08 &  0.07$\pm$0.11  &  4919$\pm$98 & 2.52$\pm$0.05 &  0.08$\pm$0.04 &  0.17$\pm$0.01 \\ 
 652 &   HD175305 & 0.082 &  1  & 1.000 & 5053$\pm$140 & 2.49$\pm$0.26 & -1.43$\pm$0.07  &  4899$\pm$17 & 2.30$\pm$0.03 & -1.43$\pm$0.05 &  0.27$\pm$0.06 \\ 
 701 &   HD187111 & 0.125 &  4  & 1.149 & 4299$\pm$75 & 0.74$\pm$0.30 & -1.78$\pm$0.18  &  4343$\pm$77 & 0.79$\pm$0.21 & -1.59$\pm$0.09 &  0.47$\pm$0.04 \\ 
 763 &   HD198149 & 0.078 &  2  & 1.015 & 4956$\pm$177 & 3.35$\pm$0.22 & -0.12$\pm$0.18  &  5027$\pm$10 & 3.12$\pm$0.06 & -0.12$\pm$0.06 &  0.12$\pm$0.01 \\ 
 791 &   HD204543 & 0.026 &  1  & 1.000 & 4667$\pm$68 & 1.30$\pm$0.23 & -1.80$\pm$0.10  &  4617$\pm$43 & 1.31$\pm$0.08 & -1.76$\pm$0.10 &  0.24$\pm$0.07 \\ 
 792 &   HD204613 & 0.101 &  2  & 1.208 & 5742$\pm$135 & 3.72$\pm$0.19 & -0.51$\pm$0.16  &  5614$\pm$111 & 3.45$\pm$0.09 & -0.48$\pm$0.08 &  0.21$\pm$0.15 \\ 
 803 &   HD207130 & 0.093 &  3  & 1.011 & 4760$\pm$53 & 2.63$\pm$0.15 &  0.01$\pm$0.11  &  4727$\pm$16 & 2.40$\pm$0.12 &  0.01$\pm$0.03 &  0.07$\pm$0.01 \\ 
 825 &   HD216143 & 0.045 &  2  & 1.019 & 4495$\pm$82 & 1.12$\pm$0.38 & -2.20$\pm$0.06  &  4480$\pm$8 & 1.15$\pm$0.06 & -2.12$\pm$0.01 &  0.39$\pm$0.05 \\ 
 826 &   HD216174 & 0.066 &  2  & 1.145 & 4413$\pm$23 & 2.11$\pm$0.36 & -0.55$\pm$0.02  &  4381$\pm$9 & 2.21$\pm$0.02 & -0.53$\pm$0.00 &  0.33$\pm$0.04 \\ 
 836 &   HD218857 & 0.054 &  8  & 1.098 & 5119$\pm$44 & 2.50$\pm$0.37 & -1.91$\pm$0.09  &  5067$\pm$89 & 2.37$\pm$0.31 & -1.93$\pm$0.15 &  0.41$\pm$0.04 \\ 
 848 &   HD221345 & 0.019 & 12  & 1.224 & 4635$\pm$108 & 2.49$\pm$0.32 & -0.30$\pm$0.07  &  4666$\pm$45 & 2.50$\pm$0.09 & -0.30$\pm$0.04 &  0.18$\pm$0.07 \\ 
 849 &   HD221377 & 0.182 &  3  & 1.150 & 6176$\pm$188 & 3.61$\pm$0.17 & -0.88$\pm$0.17  &  6027$\pm$57 & 3.24$\pm$0.15 & -1.01$\pm$0.15 &  0.57$\pm$0.12 \\ 
 871 &   HD232078 & 0.101 &  1  & 1.000 & 3939$\pm$175 & 0.31$\pm$0.34 & -1.58$\pm$0.15  &  3983$\pm$186 & 0.30$\pm$0.53 & -1.73$\pm$0.76 &  0.27$\pm$0.15 \\ 
 878 &  BD+233130 & 0.033 &  1  & 1.000 & 5119$\pm$140 & 2.39$\pm$0.38 & -2.62$\pm$0.19  &  5039$\pm$20 & 2.42$\pm$0.16 & -2.55$\pm$0.01 &  0.60$\pm$0.04 \\ 
 883 &  BD+302611 & 0.139 &  2  & 1.013 & 4292$\pm$100 & 0.96$\pm$0.37 & -1.41$\pm$0.19  &  4421$\pm$274 & 0.83$\pm$0.72 & -1.43$\pm$0.05 &  0.46$\pm$0.13 \\ 
 927 &   HD000245 & 0.110 &  2  & 1.032 & 5490$\pm$153 & 3.48$\pm$0.15 & -0.77$\pm$0.08  &  5377$\pm$59 & 3.67$\pm$0.06 & -0.84$\pm$0.19 &  0.34$\pm$0.08 \\ 
 941 &   HD003546 & 0.035 & 15  & 1.412 & 4906$\pm$168 & 2.45$\pm$0.43 & -0.64$\pm$0.12  &  4868$\pm$71 & 2.51$\pm$0.19 & -0.65$\pm$0.09 &  0.27$\pm$0.03 \\ 
1395 &   HD105546 & 0.047 &  2  & 1.018 & 5234$\pm$79 & 2.38$\pm$0.14 & -1.39$\pm$0.15  &  5387$\pm$190 & 2.30$\pm$0.14 & -1.37$\pm$0.23 &  0.54$\pm$0.06 \\ 
1452 &   HD122563 & 0.043 &  1  & 1.000 & 4565$\pm$131 & 1.17$\pm$0.24 & -2.62$\pm$0.14  &  4566$\pm$440 & 1.12$\pm$1.31 & -2.63$\pm$0.37 &  0.60$\pm$0.22 \\ 
1483 &   HD136512 & 0.098 &  2  & 1.060 & 4719$\pm$66 & 2.72$\pm$0.04 & -0.33$\pm$0.16  &  4747$\pm$46 & 2.62$\pm$0.25 & -0.30$\pm$0.00 &  0.08$\pm$0.03 \\ 
1484 &   HD136512 & 0.110 &  4  & 1.072 & 4719$\pm$66 & 2.72$\pm$0.04 & -0.33$\pm$0.16  &  4754$\pm$29 & 2.53$\pm$0.17 & -0.30$\pm$0.08 &  0.14$\pm$0.11 \\ 
1485 &   HD136512 & 0.111 &  4  & 1.072 & 4719$\pm$66 & 2.72$\pm$0.04 & -0.33$\pm$0.16  &  4754$\pm$29 & 2.53$\pm$0.17 & -0.30$\pm$0.08 &  0.14$\pm$0.11 \\ 
1486 &   HD136512 & 0.070 & 14  & 1.149 & 4719$\pm$66 & 2.72$\pm$0.04 & -0.33$\pm$0.16  &  4824$\pm$25 & 2.54$\pm$0.08 & -0.33$\pm$0.05 &  0.21$\pm$0.06 \\ 
1487 &   HD136512 & 0.080 & 13  & 1.144 & 4719$\pm$66 & 2.72$\pm$0.04 & -0.33$\pm$0.16  &  4821$\pm$27 & 2.52$\pm$0.08 & -0.32$\pm$0.06 &  0.21$\pm$0.06 \\ 
1576 &   HD162211 & 0.058 &  4  & 1.210 & 4568$\pm$74 & 2.74$\pm$0.11 &  0.04$\pm$0.06  &  4581$\pm$72 & 2.59$\pm$0.17 &  0.04$\pm$0.02 &  0.04$\pm$0.07 \\ 
1811 &   HD188326 & 0.172 &  3  & 1.092 & 5272$\pm$40 & 3.80$\pm$0.01 & -0.18$\pm$0.00  &  5074$\pm$39 & 3.34$\pm$0.15 & -0.20$\pm$0.21 &  0.12$\pm$0.03 \\ 
1812 &   HD188326 & 0.161 &  3  & 1.084 & 5272$\pm$40 & 3.80$\pm$0.01 & -0.18$\pm$0.00  &  5074$\pm$41 & 3.34$\pm$0.16 & -0.20$\pm$0.21 &  0.12$\pm$0.03 \\ 
1876 &   HD212943 & 0.081 &  4  & 1.085 & 4625$\pm$67 & 2.79$\pm$0.05 & -0.29$\pm$0.09  &  4656$\pm$35 & 2.61$\pm$0.10 & -0.30$\pm$0.04 &  0.17$\pm$0.07 \\ 
1893 &   HD216219 & 0.093 &  1  & 1.000 & 5628$\pm$106 & 3.12$\pm$0.22 & -0.41$\pm$0.10  &  5727$\pm$152 & 3.36$\pm$0.50 & -0.39$\pm$0.21 &  0.06$\pm$0.19 \\ 
1916 &   HD219449 & 0.070 & 16  & 1.259 & 4647$\pm$75 & 2.56$\pm$0.26 & -0.03$\pm$0.07  &  4626$\pm$35 & 2.39$\pm$0.07 & -0.03$\pm$0.03 &  0.03$\pm$0.03 \\ 
\end{longtable}
}% End \longtab

\longtabL{6}{
\begin{landscape}
\tiny
\begin{longtable}{l|p{1cm}p{1cm}|l|p{1cm}p{1cm}|l|p{1.27cm}p{1.27cm}|c|p{1.27cm}p{1.27cm}|c}
\caption{\label{finalparam} Atmospheric parameters for all stars analysed in the six
  clusters: T$_{eff}$, log($g$), [Fe/H], [Mg/Fe] and
  [$\alpha$/Fe]. Membership identification is copied from Table
  \ref{starinfo} to guide the reader.}\\
\hline\hline
\noalign{\smallskip}
 {\rm NGC ID} & T$_{eff}^{(a)}$ (K) & T$_{eff}^{(b)}$ (K) & T$_{eff}^{(avg)}$ (K) & log($g$)$^{(a)}$ & log($g$)$^{(b)}$ & log($g$)$^{(avg)}$ & [Fe/H]$^{(a)}$ & [Fe/H]$^{(b)}$ & [Fe/H]$^{(avg)}$ & [Mg/Fe]$^{(a)}$ &  [$\alpha$/Fe]$^{(b)}$ & members\\
\noalign{\smallskip}
\hline
\noalign{\smallskip}
\endfirsthead
\caption{continued.}\\
\hline\hline
\noalign{\smallskip}
 {\rm NGC ID} & T$_{eff}^{(a)}$ (K) & T$_{eff}^{(b)}$ (K) & T$_{eff}^{(avg)}$ (K) & log($g$)$^{(a)}$ & log($g$)$^{(b)}$ & log($g$)$^{(avg)}$ & [Fe/H]$^{(a)}$ & [Fe/H]$^{(b)}$ & [Fe/H]$^{(avg)}$ & [Mg/Fe]$^{(a)}$ &  [$\alpha$/Fe]$^{(b)}$  & members\\
\noalign{\smallskip}
\hline
\noalign{\smallskip}
\endhead
\hline
\endfoot
6528\_2    &   3969$\pm$80   &    4074$\pm$114	&   4004$\pm$ 66	  &    1.54$\pm$0.18 	&    1.9$\pm$0.4     &   1.60$\pm$0.17      &  0.02$\pm$0.16  	  &   -0.50$\pm$0.15	&   -0.24$\pm$0.16     &   -0.04$\pm$0.18    &   0.20$\pm$0.14     &   \\  
6528\_3    &   3640$\pm$100  &    3525$\pm$75   &   3566$\pm$ 60     &    0.70$\pm$0.00 	&    2.9$\pm$0.8     &   0.96$\pm$0.28      &  -0.10$\pm$0.07 	  &   -0.70$\pm$0.25	&   -0.14$\pm$0.06     &   0.23$\pm$0.00     &   0.32$\pm$0.07     &   \\  
6528\_4    &   3900$\pm$79   &    3623$\pm$125	&   3821$\pm$ 67	  &    1.38$\pm$0.19 	&    0.8$\pm$0.7     &   1.34$\pm$0.18      &  -0.07$\pm$0.20 	  &   -1.35$\pm$0.23	&   -0.62$\pm$0.15     &   0.06$\pm$0.20     &   0.26$\pm$0.11     &   \\  
6528\_5    &   4027$\pm$79   &    4198$\pm$151	&   4064$\pm$ 70	  &    1.54$\pm$0.34 	&    2.0$\pm$0.4     &   1.74$\pm$0.26      &  -0.31$\pm$0.22 	  &   -0.85$\pm$0.23	&   -0.57$\pm$0.16     &   0.12$\pm$0.18     &   0.31$\pm$0.10     &   \\  
6528\_6    &   4344$\pm$215  &    4625$\pm$125	&   4554$\pm$108	  &    2.1$\pm$0.5   	&    2.8$\pm$0.23    &   2.71$\pm$0.21      &  0.04$\pm$0.17  	  &   -0.21$\pm$0.30	&   -0.02$\pm$0.15     &   0.03$\pm$0.12     &   0.18$\pm$0.12     &   \\  
6528\_7    &   3930$\pm$58   &    4124$\pm$125	&   3964$\pm$ 53	  &    1.51$\pm$0.01 	&    2.05$\pm$0.35   &   1.51$\pm$0.01      &  0.10$\pm$0.13  	  &   -0.25$\pm$0.25	&    0.02$\pm$0.12     &   0.16$\pm$0.06     &   0.30$\pm$0.08     &   \\  
6528\_8    &   4159$\pm$162  &    4225$\pm$175	&   4189$\pm$119	  &    1.7$\pm$0.4   	&    2.0$\pm$0.5     &   1.88$\pm$0.32      &  -0.08$\pm$0.20 	  &   -0.75$\pm$0.25	&   -0.34$\pm$0.16     &   0.02$\pm$0.19     &   0.28$\pm$0.09     & M  \\  
6528\_9    &   4244$\pm$170  &    4400$\pm$122	&   4347$\pm$ 99	  &    1.9$\pm$0.4   	&    2.6$\pm$0.4     &   2.28$\pm$0.27      &  -0.06$\pm$0.19 	  &   -0.70$\pm$0.24	&   -0.31$\pm$0.15     &   0.03$\pm$0.17     &   0.27$\pm$0.09     & M  \\  
6528\_10   &   4557$\pm$236  &    4925$\pm$115	&   4855$\pm$103	  &    2.4$\pm$0.5       &    3.30$\pm$0.24   &   3.14$\pm$0.22      &  -0.10$\pm$0.25 	  &    0.10$\pm$0.10	&    0.07$\pm$0.09     &   0.06$\pm$0.17     &   0.19$\pm$0.11     & M  \\  
6528\_11   &   3911$\pm$70   &    3800$\pm$187	&   3897$\pm$ 66	  &    1.43$\pm$0.18 	&    1.3$\pm$0.7     &   1.42$\pm$0.17      &  -0.04$\pm$0.21 	  &   -1.15$\pm$0.32	&   -0.37$\pm$0.18     &   0.06$\pm$0.17     &   0.29$\pm$0.11     & M  \\  
6528\_13   &   4853$\pm$198  &    4876$\pm$167	&   4866$\pm$128	  &    2.5$\pm$0.4       &    2.6$\pm$0.5     &   2.53$\pm$0.32      &  -0.44$\pm$0.22 	  &   -0.80$\pm$0.25	&   -0.60$\pm$0.17     &   0.24$\pm$0.20     &   0.22$\pm$0.11     &   \\  
6528\_14   &   4239$\pm$172  &    4397$\pm$164	&   4322$\pm$119	  &    1.9$\pm$0.5   	&    2.2$\pm$0.4     &   2.07$\pm$0.30      &  -0.13$\pm$0.21 	  &   -0.71$\pm$0.33	&   -0.30$\pm$0.18     &   0.03$\pm$0.11     &   0.28$\pm$0.12     &   \\  
6528\_15   &   4083$\pm$176  &    4325$\pm$114	&   4253$\pm$ 96	  &    1.61$\pm$0.19 	&    2.30$\pm$0.25   &   1.86$\pm$0.15      &  0.05$\pm$0.16  	  &   -0.35$\pm$0.23	&   -0.08$\pm$0.13     &   0.05$\pm$0.11     &   0.17$\pm$0.13     &   \\   
6528\_16   &   3938$\pm$40   &    4075$\pm$114	&   3953$\pm$ 38	  &    1.48$\pm$0.12 	&    1.95$\pm$0.35   &   1.53$\pm$0.11      &  -0.04$\pm$0.15 	  &   -0.80$\pm$0.24	&   -0.25$\pm$0.13     &   0.05$\pm$0.10     &   0.29$\pm$0.09     &   \\  
6528\_17   &   4500$\pm$147  &    4700$\pm$186	&   4577$\pm$115	  &    2.35$\pm$0.27     &    2.70$\pm$0.33   &   2.49$\pm$0.21      &  -0.37$\pm$0.13 	  &   -0.6$\pm$0.4	    &   -0.39$\pm$0.12     &   0.17$\pm$0.16     &   0.31$\pm$0.10     &   \\  
6528\_18   &   4709$\pm$194  &    4676$\pm$195	&   4692$\pm$138	  &    2.5$\pm$0.4   	&    2.5$\pm$0.5     &   2.52$\pm$0.29      &  -0.22$\pm$0.25 	  &   -0.6$\pm$0.4  	&   -0.35$\pm$0.21     &   0.12$\pm$0.18     &   0.16$\pm$0.12     &   \\  
6528\_19   &   4496$\pm$94   &    4701$\pm$99   &   4593$\pm$ 68     &    2.53$\pm$0.30 	&    2.90$\pm$0.20   &   2.79$\pm$0.17      &  0.13$\pm$0.09  	  &   -0.02$\pm$0.25	&    0.11$\pm$0.08     &   0.04$\pm$0.08     &   0.17$\pm$0.13     &   \\  
\noalign{\smallskip}
\hline
\noalign{\smallskip}
6553\_1   &   4141$\pm$107   &    4425$\pm$195  &  4207$\pm$ 94   &    1.58$\pm$0.14  &    2.2$\pm$0.5	 &    1.62$\pm$0.14     &   -0.08$\pm$0.09 	  &   -0.40$\pm$0.20	&   -0.14$\pm$0.08     &   0.08$\pm$0.06    &   0.32$\pm$0.07      & M \\ 
6553\_3   &   4384$\pm$155   &    4475$\pm$174  &  4424$\pm$116   &    1.9$\pm$0.4    &    2.2$\pm$0.5	 &    1.98$\pm$0.31     &   -0.06$\pm$0.08 	  &   -0.40$\pm$0.20	&   -0.10$\pm$0.07     &   0.12$\pm$0.12    &   0.29$\pm$0.09      & M \\ 
6553\_4   &   3837$\pm$100   &    3574$\pm$114  &  3723$\pm$ 75   &    1.17$\pm$0.18  &    1.1$\pm$0.7	 &    1.17$\pm$0.17     &   0.10$\pm$0.17  	  &   -1.35$\pm$0.23	&   -0.41$\pm$0.14     &   0.09$\pm$0.07    &   0.34$\pm$0.07      & M \\ 
6553\_5   &   3970$\pm$71    &    4076$\pm$160  &  3987$\pm$ 65   &    1.65$\pm$0.10  &    1.6$\pm$0.6	 &    1.65$\pm$0.10     &   0.06$\pm$0.13  	  &   -0.75$\pm$0.33	&   -0.04$\pm$0.12     &   0.05$\pm$0.11    &   0.34$\pm$0.07      & M \\ 
6553\_6   &   4312$\pm$135   &    4400$\pm$122  &  4360$\pm$ 91   &    1.9$\pm$0.4    &    1.9$\pm$0.4	 &    1.90$\pm$0.28     &   0.004$\pm$0.09 	  &   -0.50$\pm$0.00	&   -0.13$\pm$0.08     &   -0.01$\pm$0.17   &   0.26$\pm$0.11      & M \\ 
6553\_7   &   3730$\pm$60    &    3550$\pm$100  &  3682$\pm$ 51   &    0.90$\pm$0.13  &    1.2$\pm$0.9	 &    0.91$\pm$0.13     &   0.27$\pm$0.18  	  &   -0.95$\pm$0.27	&   -0.10$\pm$0.15     &   0.08$\pm$0.10    &   0.28$\pm$0.10      & M \\ 
6553\_8   &   3640$\pm$0     &    3500$\pm$0    &  3528$\pm$ 45   &    0.70$\pm$0.00  &    2.8$\pm$0.8	 &    0.98$\pm$0.28     &   -0.10$\pm$0.06 	  &   -0.70$\pm$0.24	&   -0.14$\pm$0.06     &   0.23$\pm$0.10    &   0.32$\pm$0.07      &  \\ 
6553\_9   &   4259$\pm$15    &    4399$\pm$165  &  4260$\pm$ 15   &    1.47$\pm$0.08  &    2.1$\pm$0.5	 &    1.49$\pm$0.08     &   -0.13$\pm$0.00 	  &   -0.45$\pm$0.15	&   -0.13$\pm$0.01     &   0.11$\pm$0.10    &   0.28$\pm$0.10      & M \\ 
6553\_10  &   4388$\pm$155   &    4525$\pm$174  &  4449$\pm$116   &    1.9$\pm$0.4    &    2.3$\pm$0.5	 &    2.02$\pm$0.31     &   -0.06$\pm$0.08 	  &   -0.35$\pm$0.23	&   -0.09$\pm$0.07     &   0.13$\pm$0.12    &   0.29$\pm$0.09      & M \\ 
6553\_11  &   3941$\pm$44    &    3801$\pm$218  &  3936$\pm$ 43   &    1.47$\pm$0.19  &    1.0$\pm$0.8	 &    1.45$\pm$0.19     &   0.01$\pm$0.14  	  &   -0.9$\pm$0.4	    &   -0.07$\pm$0.13     &   0.07$\pm$0.09    &   0.27$\pm$0.09      & M \\ 
6553\_13  &   4219$\pm$57    &    4350$\pm$122  &  4242$\pm$ 52   &    1.63$\pm$0.26  &    2.0$\pm$0.4	 &    1.74$\pm$0.22     &   -0.01$\pm$0.13	  &   -0.50$\pm$0.00	&   -0.22$\pm$0.10     &   0.02$\pm$0.10    &   0.25$\pm$0.10      & M \\ 
6553\_14  &   3889$\pm$64    &    3896$\pm$301  &  3889$\pm$ 63   &    1.38$\pm$0.16  &    1.3$\pm$1.0	 &    1.38$\pm$0.16     &   -0.01$\pm$0.17 	  &   -0.95$\pm$0.42	&   -0.15$\pm$0.16     &   0.004$\pm$0.20   &   0.33$\pm$0.10      & M \\ 
6553\_15  &   3640$\pm$0     &    3500$\pm$0    &  3528$\pm$ 45   &    0.70$\pm$0.00  &    3.1$\pm$0.8	 &    1.01$\pm$0.28     &   -0.10$\pm$0.06 	  &   -0.60$\pm$0.20	&   -0.15$\pm$0.06     &   0.23$\pm$0.10    &   0.32$\pm$0.07      &  \\ 
6553\_16  &   3932$\pm$29    &    4101$\pm$123  &  3941$\pm$ 28   &    1.54$\pm$0.13  &    1.9$\pm$0.4	 &    1.58$\pm$0.12     &   -0.20$\pm$0.05 	  &   -0.70$\pm$0.24	&   -0.22$\pm$0.05     &   0.05$\pm$0.09    &   0.33$\pm$0.06      & M \\ 
6553\_17  &   3640$\pm$100   &    3548$\pm$99   &  3594$\pm$ 70   &    0.70$\pm$0.00  &    2.4$\pm$0.7	 &    1.15$\pm$0.34     &   -0.05$\pm$0.05 	  &   -0.90$\pm$0.20	&   -0.10$\pm$0.05     &   0.23$\pm$0.10    &   0.35$\pm$0.07      &  \\ 
6553\_18  &   3640$\pm$100   &    3500$\pm$0.0  &  3528$\pm$ 45   &    0.70$\pm$0.00  &    3.4$\pm$0.5	 &    1.78$\pm$0.31     &   -0.10$\pm$0.07 	  &    0.20$\pm$0.17	&   -0.06$\pm$0.06     &   0.23$\pm$0.10    &   0.33$\pm$0.06      &  \\ 
6553\_19  &   4259$\pm$15    &    4176$\pm$114  &  4258$\pm$ 15   &    1.47$\pm$0.08  &   1.75$\pm$0.33 &    1.49$\pm$0.08      &   -0.13$\pm$0.00 	  &   -0.65$\pm$0.23	&   -0.13$\pm$0.01     &   0.11$\pm$0.10    &   0.27$\pm$0.09      & M \\ 
\noalign{\smallskip}
\hline
\noalign{\smallskip}
M71\_2   &  4743$\pm$210   &    4850$\pm$122  &  4823$\pm$106    &    2.5$\pm$0.5  	&    2.95$\pm$0.27   &  2.85$\pm$0.24    &  -0.48$\pm$0.24	  &   -0.85$\pm$0.23    &   -0.67$\pm$0.17      &    0.25$\pm$0.20    &   0.28$\pm$0.13      & M \\
M71\_4   &  4305$\pm$77    &    4449$\pm$187  &  4326$\pm$ 71    &    1.8$\pm$0.5  	&    2.2$\pm$0.6     &  1.99$\pm$0.36    &  -0.56$\pm$0.24	  &   -0.85$\pm$0.23    &   -0.71$\pm$0.17      &    0.32$\pm$0.18    &   0.28$\pm$0.10      & M \\
M71\_5   &  4722$\pm$223   &    4899$\pm$198  &  4821$\pm$148    &    2.4$\pm$0.6  	&    3.10$\pm$0.30   &  2.97$\pm$0.27    &  -0.51$\pm$0.26	  &   -0.85$\pm$0.32    &   -0.64$\pm$0.20      &    0.27$\pm$0.20    &   0.29$\pm$0.09      & M \\
M71\_6   &  4364$\pm$94    &    4523$\pm$174  &  4400$\pm$ 83    &    1.9$\pm$0.4  	&    2.2$\pm$0.6     &  2.02$\pm$0.32    &  -0.55$\pm$0.22	  &   -0.95$\pm$0.35    &   -0.66$\pm$0.19      &    0.26$\pm$0.16    &   0.30$\pm$0.08      & M \\
M71\_7   &  3991$\pm$99    &    4026$\pm$207  &  3997$\pm$ 89    &    1.5$\pm$0.4  	&    1.7$\pm$0.7     &  1.53$\pm$0.35    &  -0.32$\pm$0.21	  &   -1.25$\pm$0.33    &   -0.58$\pm$0.17      &    0.15$\pm$0.18    &   0.34$\pm$0.07      & M \\
M71\_8   &  4409$\pm$132   &    4850$\pm$122  &  4646$\pm$ 90    &    2.1$\pm$0.5  	&    3.15$\pm$0.23   &  2.95$\pm$0.21    &  0.16$\pm$0.13 	  &    0.26$\pm$0.21    &    0.18$\pm$0.11      &    0.01$\pm$0.14    &   0.13$\pm$0.10      &  \\
M71\_9   &  4303$\pm$102   &    4350$\pm$166  &  4316$\pm$ 87    &    1.9$\pm$0.5  	&    2.0$\pm$0.5     &  1.97$\pm$0.33    &  -0.54$\pm$0.23	  &   -1.05$\pm$0.27    &   -0.76$\pm$0.17      &    0.27$\pm$0.21    &   0.28$\pm$0.10      & M \\
M71\_10  &  3861$\pm$77    &    3549$\pm$99   &  3743$\pm$ 61    &    1.23$\pm$0.21	&    0.5$\pm$0.6     &  1.16$\pm$0.20    &  -0.06$\pm$0.26	  &   -1.70$\pm$0.24    &   -0.96$\pm$0.18      &    0.11$\pm$0.07    &   0.29$\pm$0.14      & M \\
M71\_13  &  4685$\pm$209   &    4850$\pm$122  &  4808$\pm$106    &    2.4$\pm$0.5  	&    2.95$\pm$0.27   &  2.82$\pm$0.24    &  -0.44$\pm$0.26	  &   -0.80$\pm$0.25    &   -0.63$\pm$0.18      &    0.23$\pm$0.20    &   0.25$\pm$0.10      & M \\
M71\_14  &  4901$\pm$212   &    5175$\pm$114  &  5113$\pm$101    &    2.7$\pm$0.5  	&    3.15$\pm$0.32   &  3.01$\pm$0.27    &  -0.44$\pm$0.25	  &   -0.65$\pm$0.23    &   -0.55$\pm$0.17      &    0.25$\pm$0.20    &   0.27$\pm$0.11      & M \\
M71\_15  &  4840$\pm$259   &    5101$\pm$165  &  5025$\pm$139    &    2.5$\pm$0.6  	&    3.10$\pm$0.30   &  2.98$\pm$0.27    &  -0.50$\pm$0.30	  &   -0.55$\pm$0.15    &   -0.54$\pm$0.13      &    0.26$\pm$0.19    &   0.27$\pm$0.11      & M \\
M71\_16  &  4694$\pm$214   &    4974$\pm$175  &  4862$\pm$135    &    2.5$\pm$0.4  	&    2.95$\pm$0.35   &  2.76$\pm$0.27    &  -0.13$\pm$0.25	  &   -0.11$\pm$0.27    &   -0.12$\pm$0.18      &    0.08$\pm$0.18    &   0.20$\pm$0.13      &  \\
\noalign{\smallskip}
\hline
\noalign{\smallskip}
 6558\_3  &  4758$\pm$228   &    4926$\pm$160  &  4871$\pm$131   &    2.5$\pm$0.5    &    3.10$\pm$0.30   &   2.94$\pm$0.25       &  -0.29$\pm$0.26       &   -0.70$\pm$0.24    &     -0.51$\pm$0.18     &   0.15$\pm$0.20    &   0.30$\pm$0.10     &    \\
 6558\_4  &  4678$\pm$172   &    4625$\pm$167  &  4651$\pm$120   &    2.55$\pm$0.33  &    3.0$\pm$0.4     &   2.75$\pm$0.26       &  -0.21$\pm$0.19       &   -1.30$\pm$0.25    &     -0.61$\pm$0.15     &   0.09$\pm$0.14    &   0.29$\pm$0.09     &    \\ 
 6558\_5  &  4921$\pm$445   &    4900$\pm$200  &  4903$\pm$182   &    2.0$\pm$0.6    &    2.00$\pm$0.32   &   2.00$\pm$0.29       &  -1.2$\pm$0.6         &   -1.80$\pm$0.24    &     -1.72$\pm$0.22     &   0.37$\pm$0.19    &   0.21$\pm$0.11     &    \\  
 6558\_6  &  4738$\pm$329   &    4951$\pm$186  &  4899$\pm$162   &    2.1$\pm$0.8    &    2.8$\pm$0.4     &   2.68$\pm$0.39       &  -0.9$\pm$0.4         &   -1.20$\pm$0.24    &     -1.11$\pm$0.20     &   0.22$\pm$0.07    &   0.26$\pm$0.13     &  M  \\ 
 6558\_7  &  4928$\pm$317   &    4926$\pm$114  &  4926$\pm$108   &    2.5$\pm$0.7    &    2.70$\pm$0.25   &   2.68$\pm$0.24       &  -0.64$\pm$0.32       &   -1.15$\pm$0.23    &     -0.98$\pm$0.19     &   0.26$\pm$0.13    &   0.26$\pm$0.13     &  M  \\ 
 6558\_8  &  3640$\pm$100   &    3524$\pm$74   &  3565$\pm$ 59   &    0.70$\pm$0.5   &    2.8$\pm$0.9     &   1.05$\pm$0.36       &  0.00$\pm$0.07        &   -0.95$\pm$0.15    &     -0.16$\pm$0.06     &   0.23$\pm$0.00    &   0.31$\pm$0.08     &  M  \\ 
 6558\_9  &  4663$\pm$336   &    5076$\pm$195  &  4972$\pm$168   &    1.8$\pm$0.9    &    2.9$\pm$0.5     &   2.68$\pm$0.46       &  -1.2$\pm$0.4         &   -1.10$\pm$0.20    &     -1.11$\pm$0.18     &   0.41$\pm$0.16    &   0.21$\pm$0.11     &  M  \\
6558\_10  &  4398$\pm$157   &    4799$\pm$100  &  4684$\pm$ 84   &    2.1$\pm$0.5    &    3.15$\pm$0.23   &   2.95$\pm$0.21       &  0.10$\pm$0.13        &    0.16$\pm$0.19    &      0.12$\pm$0.11     &   0.02$\pm$0.13    &   0.12$\pm$0.12     &    \\ 
6558\_11  &  4887$\pm$413   &    5325$\pm$114  &  5294$\pm$110   &    2.2$\pm$0.9    &    3.30$\pm$0.25   &   3.22$\pm$0.24       &  -1.1$\pm$0.5         &   -1.00$\pm$0.00    &     -1.00$\pm$0.05     &   0.37$\pm$0.19    &   0.19$\pm$0.14     &  M  \\
6558\_12  &  4516$\pm$177   &    4548$\pm$245  &  4527$\pm$143   &    2.3$\pm$0.5    &    2.2$\pm$0.6     &   2.26$\pm$0.37       &  -0.18$\pm$0.27       &   -0.58$\pm$0.34    &     -0.33$\pm$0.21     &   0.10$\pm$0.18    &   0.25$\pm$0.10     &    \\
6558\_13  &  4683$\pm$94    &    5076$\pm$115  &  4840$\pm$ 73   &    2.49$\pm$0.14  &    3.25$\pm$0.25   &   2.67$\pm$0.12       &  -0.22$\pm$0.09       &   -0.25$\pm$0.25    &     -0.22$\pm$0.08     &   0.09$\pm$0.15    &   0.31$\pm$0.08     &    \\
6558\_14  &  4643$\pm$454   &    4675$\pm$115  &  4673$\pm$111   &    3.0$\pm$1.0    &    4.4$\pm$0.5     &   4.12$\pm$0.44       &  0.30$\pm$0.13        &   -0.25$\pm$0.25    &      0.18$\pm$0.12     &   0.18$\pm$0.07    &   0.12$\pm$0.07     &    \\
6558\_15  &  4885$\pm$409   &    5500$\pm$0    &  5491$\pm$ 50   &    1.9$\pm$0.8    &    3.0$\pm$0.4     &   2.79$\pm$0.35       &  -1.40$\pm$0.25       &   -1.00$\pm$0.00    &     -1.02$\pm$0.05     &   0.44$\pm$0.15    &   0.20$\pm$0.13     &    \\ 
6558\_16  &  4690$\pm$186   &    4825$\pm$195  &  4754$\pm$135   &    2.50$\pm$0.35  &    2.9$\pm$0.5     &   2.64$\pm$0.28       &  -0.20$\pm$0.22       &   -0.50$\pm$0.32    &     -0.29$\pm$0.18     &   0.10$\pm$0.16    &   0.23$\pm$0.13     &    \\
6558\_17  &  4829$\pm$192   &    4800$\pm$187  &  4814$\pm$134   &    2.7$\pm$0.6    &    3.30$\pm$0.33   &   3.14$\pm$0.28       &  -0.56$\pm$0.26       &   -1.4$\pm$0.4      &     -0.84$\pm$0.21     &   0.33$\pm$0.18    &   0.36$\pm$0.05     &    \\
6558\_18  &  4888$\pm$241   &    5151$\pm$122  &  5097$\pm$109   &    2.6$\pm$0.4    &    3.15$\pm$0.23   &   3.04$\pm$0.20       &  -0.36$\pm$0.26       &   -0.50$\pm$0.00    &     -0.50$\pm$0.05     &   0.20$\pm$0.22    &   0.26$\pm$0.11     &    \\ 
6558\_19  &  4912$\pm$435   &    5175$\pm$160  &  5144$\pm$150   &    2.1$\pm$0.9    &    3.20$\pm$0.25   &   3.12$\pm$0.24       &  -1.5$\pm$0.4         &   -1.45$\pm$0.15    &     -1.46$\pm$0.14     &   0.46$\pm$0.15    &   0.22$\pm$0.12     &    \\
\noalign{\smallskip}
\hline
\noalign{\smallskip}
6426\_1 	  &    4849$\pm$25   	&    5026$\pm$175	  	& 	4853$\pm$ 25		& 	  3.45$\pm$0.04		& 	  4.5$\pm$0.4	  	&   3.46$\pm$0.04  		&   0.09$\pm$0.14 	 	&    -0.75$\pm$0.25	 	&   -0.11$\pm$0.12 	& 	     -0.02$\pm$0.01 	&     0.34$\pm$0.07		 &    		     \\	
6426\_2 	  &    4643$\pm$51      &    5075$\pm$115	    & 	4714$\pm$ 47	    &  	  3.05$\pm$0.21		&  	  4.7$\pm$0.4	    &   3.41$\pm$0.19       &   0.30$\pm$0.05 	 	&    -0.11$\pm$0.27	    &    0.28$\pm$0.05 	&        0.18$\pm$0.10  	&     0.14$\pm$0.09	     &  		     \\ 
6426\_3 	  &    4250$\pm$163  	&    4625$\pm$125	  	& 	4486$\pm$ 99		& 	  3.50$\pm$0.34	    & 	  3.95$\pm$0.35	  	&   3.72$\pm$0.24  		&   0.10$\pm$0.06 	 	&    -0.60$\pm$0.20	 	&    0.05$\pm$0.05 	& 	     -0.02$\pm$0.10 	&     0.17$\pm$0.12		 &  		     \\	
6426\_4 	  &    4881$\pm$350  	&    4900$\pm$122	  	& 	4898$\pm$116		& 	  2.0$\pm$0.8  	    & 	  1.9$\pm$0.4	  	&   1.92$\pm$0.39  		&   -1.93$\pm$0.35	 	&    -2.5$\pm$0.1	 	&   -2.46$\pm$0.10 	& 	     0.37$\pm$0.12  	&     0.28$\pm$0.10		 &  	M	     \\	
6426\_7 	  &    5008$\pm$435  	&    5250$\pm$10	  	&   5250$\pm$ 10	 	&  	  2.1$\pm$0.8  	    &  	  2.1$\pm$0.4	  	&   2.11$\pm$0.34  		&   -1.9$\pm$0.4  	 	&    -2.5$\pm$0.1	 	&   -2.47$\pm$0.10 	&        0.41$\pm$0.17  	&     0.23$\pm$0.13		 &  	M	     \\  
6426\_9 	  &    5071$\pm$374  	&    5250$\pm$10	  	&   5250$\pm$ 10	 	&  	  2.3$\pm$0.9  	    &  	  2.30$\pm$0.33	  	&   2.30$\pm$0.31  		&   -2.1$\pm$0.4  	 	&    -2.5$\pm$0.1	 	&   -2.47$\pm$0.10 	&        0.38$\pm$0.19  	&     0.20$\pm$0.14		 &  	M	     \\  
6426\_10	  &    4487$\pm$70      &    4951$\pm$149	    & 	4571$\pm$ 63	    &  	  1.15$\pm$0.12	    &  	  2.1$\pm$0.4	    &   1.22$\pm$0.12  	    &   -2.06$\pm$0.14	 	&    -2.05$\pm$0.15	    &   -2.05$\pm$0.10 	&        0.39$\pm$0.11  	&     0.21$\pm$0.14	     &  	M	     \\ 
6426\_11	  &    4967$\pm$156 	&    5475$\pm$207	  	& 	5150$\pm$125		&     3.1$\pm$0.4  		&     4.65$\pm$0.32	  	&   4.13$\pm$0.26  		&   -0.44$\pm$0.32	 	&    -0.70$\pm$0.24	 	&   -0.61$\pm$0.19 	&  	     0.25$\pm$0.16      &     0.30$\pm$0.08		 &  		     \\ 	
6426\_13	  &    5015$\pm$331  	&    5000$\pm$10	   	&   5000$\pm$ 10	 	&  	  2.2$\pm$0.8  	    &  	  2.0$\pm$0.4	  	&   2.08$\pm$0.37  		&   -2.0$\pm$0.4  	 	&    -2.5$\pm$0.1	 	&   -2.47$\pm$0.10 	&        0.37$\pm$0.19  	&     0.25$\pm$0.10		 &  	M	     \\  
6426\_18	  &    4824$\pm$52      &    4875$\pm$125	    & 	4831$\pm$ 48	    &  	  3.24$\pm$0.23	    &  	  3.5$\pm$0.4	    &   3.31$\pm$0.20  	    &   0.01$\pm$0.07 	 	&    -0.5$\pm$0.1	    &   -0.15$\pm$0.06 	&        0.05$\pm$0.07  	&     0.24$\pm$0.12	     &  		     \\ 
\noalign{\smallskip} 
\hline
\noalign{\smallskip}
 Ter8\_1	 &  4885$\pm$314   &    5250$\pm$100    &  5067$\pm$100   &    2.0$\pm$0.7    &    2.30$\pm$0.33   &   2.24$\pm$0.30    &  -1.82$\pm$0.34     &   -2.00$\pm$0.15    &    -1.97$\pm$0.14  &    0.40$\pm$0.13     &   0.20$\pm$0.14  &  M  \\
 Ter8\_4	 &  4565$\pm$193   &    4299$\pm$99   &  4354$\pm$ 88   &    1.3$\pm$0.5    &    0.4$\pm$0.4     &   0.67$\pm$0.31    &  -1.78$\pm$0.23     &   -2.50$\pm$0.15    &    -2.28$\pm$0.13  &    0.40$\pm$0.14     &   0.20$\pm$0.14  &  M  \\
 Ter8\_5	 &  4952$\pm$326   &    5000$\pm$100    &  4976$\pm$100   &    2.1$\pm$0.8    &    1.9$\pm$0.4     &   1.98$\pm$0.37    &  -1.9$\pm$0.4       &   -2.50$\pm$0.15    &    -2.42$\pm$0.14  &    0.40$\pm$0.17     &   0.25$\pm$0.10  &  M  \\
 Ter8\_6	 &  4956$\pm$337   &    5250$\pm$100    &  4978$\pm$100   &    2.1$\pm$0.8    &    2.6$\pm$0.4     &   2.50$\pm$0.39    &  -1.9$\pm$0.4       &   -2.00$\pm$0.15    &    -1.98$\pm$0.14  &    0.40$\pm$0.17     &   0.15$\pm$0.13  &  M  \\
 Ter8\_8	 &  4947$\pm$333   &    5175$\pm$114  &  5151$\pm$108   &    2.1$\pm$0.8    &    2.7$\pm$0.4     &   2.56$\pm$0.36    &  -1.9$\pm$0.4       &   -2.15$\pm$0.23    &    -2.06$\pm$0.19  &    0.40$\pm$0.18     &   0.19$\pm$0.15  &  M  \\
 Ter8\_9	 &  4694$\pm$293   &    4549$\pm$99   &  4564$\pm$ 94   &    1.6$\pm$0.5    &    0.9$\pm$0.4     &   1.12$\pm$0.30    &  -1.80$\pm$0.33     &   -2.40$\pm$0.20    &    -2.24$\pm$0.17  &    0.42$\pm$0.13     &   0.19$\pm$0.14  &  M  \\
Ter8\_10	 &  4849$\pm$195   &    4825$\pm$114  &  4831$\pm$ 99   &    3.5$\pm$0.5    &    4.30$\pm$0.25   &   4.13$\pm$0.22    &  0.09$\pm$0.09      &   -0.65$\pm$0.23    &    -0.01$\pm$0.08  &    -0.02$\pm$0.03    &   0.23$\pm$0.13  &    \\
Ter8\_11	 &  4655$\pm$247   &    4951$\pm$100  &  4909$\pm$ 92   &    1.5$\pm$0.5    &    2.05$\pm$0.35   &   1.84$\pm$0.28    &  -1.90$\pm$0.25     &   -2.00$\pm$0.15    &    -1.97$\pm$0.13  &    0.45$\pm$0.12     &   0.16$\pm$0.14  &  M  \\
Ter8\_13	 &  4967$\pm$367   &    5250$\pm$100    &  5108$\pm$100   &    2.1$\pm$0.8    &    2.4$\pm$0.4     &   2.34$\pm$0.34    &  -1.8$\pm$0.4       &   -2.00$\pm$0.15    &    -1.98$\pm$0.14  &    0.41$\pm$0.17     &   0.22$\pm$0.12  &  M  \\
Ter8\_14	 &  4756$\pm$142   &    4324$\pm$114  &  4493$\pm$ 89   &    1.8$\pm$0.4    &    0.4$\pm$0.4     &   1.16$\pm$0.28    &  -1.68$\pm$0.11     &   -2.50$\pm$0.15    &    -1.98$\pm$0.09  &    0.43$\pm$0.14     &   0.19$\pm$0.14  &  M  \\
Ter8\_15	 &  4984$\pm$344   &    5225$\pm$75   &  5214$\pm$ 73   &    2.1$\pm$0.8    &    2.20$\pm$0.25   &   2.19$\pm$0.24    &  -1.8$\pm$0.4       &   -2.00$\pm$0.15    &    -1.97$\pm$0.14  &    0.39$\pm$0.17     &   0.19$\pm$0.15  &  M  \\
Ter8\_16	 &  4871$\pm$333   &    5050$\pm$100  &  5035$\pm$ 96   &    1.9$\pm$0.6    &    2.20$\pm$0.33   &   2.14$\pm$0.29    &  -1.75$\pm$0.34     &   -2.00$\pm$0.15    &    -1.96$\pm$0.14  &    0.42$\pm$0.12     &   0.21$\pm$0.14  &  M  \\
Ter8\_18	 &  4722$\pm$195   &    4700$\pm$245  &  4713$\pm$153   &    1.7$\pm$0.5    &    1.6$\pm$0.5     &   1.68$\pm$0.38    &  -1.74$\pm$0.30     &   -2.30$\pm$0.24    &    -2.08$\pm$0.19  &    0.37$\pm$0.13     &   0.27$\pm$0.09  &  M  \\ 
\noalign{\smallskip}
\end{longtable}
\tablefoot{ 
\tablefoottext{a}{MILES library}
\tablefoottext{b}{COELHO library}
\tablefoottext{avg}{Average of MILES and COELHO results}
}
\end{landscape}
}% End \longtabL

\end{document}